\documentstyle[11pt,axodraw]{article}

\newlength{\dinwidth} 
\newlength{\dinmargin}
\def\sign(#1){(\!-\!1)^{#1}}
\def\binom(#1,#2){ (\!\!
         \begin{array}{c} #1 \\ #2 \end{array}\!\! ) }
\def\plus{\!+\!}
\def\minus{\!-\!}
\def\mydot{\!\cdot\!}
\def\nn{\nonumber \\ &&}

\def\TA(#1,#2,#3,#4,#5,#6){
        \raisebox{-19.1pt}{
        \SetScale{0.5} \SetPFont{Helvetica}{14}
        \hspace{-15pt}
        \begin{picture}(50,39)(0,-4)
        \SetColor{Blue}
        \CArc(40,35)(25,90,270) \CArc(60,35)(25,270,90)
        \Line(40,60)(60,60) \Line(40,10)(60,10) \Line(50,10)(50,60)
        \Line(0,35)(15,35) \Line(85,35)(100,35)
        \SetColor{Black}
        \PText(53,39)(0)[lb]{#5} \PText(53,36)(0)[lt]{#6}
        \PText(35,62)(0)[rb]{#1} \PText(65,62)(0)[lb]{#2}
        \PText(65,12)(0)[lt]{#3} \PText(35,12)(0)[rt]{#4}
        \end{picture}
        \SetScale{1.0}
        \hspace{-7pt}
        }
}
\def\TB(#1,#2,#3,#4,#5){
        \raisebox{-19.1pt}{
        \SetScale{0.5} \SetPFont{Helvetica}{14}
        \hspace{-15pt}
        \begin{picture}(50,39)(0,-4)
        \SetColor{Blue}
        \CArc(40,35)(25,90,270) \CArc(60,35)(25,270,90)
        \Line(40,60)(60,60) \Line(40,10)(60,10) \Line(50,10)(50,60)
        \Line(0,35)(15,35) \Line(85,35)(100,35)
        \SetColor{Black}
        \PText(53,40)(0)[l]{#5}
        \PText(35,62)(0)[rb]{#1} \PText(65,62)(0)[lb]{#2}
        \PText(65,12)(0)[lt]{#3} \PText(35,12)(0)[rt]{#4}
        \end{picture}
        \SetScale{1.0}
        \hspace{-7pt}
        }
}
\def\TAA(#1,#2,#3,#4,#5,#6){
        \raisebox{-19.1pt}{
        \SetScale{0.5} \SetPFont{Helvetica}{14}
        \hspace{-15pt}
        \begin{picture}(50,39)(0,-4)
        \SetColor{Blue}
        \CArc(40,35)(25,90,270) \CArc(60,35)(25,270,90)
        \Line(40,60)(60,60) \Line(40,10)(60,10) \Line(50,10)(50,60)
        \Line(0,35)(15,35) \Line(85,35)(100,35)
        \SetColor{Black}
        \PText(53,39)(0)[lb]{#5} \PText(53,36)(0)[lt]{#6}
        \PText(35,62)(0)[rb]{#1} \PText(65,62)(0)[lb]{#2}
        \PText(65,12)(0)[lt]{#3} \PText(35,12)(0)[rt]{#4}
        \SetColor{Red}
        \SetWidth{3}
                \Line(50,35)(50,60)
                \Line(40,60)(50,60)
                \CArc(40,35)(25,90,180)
                \Vertex(50,60){1.3}
        \SetWidth{0.5}
        \end{picture}
        \SetScale{1.0}
        \hspace{-7pt}
        }
}
\def\TAB(#1,#2,#3,#4,#5,#6){
        \raisebox{-19.1pt}{
        \SetScale{0.5} \SetPFont{Helvetica}{14}
        \hspace{-15pt}
        \begin{picture}(50,39)(0,-4)
        \SetColor{Blue}
        \CArc(40,35)(25,90,270) \CArc(60,35)(25,270,90)
        \Line(40,60)(60,60) \Line(40,10)(60,10) \Line(50,10)(50,60)
        \Line(0,35)(15,35) \Line(85,35)(100,35)
        \SetColor{Black}
        \PText(53,39)(0)[lb]{#5} \PText(53,36)(0)[lt]{#6}
        \PText(35,62)(0)[rb]{#1} \PText(65,62)(0)[lb]{#2}
        \PText(65,12)(0)[lt]{#3} \PText(35,12)(0)[rt]{#4}
        \SetColor{Red}
        \SetWidth{3}
                \Line(50,10)(50,60)
                \Vertex(50,60){1.3}
                \Line(40,60)(50,60)
                \CArc(40,35)(25,90,180)
        \SetWidth{0.5}
        \end{picture}
        \SetScale{1.0}
        \hspace{-7pt}
        }
}
\def\TABs(#1,#2,#3,#4,#5){
        \raisebox{-19.1pt}{
        \SetScale{0.5} \SetPFont{Helvetica}{14}
        \hspace{-15pt}
        \begin{picture}(50,39)(0,-4)
        \SetColor{Blue}
        \CArc(40,35)(25,90,270) \CArc(60,35)(25,270,90)
        \Line(40,60)(60,60) \Line(40,10)(60,10) \Line(50,10)(50,60)
        \Line(0,35)(15,35) \Line(85,35)(100,35)
        \SetColor{Black}
        \PText(53,38)(0)[l]{#5}
        \PText(35,62)(0)[rb]{#1} \PText(65,62)(0)[lb]{#2}
        \PText(65,12)(0)[lt]{#3} \PText(35,12)(0)[rt]{#4}
        \SetColor{Red}
        \SetWidth{3}
                \Line(50,10)(50,60)
                \Vertex(50,60){1.3}
                \Line(40,60)(50,60)
                \CArc(40,35)(25,90,180)
        \SetWidth{0.5}
        \end{picture}
        \SetScale{1.0}
        \hspace{-7pt}
        }
}
\def\TAC(#1,#2,#3,#4,#5,#6){
        \raisebox{-19.1pt}{
        \SetScale{0.5} \SetPFont{Helvetica}{14}
        \hspace{-15pt}
        \begin{picture}(50,39)(0,-4)
        \SetColor{Blue}
        \CArc(40,35)(25,90,270) \CArc(60,35)(25,270,90)
        \Line(40,60)(60,60) \Line(40,10)(60,10) \Line(50,10)(50,60)
        \Line(0,35)(15,35) \Line(85,35)(100,35)
        \SetColor{Black}
        \PText(53,39)(0)[lb]{#5} \PText(53,36)(0)[lt]{#6}
        \PText(35,62)(0)[rb]{#1} \PText(65,62)(0)[lb]{#2}
        \PText(65,12)(0)[lt]{#3} \PText(35,12)(0)[rt]{#4}
        \SetColor{Red}
        \SetWidth{3}
                \Line(40,60)(50,60)
                \CArc(40,35)(25,90,180)
        \SetWidth{0.5}
        \end{picture}
        \SetScale{1.0}
        \hspace{-7pt}
        }
}
\def\TACs(#1,#2,#3,#4,#5){
        \raisebox{-19.1pt}{
        \SetScale{0.5} \SetPFont{Helvetica}{14}
        \hspace{-15pt}
        \begin{picture}(50,39)(0,-4)
        \SetColor{Blue}
        \CArc(40,35)(25,90,270) \CArc(60,35)(25,270,90)
        \Line(40,60)(60,60) \Line(40,10)(60,10) \Line(50,10)(50,60)
        \Line(0,35)(15,35) \Line(85,35)(100,35)
        \SetColor{Black}
        \PText(53,38)(0)[l]{#5}
        \PText(35,62)(0)[rb]{#1} \PText(65,62)(0)[lb]{#2}
        \PText(65,12)(0)[lt]{#3} \PText(35,12)(0)[rt]{#4}
        \SetColor{Red}
        \SetWidth{3}
                \Line(40,60)(50,60)
                \CArc(40,35)(25,90,180)
        \SetWidth{0.5}
        \end{picture}
        \SetScale{1.0}
        \hspace{-7pt}
        }
}
\def\TACR(#1,#2,#3,#4,#5,#6){
        \raisebox{-19.1pt}{
        \SetScale{0.5} \SetPFont{Helvetica}{14}
        \hspace{-15pt}
        \begin{picture}(50,39)(0,-4)
        \SetColor{Blue}
        \CArc(40,35)(25,90,270) \CArc(60,35)(25,270,90)
        \Line(40,60)(60,60) \Line(40,10)(60,10) \Line(50,10)(50,60)
        \Line(0,35)(15,35) \Line(85,35)(100,35)
        \SetColor{Black}
        \PText(53,39)(0)[lb]{#5} \PText(53,36)(0)[lt]{#6}
        \PText(35,62)(0)[rb]{#1} \PText(65,62)(0)[lb]{#2}
        \PText(65,12)(0)[lt]{#3} \PText(35,12)(0)[rt]{#4}
        \SetColor{Red}
        \SetWidth{3}
                \Line(50,60)(60,60)
                \CArc(60,35)(25,0,90)
        \SetWidth{0.5}
        \end{picture}
        \SetScale{1.0}
        \hspace{-7pt}
        }
}
\def\TAD(#1,#2,#3,#4,#5,#6){
        \raisebox{-19.1pt}{
        \SetScale{0.5} \SetPFont{Helvetica}{14}
        \hspace{-15pt}
        \begin{picture}(50,39)(0,-4)
        \SetColor{Blue}
        \CArc(40,35)(25,90,270) \CArc(60,35)(25,270,90)
        \Line(40,60)(60,60) \Line(40,10)(60,10) \Line(50,10)(50,60)
        \Line(0,35)(15,35) \Line(85,35)(100,35)
        \SetColor{Black}
        \PText(53,39)(0)[lb]{#5} \PText(53,36)(0)[lt]{#6}
        \PText(35,62)(0)[rb]{#1} \PText(65,62)(0)[lb]{#2}
        \PText(65,12)(0)[lt]{#3} \PText(35,12)(0)[rt]{#4}
        \SetColor{Red}
        \SetWidth{3}
                \Line(50,10)(50,60)
        \SetWidth{0.5}
        \end{picture}
        \SetScale{1.0}
        \hspace{-7pt}
        }
}
\def\TAE(#1,#2,#3,#4,#5,#6,#7){
        \raisebox{-19.1pt}{
        \SetScale{0.5} \SetPFont{Helvetica}{14}
        \hspace{-15pt}
        \begin{picture}(50,39)(0,-4)
        \SetColor{Blue}
        \CArc(40,35)(25,90,270) \CArc(60,35)(25,270,90)
        \Line(40,60)(60,60) \Line(40,10)(60,10) \Line(50,10)(50,60)
        \Line(0,35)(15,35) \Line(85,35)(100,35)
        \SetColor{Black}
        \PText(53,39)(0)[lb]{#5} \PText(53,36)(0)[lt]{#6}
        \PText(40,65)(0)[rb]{#1} \PText(15,48)(0)[rb]{#7} \PText(65,62)(0)[lb]{#2}
        \PText(65,12)(0)[lt]{#3} \PText(35,12)(0)[rt]{#4}
        \SetColor{Red}
        \SetWidth{3}
                \Line(50,35)(50,60)
                \Line(40,60)(50,60)
                \CArc(40,35)(25,90,130)
                \Vertex(50,60){1.3}
        \SetWidth{0.5}
        \end{picture}
        \SetScale{1.0}
        \hspace{-7pt}
        }
}
\def\TAF(#1,#2,#3,#4,#5,#6,#7){
        \raisebox{-19.1pt}{
        \SetScale{0.5} \SetPFont{Helvetica}{14}
        \hspace{-15pt}
        \begin{picture}(50,39)(0,-4)
        \SetColor{Blue}
        \CArc(40,35)(25,90,270) \CArc(60,35)(25,270,90)
        \Line(40,60)(60,60) \Line(40,10)(60,10) \Line(50,10)(50,60)
        \Line(0,35)(15,35) \Line(85,35)(100,35)
        \SetColor{Black}
        \PText(53,39)(0)[lb]{#5} \PText(53,36)(0)[lt]{#6}
        \PText(40,65)(0)[rb]{#1} \PText(15,48)(0)[rb]{#7} \PText(65,62)(0)[lb]{#2}
        \PText(65,12)(0)[lt]{#3} \PText(35,12)(0)[rt]{#4}
        \SetColor{Red}
        \SetWidth{3}
                \Line(50,10)(50,60)
                \Line(40,60)(50,60)
                \CArc(40,35)(25,90,130)
                \Vertex(50,60){1.3}
        \SetWidth{0.5}
        \end{picture}
        \SetScale{1.0}
        \hspace{-7pt}
        }
}
\def\TAFs(#1,#2,#3,#4,#5,#6){
        \raisebox{-19.1pt}{
        \SetScale{0.5} \SetPFont{Helvetica}{14}
        \hspace{-15pt}
        \begin{picture}(50,39)(0,-4)
        \SetColor{Blue}
        \CArc(40,35)(25,90,270) \CArc(60,35)(25,270,90)
        \Line(40,60)(60,60) \Line(40,10)(60,10) \Line(50,10)(50,60)
        \Line(0,35)(15,35) \Line(85,35)(100,35)
        \SetColor{Black}
        \PText(53,38)(0)[l]{#5}
        \PText(40,65)(0)[rb]{#1} \PText(15,48)(0)[rb]{#6} \PText(65,62)(0)[lb]{#2}
        \PText(65,12)(0)[lt]{#3} \PText(35,12)(0)[rt]{#4}
        \SetColor{Red}
        \SetWidth{3}
                \Line(50,10)(50,60)
                \Line(40,60)(50,60)
                \CArc(40,35)(25,90,130)
                \Vertex(50,60){1.3}
        \SetWidth{0.5}
        \end{picture}
        \SetScale{1.0}
        \hspace{-7pt}
        }
}
\def\TAG(#1,#2,#3,#4,#5,#6,#7){
        \raisebox{-19.1pt}{
        \SetScale{0.5} \SetPFont{Helvetica}{14}
        \hspace{-15pt}
        \begin{picture}(50,39)(0,-4)
        \SetColor{Blue}
        \CArc(40,35)(25,90,270) \CArc(60,35)(25,270,90)
        \Line(40,60)(60,60) \Line(40,10)(60,10) \Line(50,10)(50,60)
        \Line(0,35)(15,35) \Line(85,35)(100,35)
        \SetColor{Black}
        \PText(53,39)(0)[lb]{#5} \PText(53,36)(0)[lt]{#6}
        \PText(40,65)(0)[rb]{#1} \PText(15,48)(0)[rb]{#7} \PText(65,62)(0)[lb]{#2}
        \PText(65,12)(0)[lt]{#3} \PText(35,12)(0)[rt]{#4}
        \SetColor{Red}
        \SetWidth{3}
                \Line(40,60)(50,60)
                \CArc(40,35)(25,90,130)
                \Vertex(50,60){1.3}
        \SetWidth{0.5}
        \end{picture}
        \SetScale{1.0}
        \hspace{-7pt}
        }
}
\def\TAH(#1,#2,#3,#4,#5,#6,#7){
        \raisebox{-19.1pt}{
        \SetScale{0.5} \SetPFont{Helvetica}{14}
        \hspace{-15pt}
        \begin{picture}(50,39)(0,-4)
        \SetColor{Blue}
        \CArc(40,35)(25,90,270) \CArc(60,35)(25,270,90)
        \Line(40,60)(60,60) \Line(40,10)(60,10) \Line(50,10)(50,60)
        \Line(0,35)(15,35) \Line(85,35)(100,35)
        \SetColor{Black}
        \PText(40,65)(0)[rb]{#1}
        \PText(60,65)(0)[lb]{#2}
        \PText(65,12)(0)[lt]{#3}
        \PText(35,12)(0)[rt]{#4}
        \PText(53,40)(0)[l]{#5}
        \PText(15,48)(0)[rb]{#6}
        \PText(85,48)(0)[lb]{#7}
        \SetColor{Red}
        \SetWidth{3}
                \Line(40,60)(60,60)
                \CArc(40,35)(25,90,130)
                \CArc(60,35)(25,50,90)
        \SetWidth{0.5}
        \end{picture}
        \SetScale{1.0}
        \hspace{-7pt}
        }
}
\def\TAI(#1,#2,#3,#4,#5,#6,#7,#8){
        \raisebox{-19.1pt}{
        \SetScale{0.5} \SetPFont{Helvetica}{14}
        \hspace{-15pt}
        \begin{picture}(50,39)(0,-4)
        \SetColor{Blue}
        \CArc(40,35)(25,90,270) \CArc(60,35)(25,270,90)
        \Line(40,60)(60,60) \Line(40,10)(60,10) \Line(50,10)(50,60)
        \Line(0,35)(15,35) \Line(85,35)(100,35)
        \SetColor{Black}
        \PText(40,65)(0)[rb]{#1}
        \PText(60,65)(0)[lb]{#2}
        \PText(65,12)(0)[lt]{#3}
        \PText(35,12)(0)[rt]{#4}
        \PText(53,39)(0)[lb]{#5} \PText(53,36)(0)[lt]{#6}
        \PText(15,48)(0)[rb]{#7}
        \PText(85,48)(0)[lb]{#8}
        \SetColor{Red}
        \SetWidth{3}
                \Line(40,60)(60,60)
                \CArc(40,35)(25,90,130)
                \CArc(60,35)(25,50,90)
        \SetWidth{0.5}
        \end{picture}
        \SetScale{1.0}
        \hspace{-7pt}
        }
}
\def\TAIs(#1,#2,#3,#4,#5,#6,#7){
        \raisebox{-19.1pt}{
        \SetScale{0.5} \SetPFont{Helvetica}{14}
        \hspace{-15pt}
        \begin{picture}(50,39)(0,-4)
        \SetColor{Blue}
        \CArc(40,35)(25,90,270) \CArc(60,35)(25,270,90)
        \Line(40,60)(60,60) \Line(40,10)(60,10) \Line(50,10)(50,60)
        \Line(0,35)(15,35) \Line(85,35)(100,35)
        \SetColor{Black}
        \PText(40,65)(0)[rb]{#1}
        \PText(60,65)(0)[lb]{#2}
        \PText(65,12)(0)[lt]{#3}
        \PText(35,12)(0)[rt]{#4}
        \PText(53,38)(0)[l]{#5}
        \PText(15,48)(0)[rb]{#6}
        \PText(85,48)(0)[lb]{#7}
        \SetColor{Red}
        \SetWidth{3}
                \Line(40,60)(60,60)
                \CArc(40,35)(25,90,130)
                \CArc(60,35)(25,50,90)
        \SetWidth{0.5}
        \end{picture}
        \SetScale{1.0}
        \hspace{-7pt}
        }
}
\def\TAJ(#1,#2,#3,#4,#5,#6,#7){
        \raisebox{-19.1pt}{
        \SetScale{0.5} \SetPFont{Helvetica}{14}
        \hspace{-15pt}
        \begin{picture}(50,39)(0,-4)
        \SetColor{Blue}
        \CArc(40,35)(25,90,270) \CArc(60,35)(25,270,90)
        \Line(40,60)(60,60) \Line(40,10)(60,10) \Line(50,10)(50,60)
        \Line(0,35)(15,35) \Line(85,35)(100,35)
        \SetColor{Black}
        \PText(35,62)(0)[rb]{#1}
        \PText(53,39)(0)[lb]{#5} \PText(53,36)(0)[lt]{#6}
        \PText(60,65)(0)[lb]{#2}
        \PText(85,48)(0)[lb]{#7}
        \PText(65,12)(0)[lt]{#3} \PText(35,12)(0)[rt]{#4}
        \SetColor{Red}
        \SetWidth{3}
                \Line(50,60)(60,60)
                \CArc(60,35)(25,50,90)
        \SetWidth{0.5}
        \end{picture}
        \SetScale{1.0}
        \hspace{-7pt}
        }
}
\def\TAK(#1,#2,#3,#4,#5,#6,#7){
        \raisebox{-19.1pt}{
        \SetScale{0.5} \SetPFont{Helvetica}{14}
        \hspace{-15pt}
        \begin{picture}(50,39)(0,-4)
        \SetColor{Blue}
        \CArc(40,35)(25,90,270) \CArc(60,35)(25,270,90)
        \Line(40,60)(60,60) \Line(40,10)(60,10) \Line(50,10)(50,60)
        \Line(0,35)(15,35) \Line(85,35)(100,35)
        \SetColor{Black}
        \PText(35,62)(0)[rb]{#1}
        \PText(53,39)(0)[lb]{#5} \PText(53,36)(0)[lt]{#6}
        \PText(60,65)(0)[lb]{#2}
        \PText(85,48)(0)[lb]{#7}
        \PText(65,12)(0)[lt]{#3} \PText(35,12)(0)[rt]{#4}
        \SetColor{Red}
        \SetWidth{3}
                \Line(40,60)(60,60)
                \CArc(40,35)(25,90,180)
                \CArc(60,35)(25,50,90)
        \SetWidth{0.5}
        \end{picture}
        \SetScale{1.0}
        \hspace{-7pt}
        }
}
\def\TAKs(#1,#2,#3,#4,#5,#6){
        \raisebox{-19.1pt}{
        \SetScale{0.5} \SetPFont{Helvetica}{14}
        \hspace{-15pt}
        \begin{picture}(50,39)(0,-4)
        \SetColor{Blue}
        \CArc(40,35)(25,90,270) \CArc(60,35)(25,270,90)
        \Line(40,60)(60,60) \Line(40,10)(60,10) \Line(50,10)(50,60)
        \Line(0,35)(15,35) \Line(85,35)(100,35)
        \SetColor{Black}
        \PText(35,62)(0)[rb]{#1}
        \PText(53,38)(0)[l]{#5}
        \PText(60,65)(0)[lb]{#2}
        \PText(85,48)(0)[lb]{#6}
        \PText(65,12)(0)[lt]{#3} \PText(35,12)(0)[rt]{#4}
        \SetColor{Red}
        \SetWidth{3}
                \Line(40,60)(60,60)
                \CArc(40,35)(25,90,180)
                \CArc(60,35)(25,50,90)
        \SetWidth{0.5}
        \end{picture}
        \SetScale{1.0}
        \hspace{-7pt}
        }
}
\def\TAKR(#1,#2,#3,#4,#5,#6,#7){
        \raisebox{-19.1pt}{
        \SetScale{0.5} \SetPFont{Helvetica}{14}
        \hspace{-15pt}
        \begin{picture}(50,39)(0,-4)
        \SetColor{Blue}
        \CArc(40,35)(25,90,270) \CArc(60,35)(25,270,90)
        \Line(40,60)(60,60) \Line(40,10)(60,10) \Line(50,10)(50,60)
        \Line(0,35)(15,35) \Line(85,35)(100,35)
        \SetColor{Black}
        \PText(40,65)(0)[rb]{#1}
        \PText(15,48)(0)[rb]{#7}
        \PText(53,39)(0)[lb]{#5} \PText(53,36)(0)[lt]{#6}
        \PText(65,62)(0)[lb]{#2}
        \PText(65,12)(0)[lt]{#3} \PText(35,12)(0)[rt]{#4}
        \SetColor{Red}
        \SetWidth{3}
                \Line(40,60)(60,60)
                \CArc(40,35)(25,90,130)
                \CArc(60,35)(25,0,90)
        \SetWidth{0.5}
        \end{picture}
        \SetScale{1.0}
        \hspace{-7pt}
        }
}
\def\TAL(#1,#2,#3,#4,#5,#6,#7){
        \raisebox{-19.1pt}{
        \SetScale{0.5} \SetPFont{Helvetica}{14}
        \hspace{-15pt}
        \begin{picture}(50,39)(0,-4)
        \SetColor{Blue}
        \CArc(40,35)(25,90,270) \CArc(60,35)(25,270,90)
        \Line(40,60)(60,60) \Line(40,10)(60,10) \Line(50,10)(50,60)
        \Line(0,35)(15,35) \Line(85,35)(100,35)
        \SetColor{Black}
        \PText(53,40)(0)[l]{#5}
        \PText(40,65)(0)[rb]{#1} \PText(15,48)(0)[rb]{#6}
        \PText(65,62)(0)[lb]{#2}
        \PText(60,9)(0)[lt]{#3} \PText(85,26)(0)[lt]{#7}
        \PText(35,12)(0)[rt]{#4}
        \SetColor{Red}
        \SetWidth{3}
                \Line(50,10)(50,60)
                \Line(40,60)(50,60)
                \CArc(40,35)(25,90,130)
                \Vertex(50,60){1.3}
                \Line(50,10)(60,10)
                \CArc(60,35)(25,270,310)
                \Vertex(50,10){1.3}
        \SetWidth{0.5}
        \end{picture}
        \SetScale{1.0}
        \hspace{-7pt}
        }
}
\def\TAbasic(#1,#2,#3,#4,#5,#6){
        \raisebox{-19.1pt}{
        \SetScale{0.5} \SetPFont{Helvetica}{14}
        \hspace{-15pt}
        \begin{picture}(50,39)(0,-4)
        \SetColor{Blue}
        \CArc(40,35)(25,90,270) \CArc(60,35)(25,270,90)
        \Line(40,60)(60,60) \Line(40,10)(60,10) \Line(50,10)(50,60)
        \Line(0,35)(15,35) \Line(85,35)(100,35)
        \SetColor{Black}
        \PText(40,65)(0)[rb]{#1}
        \PText(65,12)(0)[lt]{#2}
        \PText(35,12)(0)[rt]{#3}
        \PText(53,40)(0)[l]{#4}
        \PText(15,48)(0)[rb]{#5}
        \PText(65,62)(0)[lb]{#6}
        \SetColor{Red}
        \SetWidth{3}
                \Line(40,60)(50,60)
                \CArc(40,35)(25,90,130)
        \SetWidth{0.5}
        \end{picture}
        \SetScale{1.0}
        \hspace{-7pt}
        }
}
\def\TEbasic(#1,#2,#3,#4,#5,#6){
        \raisebox{-19.1pt}{
        \SetScale{0.5} \SetPFont{Helvetica}{14}
        \hspace{-15pt}
        \begin{picture}(50,39)(0,-4)
        \SetColor{Blue}
        \CArc(40,35)(25,90,270) \CArc(60,35)(25,270,90)
        \Line(40,60)(60,60) \Line(40,10)(60,10) \Line(50,10)(50,60)
        \Line(0,35)(15,35) \Line(85,35)(100,35)
        \SetColor{Black}
        \PText(53,39)(0)[lb]{#4} \PText(53,36)(0)[lt]{#5}
        \PText(35,62)(0)[rb]{#6} \PText(65,62)(0)[lb]{#1}
        \PText(65,12)(0)[lt]{#2} \PText(35,12)(0)[rt]{#3}
        \SetColor{Red}
        \SetWidth{3}
                \Line(50,35)(50,60)
        \SetWidth{0.5}
        \end{picture}
        \SetScale{1.0}
        \hspace{-7pt}
        }
}
\def\TCa(#1,#2,#3,#4){
        \raisebox{-19.1pt}{
        \SetScale{0.5} \SetPFont{Helvetica}{14}
        \hspace{-15pt}
        \begin{picture}(80,39)(0,-4)
        \SetColor{Blue}
        \CArc(40,35)(25,90,270) \CArc(120,35)(25,270,90)
        \Line(40,60)(120,60) \Line(40,10)(120,10)
        \Line(50,60)(70,30) \Line(110,60)(90,30)
        \Line(0,35)(15,35) \Line(145,35)(160,35)
        \SetColor{Black}
        \PText(40,65)(0)[rb]{#2}
        \PText(120,65)(0)[lb]{#3}
        \PText(15,48)(0)[rb]{#1}
        \PText(145,48)(0)[lb]{#4}
        \SetColor{Red}
        \SetWidth{3}
                \Line(40,60)(120,60)
                \CArc(40,35)(25,90,130)
                \CArc(120,35)(25,50,90)
        \SetWidth{0.5}
        \SetColor{Blue}
        \GOval(80,35)(35,20)(0){0.75}
        \end{picture}
        \SetScale{1.0}
        \hspace{-7pt}
        }
}

\def\a{\alpha} \def\b{\beta}   \def\c{\chi} \def\d{\delta} \def\e{\epsilon} 
\def\f{\phi}  \def\g{\gamma}  \def\j{\psi}  \def\m{\mu}  
\def\n{\nu} \def\p{\pi} \def\q{\theta}  \def\r{\rho} \def\s{\sigma} 
\def\t{\tau} \def\x{\xi} \def\z{\zeta}   
\def\ve{\epsilon} 

\def\co{{\cal O}} \def\cp{{\cal P}}

\def\slash#1{\rlap{\hbox{$\mskip 1 mu /$}}#1}  
\def\Bar#1{\overline{#1}} 
\def\Li{\hbox{Li}}

\begin{document}
\setcounter{page}{0}
\thispagestyle{empty}
\rightline{NIKHEF 99-030}
\rightline{December 1999}
\vspace{20mm}

\begin{center}
{\LARGE\bf\sc
Deep Inelastic Structure Functions at two Loops}
\end{center}
\vspace{5mm}
\begin{center}
{\large S. Moch and J.A.M. Vermaseren}\\[1ex]
{\it NIKHEF Theory Group\\
        Kruislaan 409, 1098 SJ Amsterdam, The Netherlands} \\
\end{center}
\vspace{5mm}

\begin{abstract}
We present the analytic calculation of the Mellin moments of the
structure functions $F_2$, $F_3$ and $F_L$ in perturbative QCD
up to second order corrections and in leading twist approximation.
We calculate the 2-loop contributions to the anomalous dimensions
of the singlet and non-singlet operator matrix elements and the
2-loop coefficient functions of $F_2$, $F_3$ and $F_L$.
We perform the inverse Mellin transformation analytically and find
our results in agreement with earlier calculations 
in the literature by Zijlstra and van Neerven.
\end{abstract}

\newpage
\section{Introduction}

As one of the best studied reactions today, 
deep inelastic lepton-hadron scattering establishes the scale evolution
of the structure functions, one of the most important precision tests
of perturbative QCD. 
It provides unique information about the deep structure of the hadrons and 
most importantly, 
measurements of these structure functions in deep inelastic scattering (DIS) 
allow to extract the parton densities, 
which are subsequently used as input for many other hard scattering processes.

Over the years, the ever increasing accuracy of deep inelastic and 
other hard scattering experiments has created a steady demand for more 
accurate theoretical predictions. 
The qualitative quest for understanding scaling violations  
of the structure functions in terms of asymptotic freedom in QCD~\cite{Gross:1973rr} 
soon changed to the quantitative task of reducing 
theoretical uncertainties due to higher order QCD corrections 
and the necessary calculations to obtain 
next-to-leading order (NLO) perturbative QCD predictions were performed in 
refs.\cite{Bardeen:1978yd,Floratos:1977au,Gonzalez-Arroyo:1979df,
Curci:1980uw,Furmanski:1980cm,Hamberg:1992qt,Ellis:1996nn,
Duke:1982ga,Kazakov:1988jk,Kazakov:1990fu,SanchezGuillen:1991iq}.

Today, high precision analyses of experimental data, 
such as the determination of the strong coupling constant $\alpha_s$ 
and the parton densities call for 
complete next-to-next-to-leading order (NNLO) perturbative QCD predictions. 
This has motivated the calculation of the 2-loop coefficient functions 
of all DIS structure functions by Zijlstra and 
van Neerven~\cite{vanNeerven:1991nn,Zijlstra:1991qc,Zijlstra:1992kj}  
and they could check their results for a number of fixed 
Mellin moments~\cite{Larin:1993fv}. 
However, to complete the NNLO analysis of DIS structure functions 
one still has to know the perturbative QCD predictions at 3-loops 
for the anomalous dimensions of these structure functions 
and also the coefficient functions for the longitudinal structure function entering 
in the ratio of the longitudinal over the transverse cross section, $R = \s_L / \s_T$.
These quantities are still unknown, 
except for a small number of fixed Mellin moments~\cite{Larin:1991zw,Larin:1991tj,Larin:1997wd}, 
some of them related to sum rules.
The results of refs.\cite{Larin:1997wd}, 
have already been used in NNLO analyses~\cite{Larin:1997wd,Kataev:1997nc}
Unfortunately, this limited information about some fixed Mellin moments 
is generally not sufficient to allow for NNLO analyses of all data of 
DIS and related hard scattering experiments, such as the Drell-Yan process.
As a consequence, there are still considerable uncertainties 
on the parton densities, a prominent example being the gluon density at small $x$. 

The aim of this paper is first of all to provide an independent check on the results of 
refs.\cite{vanNeerven:1991nn,Zijlstra:1991qc,Zijlstra:1992kj}, 
by means of a completely different method. 
At the same time, we also wish to demonstrate the power of our method, 
which we believe to be flexible enough and most promising in view of the 
ultimate challenge, the calculation of the anomalous dimensions at 3-loops. 
Our approach, that actually dates back to the origins of 
QCD~\cite{Gross:1973rr,Gonzalez-Arroyo:1979df} is to calculate the Mellin moments of the 
DIS structure functions analytically as a general function of $N$. 
This idea was further pioneered by Kazakov and Kotikov~\cite{Kazakov:1988jk}
to obtain the longitudinal structure function $F_L$ at two loops. 
In the present paper, we calculate  in this way the Mellin moments of all unpolarized 
DIS structure functions $F_2$, $F_3$ and $F_L$ up to two loops. 
Subsequently, we perform the inverse Mellin transformation 
to express our results in momentum space as functions of $x$ to compare with 
refs.\cite{vanNeerven:1991nn,Zijlstra:1991qc,Zijlstra:1992kj}.

A remark about other strategies towards completing the NNLO perturbative 
QCD predictions for DIS structure functions is in order here.
A straightforward extension of the methods of ref.\cite{Larin:1997wd} 
to simply calculate sufficiently more fixed Mellin moments 
for a combined NNLO analyses of all data of, say, 
DIS and Drell-Yan experiments is not feasible. 
One has to know the Mellin moments at least up to $N=100$ which would 
result in analytical expressions consisting of huge rational numbers. 
Unfortunately, this is beyond the capabilities of present computer algebra programs.
A different approach is the direct calculation of the anomalous dimensions of 
DIS operator matrix elements, 
extending the work of ref.\cite{Curci:1980uw,Furmanski:1980cm} to 3-loops.
First steps in this direction have been achieved in ref.\cite{Matiounine:1998ky},
where the finite terms of DIS operator matrix elements at two loops have been 
calculated.

The outline of this paper is as follows. 
In section~\ref{sec:Formalism} we briefly discuss 
the operator product expansion (OPE) and some issues of the renormalization procedure. 
Section~\ref{sec:Method} gives a detailed explanation of the method 
to calculate the Mellin moments as an analytical function of $N$ 
for a given diagram. It also contains a short summary of properties of 
harmonic sums~\cite{Gonzalez-Arroyo:1979df,Vermaseren:1998uu,Blumlein:1998if} 
and harmonic polylogarithms~\cite{Remiddi:1999ew}.
Section~\ref{sec:CalculationResults} describes our calculation and lists our results 
for the structure functions $F_2$, $F_3$ and $F_L$ up to two loops, 
both in Mellin space as a function of $N$ and in momentum space as a function of $x$ 
expressed in terms of harmonic polylogarithms. 
Finally, section~\ref{sec:Conclusions} gives our conclusions.
The appendices contain some relations between the standard polylogarithms and 
Nielsen functions and the harmonic polylogarithms and all explicit expressions 
for the 2-loop coefficient functions.

\section{Formalism
\label{sec:Formalism}}

In this section we set up the stage for the calculation of 
Mellin moments of the deep inelastic structure functions. 
We briefly recall the operator product expansion and, 
in particular, pay attention to the renormalization procedure. 
For reviews see refs.\cite{Buras:1980yt,Reya:1981zk}.

We wish to calculate the hadronic part of the amplitude for unpolarized 
deep inelastic lepton-nucleon scattering which is given by the hadronic tensor
\begin{eqnarray}
\label{htensor}
W_{\m\n}(p,q) 
&=& \frac{1}{4\pi}
     \int d^4z\, {\rm{e}}^{{\rm{i}}q \cdot z}\langle  {\rm{P}}\vert
 J^{\dagger}_{\m}(z)J_{\n}(0)\vert {\rm{P}}\rangle  
\\ 
&=& e_{\m\n}\, \frac{1}{2x}F_{L}(x,Q^2) + 
    d_{\m\n}\, \frac{1}{2x}F_{2}(x,Q^2) 
  + {\rm{i}} \e_{\m\n\a\b} \frac{p^\a q^\b}{p\mydot q} F_{3}(x,Q^2)\, ,\nonumber
\end{eqnarray}
where $J_{\m}$ is either an electromagnetic or a weak current 
and $\vert{\rm{P}}\rangle$ is the unpolarized hadronic state. 
The boson transfers momentum $q$, $Q^2=-q^2$, the hadron carries  momentum $p$ and 
the Bjorken scaling variable is defined as $x=Q^2/ (2p\cdot q)$ with $0 < x \leq 1$.
The tensors $e_{\m\n}$ and $d_{\m\n}$ are given by  
\begin{eqnarray}
\label{eq:tensordef}
e_{\m\n} &=& g_{\m \n}-\frac{q_{\m} q_{\n}}{q^2} \, ,\\
d_{\m\n} &=& -g_{\m \n}-p_{\m}p_{\n}\frac{4x^2}{q^2}
              -(p_{\m}q_{\n}+p_{\n}q_{\m})\frac{2x}{q^2} \, .
\end{eqnarray} 
The longitudinal structure function $F_{L}$ is related to the structure function 
$F_{1}$,
\begin{eqnarray}
F_{L}(x,Q^2) = F_{2}(x,Q^2) - 2xF_{1}(x,Q^2)\, .
\end{eqnarray}
The structure function $F_{3}$ describes parity-violating effects that 
arise from vector and axial-vector interference. It vanishes for pure 
electromagnetic interactions.

We are interested in the Mellin moments of the structure functions,
defined as 
\begin{eqnarray}
\label{eq:mellindef}
\displaystyle
F_{i}^N(Q^2) &=&
\int\limits_0^1 dx\, x^{N-2} F_{i}(x,Q^2)\, ,\quad\quad
i = 2,L\, .
\end{eqnarray}
A similar relation defines $F_{3}^N$ with $N$ replaced by $N+1$ 
in the integral on the right hand side in eq.(\ref{eq:mellindef}).

\subsection{Operator product expansion}

In the Bjorken limit, $Q^2 \rightarrow \infty$, $ x$ fixed,
the integral in eq.(\ref{htensor}) is dominated by 
the integration region near the lightcone $z^2 \sim 0$, 
because in this region the phase in the exponent in eq.(\ref{htensor}) 
becomes stationary. 
The external momentum $q$ of the hadronic scattering amplitude  
is highly virtual, that is to say in the unphysical Euclidean region. 
Thus, we can use dispersion relations together with the 
operator product expansion for a formal expansion of the current product 
in eq.(\ref{htensor}) around the lightcone $z^2 \sim 0$ 
into a series of local composite operators of leading twist.

The optical theorem relates the hadronic tensor in eq.(\ref{htensor}) 
to the imaginary part of the forward scattering amplitude of 
boson-nucleon scattering, $T_{\m\n}$,
\begin{eqnarray}
\label{opticaltheorem}
W_{\m \n}(p,q) &=&  \frac{1}{2\pi}\, {\rm{Im}}\, T_{\m \n}(p,q)\, .
\end{eqnarray}
The forward Compton amplitude $T_{\m\n}$ has a time-ordered product of two local currents, 
to which standard perturbation theory applies,
\begin{eqnarray}
\label{forwardcompton}
T_{\m\n}(p,q) &=& {\rm{i}} \int d^4z\, {\rm{e}}^{{\rm{i}}q \cdot z}
\langle {\rm{P}} \vert\,
 T \left( J^{\dagger}_{\m}(z)J_{\n}(0) \right) \vert {\rm{P}}\rangle\, .
\end{eqnarray}
In terms of local operators for a time ordered product of 
the two electromagnetic or weak hadronic currents 
such as in eq.(\ref{forwardcompton}) the OPE reads
\begin{eqnarray}
\label{OPE}
{\lefteqn{
 {\rm{i}} \int d^4z\, {\rm{e}}^{{\rm{i}}q \cdot z}\,
T\left( J^{\dagger}_{\n_1}(z)J_{\n_2}(0)\right) 
         \,=}} \\
& & 
 \sum_{N,j} \left(\frac{1}{Q^2}\right)^N 
\left[ \left(g_{\n_1\n_2}-\frac{q_{\n_1}q_{\n_2}}{q^2}\right)
 q_{\m_1}q_{\m_2}
 C_{L,j}^{N}\left(\frac{Q^2}{\m^2},\a_s\right) \right. \nonumber\\
& &
-\left(g_{\n_1 \m_1}g_{\n_2\m_2}q^2  -g_{\n_1\m_1}q_{\n_2}q_{\m_2}
   -g_{\n_2\m_2}q_{\n_1}q_{\m_1}   +g_{\n_1\n_2}q_{\m_1}q_{\m_2} \right)
      C_{2,j}^{N}\left(\frac{Q^2}{\m^2},\a_s\right)  \nonumber\\
& &
\left. 
+ {\rm{i}} \e_{\n_1 \n_2 \m_1 \n_3} g^{\n_3 \n_4} q_{\n_4}q_{\m_2} 
C_{3,j}^{N}\left(\frac{Q^2}{\m^2},\a_s\right) \right]
q_{\m_3}...q_{\m_N} O^{j,\{\m_1,...,\m_N\}}(\m^2) 
\,+\,\, {\rm{higher\,\, twists,}} 
\nonumber
\end{eqnarray}
where  $j=\alpha,{\rm{q}},{\rm{g}}$.
Here, all quantities are assumed to be renormalized, $\m$ being the renormalization scale. 
Higher twist contributions 
are less singular near the lightcone $z^2 \sim 0$ and suppressed by powers of $1/Q^2$.  
They are omitted in eq.(\ref{OPE}).
Thus, the sum over $N$ in eq.(\ref{OPE}) extends to infinity and 
runs only over the standard set of the spin-$N$ twist-2 irreducible, 
symmetrical and traceless flavour non-singlet quark operators $O^\a$,
and the singlet quark and gluon operators $O^{\rm{q}}$ and $O^{\rm{g}}$.
These are defined by,
\begin{eqnarray}
\label{defoperatorns}
O^{\alpha,\{\m_1,\cdots ,\m_N\}} & = & \overline{\psi}\lambda^{\alpha}
  \gamma^{\{\m_1}D^{\m_2}\cdots D^{\m_N\}}\psi,~~\alpha=1,2,...,(n_f^2-1)\, , \\
\label{defoperatorquark}
O^{{\rm{q}},\{\m_1,\cdots ,\m_N\}} & = & \overline{\psi}
  \gamma^{\{\m_1}D^{\m_2}\cdots D^{\m_N\}}\psi, \\
\label{defoperatorgluon}
O^{{\rm{g}},\{\m_1,\cdots ,\m_N\}} & = & F^{\{\n\m_1} D^{\m_2}\cdots
  D^{\m_{N-1}} F^{\m_N \n \}}\, .
\end{eqnarray}
Here, $\psi$ defines the quark operator and $F^{\m\n}$ the gluon operator.
The generators of the flavour group $SU(n_f)$ are denoted 
by $\lambda^{\alpha}$, and the covariant derivative by $D^\m$. 
It is understood that the symmetrical and traceless part 
is taken with respect to the indices in curly brackets.

The spin averaged matrix elements of these operators 
in eqs.(\ref{defoperatorns})--(\ref{defoperatorgluon}) 
sandwiched between some hadronic state are given by
\begin{eqnarray}
\label{OME}
\langle {\rm{P}} \vert O^{j, \{\m_1,...,\m_N\}}\vert
 {\rm{P}} \rangle =p^{\{\m_1}...p^{\m_N\}}A_{{\rm{P}},N}^j\left(\frac{p^2}{\m^2}\right)\, ,
\end{eqnarray}
where hadron mass effects have been neglected. 
The anomalous dimensions of these operator matrix elements in eq.(\ref{OME})
govern the scale evolution of the structure functions. 
They are finite quantities as well as the coefficient functions $C_{i,j}^{N}$ multiplying 
these operator matrix elements according to eq.(\ref{OPE}). 
Both are calculable order by order in perturbative QCD in an expansion 
in the strong coupling constant $\a_s$. 

The operator matrix elements themselves as given in eq.(\ref{OME})
are not calculable  in perturbative QCD. 
However, they can be related to the distributions 
$\rm{q_{i}}$, $\Bar{\rm{q}}_{i}$ 
of quark and anti-quark of flavour $i$ and to the gluon distribution $g$ in the hadron.
Let us briefly note, that the general structure of these densities allows 
for three independently evolving types of non-singlet distributions 
${\rm{q}}_{{\rm ns},ij}^{\pm}$ and ${\rm{q}}_{{\rm ns}}^{\rm{V}}$ and 
one quark singlet distribution ${\rm{q}}_{\rm{s}}$. 
The three non-singlet distributions are the flavour asymmetries
\begin{eqnarray}
\label{ns-dens}
{\rm{q}}_{{\rm ns},ij}^{\pm} = {\rm{q}}_i \pm \Bar{{\rm{q}}}_i - ({\rm{q}}_j \pm \Bar{{\rm{q}}}_j)\, ,
\end{eqnarray}
and the sum of the valence distributions of all flavours,
\begin{eqnarray}
\label{v-dens}
{\rm{q}}_{\rm ns}^{\rm{V}} = \sum\limits_{i=1}^{n_f} ({\rm{q}}_i - \Bar{{\rm{q}}}_i) \, ,
\end{eqnarray}
while the singlet distribution is simply the sum of the distributions 
of all flavours,
\begin{eqnarray}
\label{s-dens}
{\rm{q}}_{\rm s} = \sum\limits_{i=1}^{n_f} ({\rm{q}}_i + \Bar{{\rm{q}}}_i) \, .
\end{eqnarray}

With the operator matrix elements in eq.(\ref{OME})
we can write the OPE of eq.(\ref{OPE}) as 
\begin{eqnarray}
\label{OPE/OME}
{\lefteqn{
 {\rm{i}} \int d^4z\, {\rm{e}}^{{\rm{i}}q \cdot z}
\langle {\rm{P}} \vert\,
 T \left( J^{\dagger}_{\m}(z)J_{\n}(0) \right) \vert {\rm{P}}\rangle\, \,=}}\\
& &
 \sum_{N,j} \left(\frac{1}{2x}\right)^N 
\left[ e_{\m\n}\, C_{L,j}^{N}\left(\frac{Q^2}{\m^2},\a_s\right) 
     + d_{\m\n}\, C_{2,j}^{N}\left(\frac{Q^2}{\m^2},\a_s\right)  
+ {\rm{i}} \e_{\m\n\a\b} \frac{p^\a q^\b}{p\mydot q} 
C_{3,j}^{N}\left(\frac{Q^2}{\m^2},\a_s\right) \right] 
A_{{\rm{P}},N}^{j}\left(\frac{p^2}{\m^2}\right) \nonumber\\
& &\quad\quad\quad\quad
\,+\,\, {\rm{higher\,\, twists,}} 
\nonumber
\end{eqnarray}
which is an expansion in terms of the variable $(2 p\cdot q)/Q^2 = 1/x$ 
for unphysical $x \rightarrow \infty$. 
The final connection to DIS structure functions in the physical region $0 < x \leq 1$ 
is achieved by taking Mellin moments of eq.(\ref{OPE/OME}) and using the 
optical theorem eq.(\ref{opticaltheorem}), which relates the structure functions 
to invariants $T_{i,{\rm{P}}}$, $i=2,3,L$, of the forward Compton amplitude 
$T_{\m \n}$ as follows,
\begin{eqnarray}
\label{opticaltheorem2L}
\frac{1}{2\p}\,{\rm{Im}}\, T_{i,{\rm{P}}}(x,Q^2) &=& \frac{1}{2x}\, F_{i}(x,Q^2)\, ,
\quad\quad i = 2,L\, ,\\
\label{opticaltheorem3}
\frac{1}{2\p}\, {\rm{Im}}\, T_{3,{\rm{P}}}(x,Q^2) &=& F_{3}(x,Q^2)\, ,
\end{eqnarray}
where the $T_{i,{\rm{P}}}$ can be projected 
in leading twist approximation and $D=4-2\ve$ dimensions, 
\begin{eqnarray}
\label{TprojL}
 T_{L,{\rm{P}}}(x,Q^2) &=& - \frac{q^2}{(p\mydot q)^2}\, p^\m p^\n \hspace{1mm}T_{\m\n}(p,q)\, , \\
\label{Tproj2}
 T_{2,{\rm{P}}}(x,Q^2) &=& -
 \left( \frac{3-2\ve}{2-2\ve}\hspace{1mm} 
  \frac{ q^2}{(p\mydot q)^2}\, p^\m p^\n
   + \frac{1}{2-2\ve} \hspace{1mm} g^{\m\n} \right) T_{\m\n}(p,q)\, , \\
\label{Tproj3}
 T_{3,{\rm{P}}}(x,Q^2) &=& - {\rm{i}}  \frac{1}{(1-2\ve)(2-2\ve)}\hspace{1mm} 
\e^{\m\n\a\b}\, \frac{p_\a q_\b}{p\mydot q} \hspace{1mm} T_{\m\n}(p,q) \, .
\end{eqnarray}
Then, with the help of eqs.(\ref{OPE/OME})--(\ref{opticaltheorem3}) we find 
\begin{eqnarray}
\label{eq:F2mellin}
\displaystyle
\int\limits_0^1 dx\, x^{N-2} F_{i}(x,Q^2)
&=& \sum\limits_{j=\a,{\rm{q, g}}}
C_{i,j}^{N}\left(\frac{Q^2}{\m^2},\a_s\right) 
A_{{\rm{P}},N}^j\left(\frac{p^2}{\m^2}\right)\, ,\, i=2,L\, , \\
\label{eq:F3mellin}
\displaystyle
\int\limits_0^1 dx\, x^{N-1} F_{3}(x,Q^2)
&=&
\sum\limits_{j=\a}
C_{3,j}^{N}\left(\frac{Q^2}{\m^2},\a_s\right) 
A_{{\rm{P}},N}^j\left(\frac{p^2}{\m^2}\right)\, , 
\end{eqnarray}
which shows that the Mellin moments of DIS structure functions $F_{i}^N$ 
as defined in eq.(\ref{eq:mellindef}) 
can naturally be written in the parameters of the OPE eq.(\ref{OPE}).

The derivation of eqs.(\ref{eq:F2mellin}) and (\ref{eq:F3mellin}) 
in the dispersive approach uses symmetry properties 
of $T_{\m\n}$ under exchange $q \to -q$, which is $x \to -x$. 
Therefore, dependent on the process under consideration, 
eqs.(\ref{eq:F2mellin}) and (\ref{eq:F3mellin}) determine only  
either the even or the odd Mellin moments of $F_{2}$, $F_{L}$ and $F_{3}$.
However, all moments in the complex $N$ plane are fixed by analytic
continuation from either the even or the odd Mellin moments. 
This implies that the $x$-space result for the physical structure functions 
in the range $0 < x \leq 1$, can be found by means of an inverse Mellin transformation
if the infinite set of either even or odd moments is known.

In the standard case of unpolarized electron-proton scattering in the one-photon 
exchange approximation the even moments of $F_{2}$ and $F_{L}$
are fixed in eq.(\ref{eq:F2mellin}), while in the case of 
electroweak interactions with neutrino-proton scattering 
eqs.(\ref{eq:F2mellin}) and (\ref{eq:F3mellin}) determine 
the odd moments of $F_{2}^{\n{\rm{P}}-{\Bar{\n}{\rm{P}}}}$ 
and $F_{3}^{\n{\rm{P}}+{\Bar{\n}{\rm{P}}}}$ 
and the even moments of $F_{2}^{\n{\rm{P}}+{\Bar{\n}{\rm{P}}}}$ 
and $F_{3}^{\n{\rm{P}}-{\Bar{\n}{\rm{P}}}}$, see ref.\cite{Buras:1980yt}.
We are going to consider all these cases mentioned in order to extract 
the complete non-singlet coefficient functions at two loops, as will be detailed 
in section~\ref{sec:CalculationResults}.

The sum in eq.(\ref{eq:F2mellin}) extends over the flavour non-singlet 
and singlet quark and gluon contributions. 
Notice however, that in eq.(\ref{eq:F3mellin}) 
the singlet operators $O^{\rm{q}}$, $O^{\rm{g}}$ do not contribute to $F_{3}$. 
This is due to the properties of $O^{\rm{q}}$ and $O^{\rm{g}}$ under 
a charge conjugation transformation~\cite{Buras:1980yt}.
It can also be understood by looking at partonic scattering processes.
If ${\rm{p}}$ and ${\rm{\Bar{p}}}$ are partons and anti-partons, 
charge conjugation implies for the cross-sections
\begin{eqnarray}
\s_{\rm{p}}&=& -\s_{\rm{\Bar{p}}}\, ,
\end{eqnarray}
which gives zero for eigenstates under charge conjugation~\cite{Zijlstra:1992kj}.

\subsection{Renormalization}

We are interested in the calculation of the scale evolution of 
the DIS structure functions. To that end, a short discussion of 
the renormalization properties of the operators and the coefficient functions 
in eq.(\ref{OPE}) is in order.

Let us first recall, that the OPE of eq.(\ref{OPE}) is an operator statement
and therefore both the coefficient functions $C_{i,j}^N$ and the 
anomalous dimensions of the operators do not depend on the hadronic states 
to which the OPE is applied. The information on the hadronic target is only 
contained in the operator matrix elements $A_{{\rm{P}},N}^j$ in eq.(\ref{OME}). 
It is therefore standard to consider simpler Green's functions 
in an infrared regulated perturbative expansion 
with the operators $O^j$ sandwiched between parton states 
\begin{eqnarray}
\label{OMEparton}
\langle {\rm{p}} \vert O^{j, \{\m_1,...,\m_N\}}\vert
 {\rm{p}} \rangle =p^{\{\m_1}...p^{\m_N\}}A_{{\rm{p}},N}^j\left(\frac{p^2}{\m^2}\right)\, ,
\end{eqnarray}
where $\vert{\rm{p}}\rangle$ denotes a spin-averaged parton state, 
being either a flavour non-singlet or singlet combination of quarks and anti-quarks 
or a gluon. 

As they stand the bare operator matrix elements in eq.(\ref{OMEparton})
require renormalization. 
We choose dimensional regularization~\cite{'tHooft:1972fi,Bollini:1972ui}
in $D=4-2\ve$ dimensions and 
define the renormalized operators in terms of bare operators as 
\begin{eqnarray}
\label{OPrenNS}
O^{\a,\rm{bare}} &=& Z^{\a \a} \,O^{\a,\rm{ren}} \, , \\
\label{OPrenS}
O^{j,\rm{bare}} &=& 
\sum_{k = {\rm{q}},{\rm{g}}}\, Z^{j k}\,O^{k,\rm{ren}}\, ,
\quad\quad j = {\rm{q}},{\rm{g}},
\end{eqnarray}
where $Z^{\a \a}$ renormalizes the quark flavour non-singlet operator $O^\a$.
In the flavour singlet case, eq.(\ref{OPrenS}) denotes operator mixing 
under renormalization as $O^{\rm{q}}$ and $O^{\rm{g}}$ have the same quantum numbers.

The anomalous dimensions $\g$ determine the scale dependence 
of the renormalized operators, 
\begin{eqnarray}
\label{gammadef}
\frac{d }{d \ln \m^2 } O^{\rm{ren}}
 & \equiv &  \gamma \hspace{.1cm} 
 O^{\rm{ren}}\, ,
\end{eqnarray}
and they are defined as 
\begin{eqnarray}
\label{anomdef}
\g &=& \displaystyle\left( \frac{d }{d \ln \m^2 }
    Z\right) Z^{-1} \, ,
\end{eqnarray}
where in the flavour singlet case eqs.(\ref{gammadef}) and (\ref{anomdef}) 
are understood as matrix equations, $Z$ representing the 
matrix $Z^{jk}$ and $\g$ the matrix $\g_{jk}$. 
The general structure of these anomalous dimensions is constrained by
charge conjugation invariance and flavour symmetry. 
In the case of quarks and anti-quarks of flavour $i,j$ they can be split up into 
a valence and a sea part,
\begin{eqnarray}
\label{splitting-functions-1}
  \g_{{\rm{q}}_{i}{\rm{q}}_{j}} & = \, \g_{\bar{{\rm{q}}}_{i}\bar{{\rm{q}}}_{j}}
  & = \, \delta_{ij} \g_{{\rm{q}}{\rm{q}}}^{\rm{V}} + \g_{{\rm{q}}{\rm{q}}}^{\rm{S}}\, , \\
\label{splitting-functions-2}
  \g_{{\rm{q}}_{i}\bar{{\rm{q}}}_{j}} & = \, \g_{\bar{{\rm{q}}}_{i}{\rm{q}}_{j}}
  & = \, \delta_{ij} \g_{{\rm{q}}\bar{{\rm{q}}}}^{\rm{V}} + \g_{{\rm{q}}\bar{{\rm{q}}}}^{\rm{S}}
  \, . 
\end{eqnarray}

As one commonly considers matrix elements of operators $O^j$ corresponding to 
the quark non-singlet and singlet distributions 
${\rm{q}}_{{\rm ns},ij}^{\pm}$, ${\rm{q}}_{{\rm ns}}^{\rm{V}}$ and ${\rm{q}}_{\rm{s}}$ of 
eqs.(\ref{ns-dens})--(\ref{s-dens}),  
we remark that their scale evolution 
is governed by the following four linear combinations,
\begin{eqnarray}
\label{splitting-functions-ns}
{\rm{q}}_{{\rm ns},ij}^{\pm} 
        \, \longrightarrow& \d_{ij} \g_{{\rm{q}}{\rm{q}}}^{\pm,\rm{V}} & = \, 
                \d_{ij} \left( 
        \g_{{\rm{q}}{\rm{q}}}^{\rm{V}} \pm \g_{{\rm{q}}\bar{{\rm{q}}}}^{\rm{V}} \right)\, \\
\label{splitting-functions-s-1}
{\rm{q}}_{\rm ns}^{\rm{V}} \, \longrightarrow& 
                \g_{{\rm{q}}{\rm{q}}}^{-,\rm{V}} + n_f \g_{{\rm{q}}{\rm{q}}}^{-,\rm{S}} & = \, 
        \g_{{\rm{q}}{\rm{q}}}^{\rm{V}} - \g_{{\rm{q}}\bar{{\rm{q}}}}^{\rm{V}} + n_f 
\left( \g_{{\rm{q}}{\rm{q}}}^{\rm{S}} - \g_{{\rm{q}}\bar{{\rm{q}}}}^{\rm{S}} \right)\, ,\\
\label{splitting-functions-s-2}
{\rm{q}}_{\rm s} \, \longrightarrow&
        \g_{{\rm{q}}{\rm{q}}}^{+,\rm{V}} +
                n_f \g_{{\rm{q}}{\rm{q}}}^{+,\rm{S}} & = \,
        \g_{{\rm{q}}{\rm{q}}}^{\rm{V}} + \g_{{\rm{q}}\bar{{\rm{q}}}}^{\rm{V}} + 
                n_f \left(
        \g_{{\rm{q}}{\rm{q}}}^{\rm{S}} + \g_{{\rm{q}}\bar{{\rm{q}}}}^{\rm{S}} \right)\, .
\end{eqnarray}
In our case of unpolarized lepton-hadron scattering, 
the even moments of $F_2$ determine the linear combinations  
$\g_{{\rm{q}}{\rm{q}}}^{+,\rm{V}}$ and $\g_{{\rm{q}}{\rm{q}}}^{+,\rm{S}}$, 
while the odd moments of $F_3$ determine 
$\g_{{\rm{q}}{\rm{q}}}^{-,\rm{V}}$ and $\g_{{\rm{q}}{\rm{q}}}^{-,\rm{S}}$. 
The individual valence and sea contributions are identified by 
the flavour structure of the diagrams for the structure functions, so that 
eqs.(\ref{splitting-functions-ns})--(\ref{splitting-functions-s-2}) 
suffice to determine the anomalous dimensions 
$\g_{{\rm{qq}}}^{\rm{V}}, \g_{{\rm{q}}\bar{{\rm{q}}}}^{\rm{V}},\g_{{\rm{qq}}}^{\rm{S}}$ 
and $\g_{{\rm{q}}\bar{{\rm{q}}}}^{\rm{S}}$.

Both, the anomalous dimensions $\g$ and the renormalization constants $Z$ 
have a series expansion in $\a_s$. 
We rewrite eq.(\ref{anomdef}) in $D=4-2\ve$ dimensions in a form that 
is suitable for an easy extraction of the expansion coefficients,
\begin{eqnarray}
\label{anomdef-eps}
\g(\a_s) &=& 
\left(\ve \frac{\a_s}{4\p} - \b(\a_s) \right)
\left( \frac{d }{d \ln \m^2 }
    Z\left(\a_s,\frac{1}{\ve}\right)\right) Z^{-1}\left(\a_s,\frac{1}{\ve}\right) \, ,
\end{eqnarray}
where $\beta (a_s)$ denotes the beta function, that determines 
the renormalization scale dependence of the running coupling. 
The renormalization constants $Z$ obey an expansion in $1/\ve$. 
Although the $Z$ contain poles in $\ve$, the anomalous dimensions are always 
finite as $\ve\rightarrow 0$. 
Thus, to lowest order in $\a_s$ the anomalous dimensions $\g$ 
in eq.(\ref{anomdef-eps}) are simply expressed through the residue 
of $Z$ in $1/\ve$.

To solve the coupled matrix equation defined by eq.(\ref{anomdef}) 
in the singlet case, one should notice that to leading order in 
$\a_s$ the matrix $Z^{jk}$ is diagonal 
with $Z^{(0),\rm{qq}}=Z^{(0),\rm{gg}}=1$ and $Z^{(0),\rm{qg}}=Z^{(0),\rm{gq}}=0$. 
This additional information 
allows for a unique iterative determination of the anomalous dimensions 
order by order in $\a_s$. 

The relation between the bare coupling $\a^{\rm{bare}}_s$ 
and the renormalized coupling $\a_s$ is given by 
\begin{eqnarray}
\label{alpha-s-renorm}
\a^{\rm{bare}}_s &=& 
Z_{\a_s}\, \a_s^{\rm{ren}}\, ,
\end{eqnarray}
with the renormalization constant $Z_{\a_s}$ in the minimal subtraction 
scheme given by
\begin{eqnarray}
Z_{\a_s} &=& 
1 - \frac{\b_0}{\ve} \frac{\a_s}{4 \p}\, 
+ \left( \frac{\b_0^2}{\ve^2} - \frac{\b_1}{2\ve} \right) 
\left( \frac{\a_s}{4 \p} \right)^2 + 
\dots\,\, .
\end{eqnarray}
In $D=4-2\ve$ dimensions the beta function of the running coupling 
up to two loops~\cite{Caswell:1974gg} 
is given by 
\begin{eqnarray}
\frac{d \a_s/(4\p)}{d \ln \m^2} &=& - \ve \frac{\a_s}{4\p} - \b_0 \left(\frac{\a_s}{4\p}\right)^2
- \b_1 \left(\frac{\a_s}{4\p}\right)^3 - \dots\,\, , \\
\b_{0} & = & \frac{11}{3} C_A - \frac{4}{3} T_F n_f \, ,
 \nonumber \\
\b_{1} & = &
 \frac{34}{3}C_A^2 - 4 C_F T_F n_f -\frac{20}{3} C_A T_F n_f\, .  \nonumber
\end{eqnarray}

Finally, to determine the scale dependence of the coefficient functions 
it is sufficient to notice, that the anomalous dimension of the current $J_{\m}$ 
in the time-ordered product of the forward Compton amplitude $T_{\m\n}$ is zero 
due to current conservation.
It implies, that the scale evolution of the coefficient functions
and the operator matrix elements is governed by the same anomalous dimensions, 
which provides us with the renormalization group equation for the 
flavour singlet and non-singlet coefficient functions,
\begin{eqnarray}
\label{callannonsin}
\left[ \m^2 \frac{\partial}{\partial \m^2} + \beta(\a_s(\m^2)) 
 \frac{\partial}{\partial \a_s(\m^2)} 
- \g^{\rm{ns}}_{\rm qq}(\a_s(\m^2)) 
\right]
C^{N,{\rm{ns}}}_{i,\rm{q}}\left(\frac{Q^2}{\m^2},\a_s(\m^2) \right)
 &= & 0 \, ,
 \\[2ex]
\label{callansinglet} 
\sum_{k = {\rm{q,g}}}
\left[ \left\{ \m^2 \frac{\partial}{\partial \m^2} + \beta(\a_s(\m^2)) 
 \frac{\partial}{\partial \a_s(\m^2)} \right\} \d_{jk}
- \g_{jk}(\a_s(\m^2))
\right] 
C^N_{i,k} \left(\frac{Q^2}{\m^2},\a_s(\m^2) \right)
&= &   0 \, ,\\
& &\quad \mbox{ $j$ = {\rm{q,g}}} \, . \nonumber
\end{eqnarray}
In eq.(\ref{callannonsin}) we have adopted the conventional notation 
to collectively denote the non-singlet anomalous dimensions 
of eqs.(\ref{splitting-functions-ns}) and (\ref{splitting-functions-s-1}) 
with $\g^{\rm{ns}}_{\rm qq}$ and the coefficient functions with $C^{N,{\rm{ns}}}_{i,\rm{q}}$.
In particular, the $Q^2$-dependence of  the coefficient functions 
does not depend on the index $\a$ of the non-singlet operator $O^\a$ 
in eq.(\ref{defoperatorns}) anymore, see for instance ref.\cite{Larin:1994vu}.

The actual calculation of the anomalous dimensions as defined in eq.(\ref{anomdef}) and 
the coefficient functions $C_{i,j}^N$ in perturbative QCD proceeds as follows. 
We introduce partonic invariants $T_{i,{\rm{p}}}$, $i=2,3,L$, 
of the forward partonic Compton amplitude in analogy to eqs.(\ref{TprojL})--(\ref{Tproj3}).
By means of the OPE eq.(\ref{OPE}), these invariants can be written 
in terms of renormalized operator matrix elements as 
\begin{eqnarray}
\lefteqn{
\label{TmunuPartonRenNS}
T^{\rm ns}_{i,{\rm{q}}}(x,Q^2,\a_s,\ve) \,=} \\
& &
\sum_{N}
\left( \frac{1}{2x}\right)^N 
C^{N,{\rm{ns}}}_{i,j}\left(\frac{Q^2}{\m^2},\a_s,\ve\right) 
Z^{\rm qq}_{\rm{ns}}\left(\a_s,\frac{1}{\ve}\right)
A^{{\rm{ns}},{\rm{ren}}}_{{\rm q},N}\left(\a_s,\frac{p^2}{\m^2},\ve\right) 
+ O(p^2)\, , \nonumber\\[1ex]
\lefteqn{
\label{TmunuPartonRenS}
T_{i,{\rm{p}}}(x,Q^2,\a_s,\ve) \,=} \\
& &
\sum_{N}
\sum_{j,k={\rm{q}},{\rm{g}}}  \left( \frac{1}{2x}\right)^N 
C^{N}_{i,j}\left(\frac{Q^2}{\m^2},\a_s,\ve\right) 
Z^{jk}\left(\a_s,\frac{1}{\ve}\right)
A^{k,{\rm{ren}}}_{{\rm p},N}\left(\a_s,\frac{p^2}{\m^2},\ve\right) 
+ O(p^2)\, ,\hspace*{0.5cm}\mbox{{\rm{p}} = {\rm{q}},{\rm{g}}}, \nonumber
\end{eqnarray}
where we have distinguished the flavour non-singlet 
from the flavour singlet case, $i =2,3,L$. 
In the former case, we used again the collective notation $Z^{\rm qq}_{\rm{ns}}$ 
for the various non-singlet renormalization constants.
The left hand side of eqs.(\ref{TmunuPartonRenNS}) and (\ref{TmunuPartonRenS}) 
is renormalized by substituting the bare coupling constant
in terms of the renormalized one as defined by eq.(\ref{alpha-s-renorm}), 
$\a_s =\a_s(\m^2/\Lambda^2)$. 
The wave function renormalization factors for the external quark and gluon 
lines are overall factors on both sides of the equations and drop out. 
The terms $O(p^2)$ on the right hand side 
of eqs.(\ref{TmunuPartonRenNS}) and (\ref{TmunuPartonRenS}) 
indicate higher twist contributions, which we neglect.

It is known that the gauge invariant operators $O^{\rm{q}}$ and $O^{\rm{g}}$ 
mix under renormalization with unphysical operators~\cite{Hamberg:1992qt,collins:book}.
These are BRST variations of some operators or else vanish by the equations of motion. 
However, matrix elements with physical polarization and on-shell momenta of such 
unphysical operators vanish.
That is to say, these unphysical operators do not contribute to quantities related to 
physical $S$-matrix elements such as the invariants $T_{i,{\rm{p}}}$ 
which we are going to calculate. 
Therefore, they are omitted in eq.(\ref{TmunuPartonRenS}).

Starting with the partonic invariants $T_{i,{\rm{p}}}$ 
from eqs.(\ref{TmunuPartonRenNS}) and (\ref{TmunuPartonRenS}), 
the renormalization constants $Z$ and the coefficient 
functions  $C_{i,j}^N$ are calculated using the method of projection developed 
in ref.\cite{Gorishnii:1983su}.
This method consists of applying the following projection operator 
to both sides of eqs.(\ref{TmunuPartonRenNS}) and (\ref{TmunuPartonRenS}) 
\begin{eqnarray}
\label{projectionoperator}
 {\cal P}_N \equiv
  \left. \Biggl[ \frac{q^{ \{\m_1}\cdots q^{\m_N \}}}{N !}
  \frac{\partial ^N}{\partial
p^{\m_1} \cdots  \partial p^{\m_N}} \Biggr] \right|_{p=0} \, ,
\end{eqnarray}
where $q^{ \{\m_1}\cdots q^{\m_N \}}$ is the harmonic, that is to say 
the symmetrical and traceless part of the tensor $q^{\m_1}\cdots q^{\m_N }$.

On the right hand side 
of eqs.(\ref{TmunuPartonRenNS}) and (\ref{TmunuPartonRenS}), it is obvious, 
that the $N$-th order differentiation in the projection operator ${\cal P}_N$ 
singles out precisely the $N$-th moment
which is the coefficient of $1/(2x)^{N}$. 
All other powers of $1/(2x)$ vanish either by differentiation or after nullifying 
the momentum $p$. 
The operator ${\cal P}_N$ does not act on the 
renormalization constants $Z$ and the coefficient functions 
on the right hand side of eqs.(\ref{TmunuPartonRenNS}) and (\ref{TmunuPartonRenS}) 
as they are only functions of $N$, $\a_s$ and $\ve$.
However, ${\cal P}_N$ does act on the partonic matrix elements $A^{j}_{{\rm{p}},N}$. 
There, the nullification of $p$ effectively eliminates all diagrams 
containing loops, which become massless tadpole diagrams and 
are therefore put to zero in dimensional regularization.
In particular, this removes the operator matrix elements 
$A^{{\rm q},{\rm{ren}}}_{{\rm g},N}$ and $A^{{\rm g},{\rm{ren}}}_{{\rm q},N}$ 
in eq.(\ref{TmunuPartonRenS}) as they start only at 1-loop level. 
Hence, in the perturbative expansion of  $A^{j}_{{\rm{p}},N}$ only 
the tree level diagrams $A^{{\rm p},{\rm{tree}}}_{{\rm p},N}$ survive. 
Finally, the $O(p^2)$ terms in eqs.(\ref{TmunuPartonRenNS}) and (\ref{TmunuPartonRenS}),
which denote higher twist contributions,  
become proportional to the metric tensor after differentiation. 
They are removed by the harmonic tensor $q^{ \{\m_1}\cdots q^{\m_N \}}$.

On the left hand side 
of eqs.(\ref{TmunuPartonRenNS}) and (\ref{TmunuPartonRenS}),
${\cal P}_N$ is applied to the integrands of all Feynman diagrams 
contributing to the invariants $T_{i,{\rm{p}}}$. 
The momentum $p$ is nullified before taking the limit $\ve \rightarrow 0$, 
so that all infrared divergences as $p\rightarrow 0$ are dimensionally 
regularized for individual diagrams. 
Effectively this reduces the 4-point diagrams that contribute to $T_{\m\n}$ 
to 2-point diagrams with symbolic
powers of scalar products in the numerator and denominator, which we can solve 
by means of recursion relations as will be detailed in section~\ref{sec:Method}.
Notice also, that we apply ${\cal P}_N$ to the projected partonic invariants 
$T_{i,{\rm{p}}}$ rather than to $T_{\m\n}$, as ${\cal P}_N$ would destroy the tensor 
structure of $T_{\m\n}$. 

To summarize, we find after application of the projection operator
$ {\cal P}_N$ to eqs.(\ref{TmunuPartonRenNS}) and (\ref{TmunuPartonRenS}),
\begin{eqnarray}
\label{TmunuPartonMomNS}
T^{N,\rm ns}_{i,{\rm{q}}}\left(\frac{Q^2}{\m^2},\a_s,\ve\right) &=&
C^{N,{\rm{ns}}}_{i,{\rm{q}}}\left(\frac{Q^2}{\m^2},\a_s,\ve\right) 
Z^{\rm qq}_{\rm{ns}}\left(\a_s,\frac{1}{\ve}\right)
A^{{\rm{ns}},{\rm{tree}}}_{{\rm q},N}(\ve)  \, , \\
\label{TmunuPartonMomS}
T^N_{i,{\rm{p}}}\left(\frac{Q^2}{\m^2},\a_s,\ve\right) &=&
\left[
C^{N}_{i,\rm{q}}\left(\frac{Q^2}{\m^2},\a_s,\ve\right) 
Z^{\rm{qp}}\left(\a_s,\frac{1}{\ve}\right) + \right.\\
& &\hspace*{10mm} \left.
C^{N}_{i,\rm{g}}\left(\frac{Q^2}{\m^2},\a_s,\ve\right) 
Z^{\rm{gp}}\left(\a_s,\frac{1}{\ve}\right)
\right]
A^{{\rm{p}},{\rm{tree}}}_{{\rm p},N}(\ve) \, ,\,\,\,\,\,\,
\mbox{{\rm{p}} = {\rm{q}},{\rm{g}}}, \nonumber
\end{eqnarray}
where $i = 2,3,L$ and the left hand side is defined as 
\begin{eqnarray}
\left.
 T^N_{i,\rm p}\left(\frac{Q^2}{\m^2},\a_s,\ve\right) 
\equiv  \cp_N\, T_{i,\rm p}(x,Q^2,\a_s,\ve)
 \right| _{p=0}. 
\end{eqnarray}

Eqs.(\ref{TmunuPartonMomNS}) and (\ref{TmunuPartonMomS}) are central 
to our approach of the calculation of the anomalous dimensions and 
coefficient functions via the OPE and dispersion relations.
They represent a coupled system of equations when both sides are expanded 
in powers of $\a_s$ and $\ve$. 
That is to say, the $C_{i,j}^N$ are expanded in positive powers of $\ve$ and 
the $Z$ are expanded in negative powers of $\ve$.
Explicit solutions to eqs.(\ref{TmunuPartonMomNS}) and (\ref{TmunuPartonMomS}) 
in terms of anomalous dimensions and coefficient functions 
will be given in section~\ref{sec:CalculationResults}.

At this point a few remarks are in order.
First of all, eq.(\ref{TmunuPartonMomS}) does not provide 
us with the full information about the renormalization constants, 
because $Z^{\rm gq} $ and $Z^{\rm gg}$ are determined only in the order $\a_s$.
This limitation follows directly from the fact that 
$C^N_{i,{\rm{g}}}$ only starts from order $\a_s$ 
since the photon couples directly only to quarks. 
To extract the 2-loop anomalous dimension 
$\g^{(1)}_{\rm{gq}}$ and $\g^{(1)}_{\rm{gg}}$ we also calculate 
Green's functions in which the photon is replaced by an external 
scalar particle $\phi$ 
that couples directly only to gluons~\cite{Larin:1997wd}.

These Green's functions can be expressed in partonic invariants $T_{\f,\rm p}$, 
for which an OPE similar to eq.(\ref{OPE}) exists with the same 
singlet operators $O^{\rm q}$ and $O^{\rm g}$ but with different 
coefficient functions $C^N_{\f,{\rm p}}$. 
The important point here is that now 
$C^N_{\f,{\rm{g}}}$ starts already at order $\a_s^0$. 
Hence, the $T_{\f,\rm p}$, provide us with the necessary renormalization 
constants $Z^{\rm gq} $ and $Z^{\rm gg}$ of the singlet operators to two loops.
The vertices that describe the coupling between the external scalar field
$\phi$ and the gluons can be obtained by adding the simplest gauge 
invariant interaction term $\phi F_{\m\n}^{a}F^{\m\n}_a$ 
to the QCD Lagrangian, where $F_{\m\n}^{a}$ is the QCD field strength.

Repeating the steps that led to eq.(\ref{TmunuPartonMomS}) 
one finds for these partonic invariants $T_{\f,\rm p}$, 
\begin{eqnarray}
\label{TmunuPartonMomPhi}
\left(Z_{F^2}\right)^{-2} 
T^N_{\f,{\rm{p}}}\left(\frac{Q^2}{\m^2},\a_s,\ve\right) 
&=&
\left[
C^{N}_{\f,\rm{q}}\left(\frac{Q^2}{\m^2},\a_s,\ve\right) 
Z^{\rm{qp}}\left(\a_s,\frac{1}{\ve}\right) + \right.\\
& &\hspace*{5mm} \left.
C^{N}_{\f,\rm{g}}\left(\frac{Q^2}{\m^2},\a_s,\ve\right) 
Z^{\rm{gp}}\left(\a_s,\frac{1}{\ve}\right)
\right]
A^{{\rm p},{\rm{tree}}}_{{\rm p},N}(\ve)  \, ,\,\,\,\,\,\,
\mbox{{\rm{p}} = {\rm{q}},{\rm{g}}}.\nonumber
\end{eqnarray}

Beyond tree level we have to take into account the overall 
renormalization~\cite{Kluberg-Stern:1975rs}
of the operator $\f F^{a\,\m\n} F^a_{\m\n}$ 
with the renormalization constant $Z_{F^2}$,
\begin{eqnarray}
\label{scalarRen}
\left( F^{a\,\m\n} F^a_{\m\n} \right)^{\rm{bare}}\, \,=\, 
Z_{F^2} \, \left( F^{a\,\m\n} F^a_{\m\n} \right)^{\rm{ren}}\, +\, \dots ,
\quad\quad
Z_{F^2} \,=\, \frac{1}{1-\b(\a_s)/(\ve\a_s)}\, .
\end{eqnarray}
For the partonic invariants $T^N_{\f,{\rm{p}}}$ of the scalar particle 
$\f$ this implies an additional overall renormalization
\begin{eqnarray}
\label{overall-ren}
\left( T^N_{\f,{\rm{p}}} \right)^{\rm{ren}} \,=\, 
\left( Z_{F^2} \right)^{-2} \left( T^N_{\f,{\rm{p}}} \right)^{\rm{bare}}\, ,
\end{eqnarray}
as indicated on the left hand side in eq.(\ref{TmunuPartonMomPhi}). 
The dots in eq.(\ref{scalarRen}) 
indicate mixing with unphysical operators, which give vanishing contributions to 
on-shell matrix elements with physical spin projections. The only physical operator, 
that mixes with $\f F^{a\,\m\n} F^a_{\m\n}$ under renormalization 
is a quark mass term, $m_{\rm{q}}\Bar{\j} \j$, which 
vanishes in the limit of massless quarks.

A second remark concerns the  parity-violating structure function $F_{3}$, 
which is obtained from the time-ordered product of one vector current $V_\m$ and one axial 
vector current $A_\n$.
The axial current contains a $\g_5$ coupling and it is well known 
that this requires some care in the framework of dimensional regularization.
We define the axial current by the substitution~\cite{Akyeampong:1973jh}
\begin{eqnarray}
\label{axial-current-def}
\g_\m\, \g_5 &=& {\rm{i}}\ \frac{1}{6}\, \e_{\m\r\s\t}\, \g^\r \g^\s \g^\t\, , 
\end{eqnarray}
from which $F_{3}$ is projected according to eqs.(\ref{opticaltheorem3})
and (\ref{Tproj3}) . 
The sum over dummy Lorentz indices in the projection 
such as in $\e^{\m\r\s\t}\e_{\m\a\b\c}$ or $\e^{\m\n\r\s}\e_{\m\n\a\b}$ 
defines products of metric tensors, which have to be considered as 
$D$-dimensional objects~\cite{Larin:1991tj,Gorishnii:1985xm}.

The definition eq.(\ref{axial-current-def}) 
violates the axial Ward identity which is to be restored by an additional renormalization. 
The necessary renormalization constant $Z_A$ in the minimal subtraction 
scheme has been given in ref.\cite{Larin:1991tj}.
Up to two loops it is,
\begin{eqnarray}
\label{ZA-const}
Z_A &=& 1 
        + \left( \frac{\a_s}{4 \p} \right)^2\, 
                \frac{1}{\ve}\,
                \left\{ \frac{22}{3} C_A C_F - \frac{4}{3} C_F n_f \right\} \, ,
\end{eqnarray}
where the expansion is performed in terms of the renormalized 
coupling $\a_s$ at the scale $\m^2$.

In addition, the treatment of $\g_5$ in $D=4-2\ve$ dimensions introduces 
an extra finite renormalization with $Z_5$. This is derived 
in the minimal subtraction scheme from an obvious relation between the vector 
and the axial-vector current,
\begin{eqnarray}
\left( R_{\Bar{\rm{MS}}} V_\m \right) \g_5 &=& Z_5\, R_{\Bar{\rm{MS}}} A_\m\, ,
\end{eqnarray}
where $R_{\Bar{\rm{MS}}}$ denotes the $R$-operation in the ${\Bar{\rm{MS}}}$-scheme 
to remove ultraviolet divergencies.
It has also been given in ref.\cite{Larin:1991tj} and reads up to two loops,
\begin{eqnarray}
\label{Z5-const}
Z_5 &=& 1 - \frac{\a_s}{4 \p}\, 4\, C_F 
          + \left( \frac{\a_s}{4 \p} \right)^2 
        \left[ 22 C_F^2 - \frac{107}{9} C_A C_F + \frac{2}{9} C_F n_f \right] \, ,
\end{eqnarray}
where again the expansion is performed in the renormalized 
coupling $\a_s$ at the scale $\m^2$. 
These additional renormalizations in eqs.(\ref{ZA-const}) and (\ref{Z5-const})
have to be taken into account 
when calculating $T^{N,\rm ns}_{3,{\rm{q}}}$.
In eq.(\ref{TmunuPartonMomNS}) one has to substitute 
on the left hand side for $T^{N,\rm ns}_{3,{\rm{q}}}$ 
\begin{eqnarray}
\label{overall-ren-g5}
\left( T^{N,\rm ns}_{3,{\rm{q}}}\right)^{\rm{bare}} \,=\, 
\left( Z_5\, Z_A  \right)^{-1} \left( T^{N,\rm ns}_{3,{\rm{q}}} \right)^{\rm{ren}}\, .
\end{eqnarray}

This concludes our review of the OPE formalism to calculate  
Mellin moments of DIS structure functions.

\section{Methods
\label{sec:Method}}

In this section we will discuss the methods used to obtain the Mellin 
moments of the DIS structure functions.

The idea is to determine reduction identities based on sets of derivative 
equations for the $N$-th Mellin moment of a given diagram in dimensional 
regularization~\cite{'tHooft:1972fi,Bollini:1972ui}. 
This is done such, that it is possible to set up 
systematically recursion relations in the Mellin moment $N$. Solving these 
recursions leads to multiple nested sums, which successively can be 
expressed in terms of harmonic sums, the basic functions in Mellin space.
This method of recursions was previously used in ref.\cite{Kazakov:1988jk} and 
indeed, a number of the relations we give can also be found there. 

\subsection{Reduction identities}

Let us begin with a classification of the relevant topologies following the 
notations of refs.\cite{Gorishnii:1989gt,Larin:1991fz}.
We will begin with 2-point functions of external momentum $q$   
and subsequently dress them up by inserting two additional 
external legs with momentum $p$.

The basic topologies of a 2-point function up to 
two loops are shown below. They are given by the basic 1-loop diagram $L1$ 
and the 2-loop topology named $T1$, 
where the off-shell $q$-momentum flows from right to left through the diagrams. 
\begin{figure}[htb]
\begin{center} 
\begin{picture}(250,90)(-150,0)
\SetColor{Blue}
\ArrowLine(-150,45)(-180,45)    \Text(-127,77)[lb]{1} 
\ArrowLine(-70,45)(-100,45)   \Text(-125,13)[t]{2}
\ArrowArc(-125,45)(25,0,180)  \Text(-170,52)[lb]{3}  
\ArrowArcn(-125,45)(25,0,180) \Text(-85,52)[lb]{4} 
\ArrowLine(30,45)(0,45)           \Text(16,50)[b]{6}
\ArrowLine(120,45)(90,45)         \Text(106,50)[b]{7}
\ArrowLine(60,15)(60,75)          \Text(65,41)[lb]{5}
\ArrowArc(60,45)(30,0,90)         \Text(78,75)[lb]{2}
\ArrowArc(60,45)(30,90,180)       \Text(42,75)[rb]{1}
\ArrowArc(60,45)(30,180,270)      \Text(42,15)[rt]{4}
\ArrowArc(60,45)(30,270,360)      \Text(78,15)[lt]{3}
\end{picture}
\end{center}
\caption{\label{fig:basictopos}
{The basic basic topologies of a 2-point function up to two loops. 
Left: The basic 1-loop diagram $L1$. Right: The 2-loop topology $T1$.}}
\end{figure}
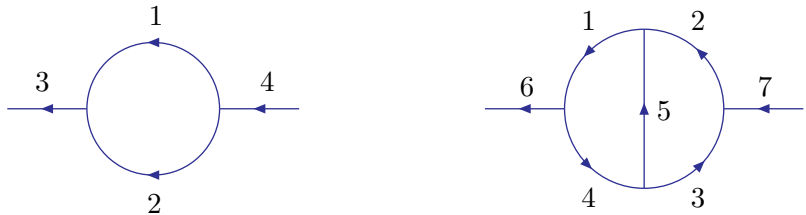

The other 2-loop topologies denoted $T2$ and $T3$ in refs.\cite{Gorishnii:1989gt,Larin:1991fz}
are just special cases of $T1$, where one of 
the lines $1,\dots,5$ is missing. They reduce to the convolution or to the 
product of two 1-loop integrals.

For the calculation of the $N$-th Mellin moment of deep inelastic structure 
functions, we need to consider 4-point diagrams with external momenta $p$, 
$p^2 = 0$, and $q$. Then, all topologies up to two loops that contribute can 
be constructed from the diagrams in fig.\ref{fig:basictopos} by attaching 
two $p$-dependent external legs in all topologically independent ways to 
the various lines. We call these composite topologies. For the calculation 
we need to construct only a single set of programs that can deal with the 
$N$-th Mellin moment of the various composite topologies.

Because the method relies on expressing composite topologies into simpler 
composite topologies we introduce also the notion of basic building blocks. 
These are composite topologies in which both $p$-lines attach to the same 
line in the basic topology. At the two loop level there are two basic 
building blocks:
\begin{eqnarray}
\label{eq:t1v11}
                T1_{11} =
        \TAbasic({A},{c},{d},{e},{a},{b}) & = & \int d^Dp_1\ d^Dp_2
                \frac{1}{(p_1^2)^a((p\plus p_1)^2)^A(p_2^2)^b
                (p_3^2)^c(p_4^2)^d(p_5^2)^e} \, ,
\\
\label{eq:t1v55}
                T1_{55} =
        \TEbasic({b},{c},{d},{E},{e},{a}) & = & \int d^Dp_1\ d^Dp_2
                \frac{1}{(p_1^2)^a(p_2^2)^b
                (p_3^2)^c(p_4^2)^d(p_5^2)^e((p\plus p_5)^2)^E}\, .
\end{eqnarray}
In the pictorial representation that we use for the diagrams we indicate 
the $p$-flow through the diagrams by fat lines. The numbers indicate the 
number of powers of denominators in the indicated momentum. If the same 
line has two numbers, the second one indicates the number of powers of 
$2p\cdot p_i$ in which $i$ is the number of the line in the corresponding 
basic topology. Hence
\begin{eqnarray}
\label{eq:TA}
        \TA({a,m},{b},{c},{d},e,n) & = & \int d^Dp_1\ d^Dp_2
                \frac{(2p\mydot p_1)^m\ (2p\mydot p_5)^n}{(p_1^2)^a(p_2^2)^b
                (p_3^2)^c(p_4^2)^d(p_5^2)^e}\, .
\end{eqnarray}
We will need the composite topologies $T1_{11}$, $T1_{12}$, $T1_{13}$, 
$T1_{14}$, $T1_{15}$, $T1_{16}$, $T1_{17}$, $T1_{55}$, $T1_{56}$ and 
$T1_{67}$. All others are related to these by symmetry operations.

The first observation is that when the external legs $6$ and/or $7$ are 
involved, one can expand the propagator(s) in $p+q$ with the formula
\begin{eqnarray}
        \frac{1}{(p+q)^{2n}} & = & \sum_{i=0}^\infty \frac{1}{(q^2)^n}
                        \sign(i)\binom(n+i-1,i)\frac{(2p\cdot q)^i}{(q^2)^i}\, .
\end{eqnarray}
If we use $n$ powers of $2p\cdot q$ from this expansion, we need only $N-n$ 
powers of $2p\cdot q$ from the rest of the diagram. Hence such a composite 
topology becomes a single sum over a simpler topology in which $p$ flows 
through one line fewer. This solves 4 of the 10 composite two loop 
topologies.

The way to deal with the other 6 composite topologies is to use all 
different variations of the integration by parts identities and some 
scaling identities. The integration by parts 
identities~\cite{'tHooft:1972fi,Chetyrkin:1981qh}
are of the type
\begin{eqnarray}
        \int d^Dp_1 d^Dp_2\,\, \frac{\partial}{\partial p_i^\mu}\,\,
                        p_j^\mu\,\, T1_{kl} & = & 0\, ,
\end{eqnarray}
because we deal with a total derivative here. The momenta $p_i$ and $p_j$ 
can be equal to any of the internal momenta. Additionally $p_j$ can be 
equal to $p$ or $q$. The scaling identities involve applying one of the 
operators
\begin{eqnarray}
\label{sde2}
q^\mu \frac{\partial}{\partial q^\mu}\, ,\quad\quad\quad
p^\mu \frac{\partial}{\partial q^\mu}\, ,\quad\quad\quad
p^\mu \frac{\partial}{\partial p^\mu}\, . 
\end{eqnarray}
both inside the integral and to the integrated result in Mellin space. The 
fourth operator of this kind cannot be used, because it leads to an 
inequality when applied to $p^2 = 0$. Sometimes a third type of identities 
can be useful. It is along the lines of the Passarino--Veltman 
decomposition in formfactors~\cite{Passarino:1979jh}:
\begin{eqnarray}
\label{eq:formfac}
        \int d^Dp_1 d^Dp_2\,\, p_i^\mu\,\, T1_{kl} & = &
                        q^\mu I_{kl}^{(q)} + p^\mu I_{kl}^{(p)}\, .
\end{eqnarray}
First the two formfactors $I^{(q)}$ and $I^{(p)}$ are determined by 
contracting eq.(\ref{eq:formfac}) either with $q^\mu$ or $p^\mu$. Then 
the relevant identity can be obtained by taking the derivative with respect 
to $q^\mu$. This can be done because the formfactors are just combinations 
of powers of $2p\cdot q$, $q^2$ and a scalar factor. For these kind of 
integrals this method was first introduced in ref.\cite{Larin:thesis}. It is 
especially useful for the three loop integrals. Here we will not need it.

Using all the above equations, and eliminating all integrals that we do not 
need, for each topology one is left with a solvable equation. 
Sometimes however some intermediate integrals occur that need to 
be solved by the same methods. First we give the two basic building blocks 
$T1_{11}$ and $T1_{55}$:
\begin{eqnarray}
\label{eq:t11}
        (N\plus 1\minus 2\ve)\TB({n,N},1,1,1,1) & = &
                 N\frac{2p\mydot q}{q\mydot q}\ \TB({n-1,N-1},1,1,1,1)
        \nonumber \\ &&
                + \frac{n\minus 1}{q\mydot q}\TB({n,N},1,1,0,1)
                - (1\plus\nu/2)\TB({n,N},1,2,0,1)
                + (2\plus\nu)\TB({n,N},2,1,1,0)
        \nonumber \\ &&
                -\frac{1}{q\mydot q}\TB({n-1,N},1,0,1,2)
                +\frac{2\plus\nu/2}{q\mydot q}\TB({n-1,N},1,1,0,2)
                -\frac{2\plus\nu}{q\mydot q}\TB({n-1,N},2,0,1,1) \, ,
        \\
        \nu & = & (n-N-2)/\ve \nonumber\, ,
\end{eqnarray}
\begin{eqnarray}
\label{eq:t55}
\TA(1,1,1,1,n,N) & = &
\frac{1+\sign(N)}{2(N\plus 1\minus 2\ve)(N\minus 2n\plus 2\minus 2\ve)}
 (
\frac{(2p\mydot q)^2}{q\mydot q^2}(N\minus 1)
        (N\minus 2n\plus 4\minus 4\ve)\TA(1,1,1,1,{n-2},{N-2})
\nonumber \\ &&
-2\frac{(2p\mydot q)^2}{q\mydot q^2}(N\minus 1)\TA(0,1,1,2,{n-2},{N-2})
+2\frac{(2p\mydot q)^2}{q\mydot q}(N\minus 1)\TA(0,2,1,1,{n-1},{N-2})
\nonumber \\ &&
+8p\mydot p_3\frac{2p\mydot q}{q\mydot q}(N\minus 1)\TA(0,1,1,2,{n-1},{N-2})
+\frac{2p\mydot q}{q\mydot q}(6N\minus 4n\minus 4\ve)
        \TA(0,1,1,2,{n-1},{N-1})
\nonumber \\ &&
+\frac{2p\mydot q}{q\mydot q}(4(n\minus 1)(3N\minus 4n\plus 3\minus 8\ve)
                \plus 8N\ve \minus 16\ve^2)\TA(0,1,1,1,n,{N-1})
\nonumber \\ &&
+4p\mydot q (2n\minus N\minus 2\plus 2\ve)\TA(0,2,1,1,{n},{N-1})
) \, .
\end{eqnarray}
Because in both cases the first term on the right hand side is of the 
same type as the term on the left hand side, these equations define proper 
recursions. All other terms on the right hand sides can be computed 
directly by conventional methods, see for instance refs.\cite{Gorishnii:1989gt,Larin:1991fz}.

Of course there could be powers of $2p\mydot p_i$ present, or different 
powers of the various denominators, but straightforward application of the 
various relations allows one to reduce all these to the above integrals. In 
the case of $T1_{55}$ one may have to use symmetry considerations.

The composite topology $T1_{12}$ is the easiest topology with $p$-momentum 
flowing through more than one line.
Here we start with applying the scaling equation that is based on 
taking $p^\mu\partial/\partial p^\mu$ of the integral.
\begin{eqnarray}
 (\tilde{N}+2)\TAIs(1,1,1,1,1,1,1) & = &
                2 \TAKs(2,1,1,1,1,1) \, .
\end{eqnarray}
$\tilde{N}$ is the number of powers of $p$ that needs to 
be generated in the expansion of the denominators. If there are $m$ extra 
powers of $2p\mydot q$ it is equal to $N-m$. 
This last diagram can be simply dealt with, using integration by parts in 
the left triangle of the diagram. The final result is
\begin{eqnarray}
 (\tilde{N}+2)\frac{D\minus 5}{2}\TAIs(1,1,1,1,1,1,1) & = &
        2 \TAKs(3,1,1,1,0,1) - 2 \TACs(3,1,1,1,1) \nonumber \\ &&
        + \TAKs(2,1,1,2,0,1) - \TAKs(2,1,0,2,1,1)\, ,
\end{eqnarray}
and we see three trivial integrals and one integral of the type $T1_{11}$, 
albeit with different coefficients after the expansion in $p$.


For the $T1_{14}$ topology we have to consider two steps. The first is to reduce the 
integrals containing $1/(p_1^{2}\,(p+p_1)^{2}p_4^{2}\,(p+p_4)^{2})$. This is 
trivial, because we can use the scaling identity based on $p^\mu
\partial/\partial p^\mu$ which removes either the $1/p_1^{2}$ or the 
$1/p_4^{2}$.
After that we can reroute the momentum $p$ and we are left with a diagram of the 
type $\TAF(a,1,1,d,E,n,1)$ in which $E$ is normally 1 and $d$ is 2. We have to 
be careful with the signs if the diagram has to be turned upside down. We 
are now left with basically one diagram with the $p$ momentum flowing through 
the lines 1 and 5.

Further reduction can happen in a variety of ways. The important thing to pay attention 
to is that during the reduction, one does not generate simpler integrals with an 
additional factor of $1/\ve$. In that case these simpler integrals would have to be expanded to 
higher powers in $\ve$ which may lead to relatively difficult sums. 
Hence we follow a rather elaborate scheme that is free of these problems by 
combining all the equations we can write down for the diagram. It results 
in:
\begin{eqnarray}
  \TAF(1,1,1,2,E,n,1)q\mydot q & = &
                (\tilde{N}+E-n-D+5)\TAF(1,1,1,1,E,n,1)
                +n\TAB(1,1,1,1,E,{n-1})
                \nonumber \\ &&
                +\TAB(1,1,1,2,E,n)
                +E\TAD(1,1,1,1,{E+1},n)
                -n\TAD(1,1,1,1,E,{n-1})
                -E\TAF(1,0,1,1,{E+1},n,1) \, ,
                \\
\label{eq:redu15b}
  \TAF(1,1,2,1,E,n,1)q\mydot q & = &
                (E-n-D+3)\TAF(1,1,1,1,E,n,1)
                -n\TAB(1,1,1,1,E,{n-1})  \nn
                -E\TAD(1,1,1,1,{E+1},n)
                +n\TAD(1,1,1,1,E,{n-1})
                +E\TAF(1,0,1,1,{E+1},n,1)
                +\TAF(1,0,2,1,E,n,1) \, ,
        \\
\label{eq:redu15}
  \TAF(1,1,1,1,E,n,1) & = & \frac{1}{\tilde{N}\plus 1\minus n}(
                \TAD(1,1,1,2,E,n)
                +\TAD(2,1,1,1,E,n)
                -\TAB(1,1,1,2,E,n) \nn
                +\frac{2p\mydot q}{q\mydot q}(
                (D\minus 4\minus\tilde{N}\minus E\plus n)\TAF(1,1,1,1,E,n,1)
                +E\TAF(1,0,1,1,{E+1},n,1)
                -E\TAD(1,1,1,1,{E+1},n) \nn
                -\TAB(1,1,1,2,E,n)
                +n\TAD(1,1,1,1,E,{n-1})
                -n\TAB(1,1,1,1,E,{n-1}) 
        )) \, .
\end{eqnarray}
The second of these equations we will need when we treat the $T1_{13}$ 
topology. 
The third equation leads to a recursion which for the fourth term in the 
right hand side is terminated when enough 
powers of $2p\mydot q/q^2$ have been pulled out. All other terms in 
the right hand side are of a simpler nature. There is no $\ve$ in any 
denominator.


The derivation of the reduction of the $T1_{15}$ topology is relatively 
simple. We use the scaling identity based on $p^\mu
\partial/\partial p^\mu$ to obtain
\begin{equation}
        (\tilde{N}\plus 2)\TAE(1,1,1,1,1,1,1) =
                \TAA(2,1,1,1,1,1)
                +\TAFs(1,1,1,1,2,1)\ .
\end{equation}
The second diagram in the right hand side can be treated with 
eq.(\ref{eq:redu15}).
Hence we have to 
consider only the topology \TAA({ },{ },{ },{ },{ },{ }).
The crucial identity for this diagram is 
obtained in a way similar to the derivation of eq.(\ref{eq:redu15}): 
i.e. we combine a number of triangle and other relations to form:
\begin{eqnarray}
\label{eq:redu56}
        \TAA({n,m},1,1,1,1,1) & = & \frac{1}{\tilde{N}\plus 5\plus n\minus 
        m\minus D} (n\TAA({n+1,m},0,1,1,1,1)
                -n\TACs({n+1,m},1,1,1,1)  \nn
                +\TAA({n,m},1,0,2,1,1)
                -\TABs({n,m},1,1,2,1)
                +m\TACs({n,m-1},1,1,1,1)
                -m\TABs({n,m-1},1,1,1,1)) \, .
\end{eqnarray}
After this there are no more serious problems. The fourth and the sixth 
diagrams are done by rerouting the momentum $p$ through the 4-line and then 
turning the diagrams upside down. This gives them the same topology as the 
fifth diagram. We can do these diagrams by expanding the propagator and 
applying the tables we constructed for the topology $T1_{11}$.

One may note also that because $\tilde{N}-m$ is always positive or zero, 
the denominator never becomes proportional to $\ve$ and hence there 
are no complications with extra singularities in this equation.


In principle $T1_{13}$ is the most complicated topology, but because of all the 
work we have done already things turn out to be rather easy. First we 
notice that the only powers of $p$ in the numerator occur in terms of 
$2p\mydot q$. All others can be eliminated leaving terms that could be of a 
simpler topology. Then we write again 
all the various equations that can be obtained with integration by parts 
and with scaling arguments. 
In this set of equations we reduce all terms that are not of a 
simpler topology and that contain one of the denominators to a higher power.
Eventually we have a rather simple relation left:
\begin{eqnarray}
        \TAL({1},{1},{1},{1},{1},{1},{1})(\frac{2 p\mydot q}{q\mydot q})^k & = &
                -\frac{N\minus k\minus D\plus 6}{N\minus k\plus 2}
                                \TAL(1,1,1,1,1,1,1)(\frac{2 p\mydot q}{q\mydot q})^{k\plus 1}
                \nn
        +\frac{2}{N\minus k\plus 2}\TAFs(1,1,2,1,1,1)(\frac{2 p\mydot q}{q\mydot q})^k \, .
\label{eq:t1v13rec}
\end{eqnarray}
In the first term we have an extra power of $p$ and the second term is of a 
simpler topology. It has already been evaluated in 
eq.(\ref{eq:redu15b}).

Actually, this result in eq.(\ref{eq:t1v13rec}) is a special case of a more general relation. 
Assume we have a general diagram with the structure
\begin{eqnarray}
        \TCa(a,A,B,b) & = & \frac{(2p\mydot q)^N}{(q\mydot q)^{N+n}}c_N \, ,
\end{eqnarray}
in which the blob in the center of the diagram can be anything and the
parameter $n$ is determined by dimensional considerations.
By rerouting the momentum $p$ through the outside line and applying the 
scaling operator $p^\mu\partial/\partial p^\mu$, after which the $p$-flow 
is restored to how it was in the beginning we obtain the relation
\begin{eqnarray}
        (N\plus a\plus b)\TCa(a,A,B,b)
        & = & a\TCa({a+1},{A-1},B,b)
    + b\TCa(a,A,{B-1},{b+1}) \nonumber \\ &&
    -(N\minus 1\plus n)\frac{2p\mydot q}{q\mydot q}\TCa(a,A,B,b) \, .
\end{eqnarray}
It is clear that in the case of $A$ and $B$ being one, the number of lines 
that contain the momentum $p$ will diminish by one. Hence this relation 
will be very useful in the future. In the special case of $T1_{13}$ 
in eq.(\ref{eq:t1v13rec}) the 
first two terms in the right hand side are identical and $n = 7\minus D$.

Many of these equations involve recursion relations and their solution can 
be written down as a single sum over the simpler diagram(s). Each of these 
sums is a socalled single parameter sum. With this we mean that the summand 
after expansion in $\ve$ is only a function of the summation 
parameter and the upper limit of the summation, the lower limit being 
either zero or one. In this hides the power of the method, because this 
avoids multiple sums that cannot be done with current techniques.
 
\subsection{Solutions}

The equations for the various composite topologies are mostly recursion 
relations, or in another terminology: first order difference equations. 
Suppose we have the equation
\begin{eqnarray}
        a(N)F(N)-b(N)F(N\minus 1)-G(N) & = & 0\, ,
\end{eqnarray}
then its solution will be
\begin{eqnarray}
\label{eq:firstsol}
        F(N) & = & \frac{\prod_{j=1}^N b(j)}{\prod_{j=1}^N a(j)}F(0)
                +\sum_{i=1}^N\frac{\prod_{j=i+1}^N b(j)}{\prod_{j=i}^N a(j)}G(i)
                \, .
\end{eqnarray}
In the case that the functions $a(N)$ and $b(N)$ can be factorized in 
linear polynomials in $N$ with the coefficient of $N$ being one and the 
other coefficients being integers, the products can be written as 
combinations of $\Gamma$-functions. Because $a$ and $b$ may also depend on 
the parameter $\ve$ the $\Gamma$-functions should be expanded around 
$\ve = 0$. This will lead to factorials and harmonic sums. 
The diagram indicated by $F(0)$ in eq.(\ref{eq:firstsol}) 
can be evaluated with standard MINCER 
techniques~\cite{Gorishnii:1989gt,Larin:1991fz}. If the 
function $G(N)$ is expressed as a power series in $\ve$ with the 
coefficients being combinations of harmonic sums in $N+m$ and powers of 
$N+m$, $m$ being a fixed integer,  the sum in eq.(\ref{eq:firstsol}) 
can be done and $F(N)$ will be a combination of harmonic sums in $N+k$ and 
powers of $N+k$ with $k$ being a fixed integer.

Eventually, all solutions to the recursion relations, which were derived above, 
can be written in terms of harmonic sums. 
Thus, before we continue we will give some attention to these functions that 
describe the solutions in Mellin space.

\subsection{Harmonic sums}

Harmonic sums are the basic functions in Mellin space and much about them can be found in 
refs.\cite{Gonzalez-Arroyo:1979df,Vermaseren:1998uu,Blumlein:1998if}.
We follow here the notations and the definitions of 
ref.\cite{Vermaseren:1998uu}. Harmonic sums are defined by
\begin{eqnarray}
\label{eq:basicharmo}
S_m(N) = \sum\limits_{i=1}^N  \frac{1}{i^m}\, , \quad 
S_{-m}(N) = \sum\limits_{i=1}^N  \frac{(-1)^m}{i^m}\, ,
\end{eqnarray}
while higher functions can be defined recursively
\begin{eqnarray}
\label{eq:higherharmo1}
S_{m_1,...,m_k}(N) &=& 
        \sum\limits_{i=1}^N  \frac{1}{i^{m_1}} S_{m_2,...,m_k}(i)\, , \\
\label{eq:higherharmo2}
S_{-m_1,...,m_k}(N) &=& 
        \sum\limits_{i=1}^N  \frac{(-1)^{m_1}}{i^{m_1}} S_{m_2,...,m_k}(i)\, .
\end{eqnarray}
These functions appear, amoung others, when $\Gamma$-functions are expanded 
in terms of $\ve$. But they also show up in sums of the type
\begin{eqnarray}
        \sum_{i=1}^n \sign(i) \binom(n,i) \frac{1}{n^3} & = &
                        -S_{1,1,1}(n)\, ,
\end{eqnarray}
which are rather common in our calculations. 

The weight of a sum is defined as the sum of the absolute values of all the 
indices $m_i$.
These sums form an algebra. Hence the product of two sums with the same 
argument can be written as a sum of terms, each with a single sum of which 
the weight is the sum of the weights of the original two sums.
Many sums involving these harmonic sums can be done by automatic algorithms 
to any level of complexity. For us the important sums are
\begin{eqnarray}
\sum_{i=1}^n\frac{S_{\vec{p}}(i)S_{\vec{q}}(n-i)}{i^m},\nonumber \\
\sum_{i=1}^n\sign(i)\frac{S_{\vec{p}}(i)S_{\vec{q}}(n-i)}{i^m},\nonumber \\
\sum_{i=1}^n\sign(i)\binom(n,i)\frac{S_{\vec{p}}(i)}{i^m},
                \nonumber \\
\sum_{i=1}^n\sign(i)\binom(n,i)\frac{S_{\vec{p}}(i)S_{\vec{q}}(n-i)}{i^m},
                \nonumber
\end{eqnarray}
in which we interpret an $S$ without indices as $1$ and
$\vec{p}=p_1,\dots,p_v$; $\vec{q}=q_1,\dots,q_w$.  
The evaluation of these sums is described in~\cite{Vermaseren:1998uu} and 
has been programmed in the language of FORM~\cite{FORM}. The important 
thing to note is that these sums all evaluate into harmonic sums and 
denominators of the same argument as the harmonic sums. In practise this 
means that with the sums we run into we will remain inside the space of 
harmonic sums and hence the harmonic sums will span the solution space of 
our integrals in Mellin space.

\subsection{Harmonic Polylogarithms}

In $x$-space the results will be presented in terms of harmonic 
polylogarithms~\cite{Remiddi:1999ew}. These functions are related to the 
multi-dimensional polylogarithms of ref.\cite{Broadhurst:1998}.
They form a natural basis in the space of inverse Mellin transformations of 
harmonic sums. We will mention here their most important properties. For 
more details the reader should consult ref.\cite{Remiddi:1999ew}.

The harmonic polylogarithms of weight $w$ and with argument $x$ 
are identified by a set of $m_1,...,m_w$ indices 
which can take one of the three values $0, 1, -1$. 
Harmonic polylogarithms are denoted by $H_{m_1,...,m_w}(x)$ and explicitly, 
for lowest weight one defines 
\begin{eqnarray}
  H_{0}(x) &=& \ln{x} \, ,          \\ 
  H_{1}(x) &=& \int_0^x \frac{dx'}{1-x'} = - \ln(1-x) \, ,\\ 
  H_{-1}(x) &=& \int_0^x \frac{dx'}{1+x'} = \ln(1+x) \, . 
\label{eq:defineh1}
\end{eqnarray}
For their derivatives, one has 
\begin{equation}
  \frac{d}{dx} H_m(x) = f_m(x) \ , 
\label{eq:derive1} 
\end{equation}
where again the index $m$ can take the values $0, +1, -1$ and the 
three rational fractions $f_m(x)$ are given by

\begin{eqnarray}
   f_0(x) = \frac{1}{x} \, ,\quad\quad\quad
   f_1(x) = \frac{1}{1-x} \, ,\quad\quad\quad
   f_{-1}(x) = \frac{1}{1+x} \, . 
\label{eq:definef}
\end{eqnarray}
In general, the harmonic polylogarithms of weight $w$ 
are then defined as follows,
\begin{eqnarray}
H_{m_1,...,m_w}(x) = \frac{1}{w!} \ln^w{x} \, ,    \quad\quad\quad 
{\rm{if}}\quad m_1,...,m_w = 0,...,0\, ,
\label{eq:defh0}
\end{eqnarray}
while, if $ m_1,...,m_w \neq 0,...,0$
\begin{eqnarray}
H_{m_1,...,m_w}(x) = \int_0^x dz \ f_{m_1}(z) \ H_{m_2,...,m_w}(z) \ .
\label{eq:defn0}
\end{eqnarray}
To provide a link to the standard literature, we give the results 
for harmonic polylogarithms up to weight three in terms of common 
polylogarithms in the appendix A.

Just like the harmonic sums the $H$-functions form an algebra. This is to 
say:
products of two $H$-functions, $H_{\vec{m}_w}(x) H_{\vec{n}_v}(x)$ 
of weight $w$ and $v$ respectively, 
where $\vec{m}_w = m_1,...,m_w$ 
and $\vec{n}_v = n_1,...,n_v$ can be expressed in a sum of 
$H$-function of weight $w+v$,
\begin{eqnarray}
 H_{\vec{m}_w}(x)H_{\vec{n}_v}(x) & = &
  \sum_{{\vec{l}_{w+v}} = \vec{m}_w \uplus \vec{n}_v} H_{\vec{l}_{w+v}}(x) \, ,  
\label{eq:halgebra} 
\end{eqnarray}
in which $\vec{m}_w \uplus \vec{n}_v$ represents all mergers of 
$\vec{m}_w$ and $\vec{n}_v$ 
in which the relative orders of the elements of 
$\vec{m}_w$ and $\vec{n}_v$ are preserved. Hence the sum consists of 
$(v+w)!/v!w!$ terms. This can be shown by induction using 
integration by parts and eq.(\ref{eq:defn0}). It should be realized that 
the rules of this algebra are complementary to those of the sums in the 
sense that the algebra of $H$-functions in $x$ is related to the extra 
algebraic rules for $S$-sums in infinity, while the extra algebraic rules 
for $H$-functions in 1 are related to the algebraic rules for $S$-sums in 
$N$.

The harmonic polylogarithms may be expanded in a Taylor series 
with the expansion coefficients being harmonic sums,
\begin{eqnarray}
 H_{\vec{m}}(x) & = &
  \sum_{i=1}^{\infty} \sum_{\vec{n}} c_{\vec{m}\vec{n}}
                 \frac{\sigma_{\vec{n}}^i\, x^i}{i}\, S_{\vec{n}}(i)\, ,
\label{eq:hexpansion} 
\end{eqnarray}
where $\sigma = \pm 1$ and $S_{\vec{n}}$ is a harmonic sum of weight $v$, 
$\vec{n} = n_1,...,n_w$. In general eq.(\ref{eq:hexpansion}) 
only holds, if $\vec{m} = m_1,...,m_w$ has no trailing zeroes in the 
index field. Those correspond to factors $\ln(x)$, which do not admit 
a regular Taylor expansion. However, trailing zeroes in the 
index field can be factored out by repeated use of the product identity 
eq.(\ref{eq:halgebra}), such that eq.(\ref{eq:hexpansion}) can safely be 
applied to the left-over $H$-function, which by construction does not 
contain powers of $\ln(x)$ anymore.

Finally, we define the Mellin transformation of regular functions as 
\begin{eqnarray}
\label{eq:mellin}
 f(N) & = & \int_0^1dx\ x^{N-1} f(x) \, .
\end{eqnarray}
The Mellin transformation of $1/(1-x)$ and possible powers of 
logarithms $\ln(1-x)$ are regularized in the sense of $+$-distributions. 
For this we have to extract first the powers of $\ln(1-x)$ which correspond 
to leading indices with the value $1$. The extraction is similar to the 
extraction of trailing zeroes. Hence:
\begin{eqnarray}
\label{eq:mellinplus}
 \int_0^1dx\ x^{N-1}\, \left[\frac{H_{1,...,1}(x)H_{\vec{m}}(x)}{1-x}\right]_+ 
&=&
   \int_0^1dx\ \left(x^{N-1}H_{\vec{m}}(x) -
                        H_{\vec{m}}(1) \right) \frac{H_{1,...,1}(x)}{1-x}\, ,
\end{eqnarray}
in which $\vec{m}$ does not have a $1$ for its first element. 

Together the above relations allow the construction of the Mellin transform 
of any $H$-function. It turns out that there is a one to one relationship 
between functions of the type $H_{\vec{m}}(x)/(1\pm x)$ with $\vec{m}$ having 
weight $w$ and $S$-sums of weight $w+1$. Hence it is possible to construct 
the inverse Mellin transform of any result in $N$ space that can be 
expressed in terms of harmonic sums. The complete algorithm has been coded 
in the language of FORM~\cite{FORM}.

\section{Calculation and Results
\label{sec:CalculationResults}}

In this section, we will discuss our calculation of the DIS structure functions 
and list our results obtained for 
the complete set of anomalous dimensions, the flavour non-singlet 
and singlet quark and the gluon coefficient functions up to two loops.

However, first a few remarks on details are in order.
All Feynman diagrams contributing to the structure functions 
have been generated with the help of QGRAF~\cite{Nogueira:1991ex} 
and have been stored in a database. 
For the calculation of the even Mellin moments of the partonic invariants 
$T^N_{2,{\rm{p}}}$ and $T^N_{L,{\rm{p}}}$ we used a database identical 
with the one in ref.\cite{Larin:1997wd}, which contains 425 diagrams up to 
two loops. 
To obtain the full information about $T^N_{3,{\rm{p}}}$ and the complete non-singlet 
contribution to $T^N_{2,{\rm{p}}}$, we have generated a new database of 360 diagrams 
corresponding to the four different structure functions $F_{2}^{\n{\rm{P}}\pm{\Bar{\n}{\rm{P}}}}$ 
and $F_{3}^{\n{\rm{P}}\pm{\Bar{\n}{\rm{P}}}}$.

Both databases have been checked by calculating some lower fixed Mellin moments 
in an arbitrary covariant gauge 
with the MINCER algorithm~\cite{Larin:1991fz}, 
keeping the gauge parameter $\x$ in the 
gluon propagator, that is to say,
\begin{eqnarray}
{\rm{i}}\  \frac{-g^{\m\n} + (1-\x)q^\m q^\n}{q^2 - {\rm{i}} \e}\, ,
\end{eqnarray}
and in the final result all dependence on $\x$ does cancel.

The calculation of the partonic invariants $T^N_{i,{\rm{p}}}$,  
$i=2,3,L$ and  $T^N_{\f,{\rm{p}}}$ requires to consider only 
external quarks and gluons with physical polarization. 
In the case of external quarks, the sum over all polarizations leads 
to the projection operator $\slash{p}$ that closes the open string 
of Dirac matrices associated with the external quark line. 
For external gluons on the other hand, the sum over all physical polarizations 
can be done by contracting the external gluon lines with 
\begin{eqnarray}
\label{gluon-pol-proj}
-g^{\a\b} + \frac{p^\a q^\b + p^\b q^\a}{p \mydot q} - \frac{p^\a p^\b q^2}{(p \mydot q)^2}\, ,
\end{eqnarray}
where $p$ is on-shell, $p^2 = 0$. 
An alternative approach contracts 
the external gluon lines only with $-g^{\a\b}$. 
To remove the unphysical contributions, one adds extra classes of diagrams, 
with external ghosts instead of external gluons. 
The latter approach has been used in ref.\cite{Larin:1997wd} 
while we have checked that both methods agree for our calculation 
of the $N$-th Mellin moment up to two loops.

The explicit calculation of the DIS structure functions 
has been done with the symbolic manipulation program FORM~\cite {FORM}.
To that end, all recursion relations given in section~\ref{sec:Method} 
have been implemented in a program, 
that reduces the Feynman diagrams of DIS structure functions 
up to two loops to multiple nested sums over the basic building blocks.
Subsequently the program calls the SUMMER algorithm~\cite{Vermaseren:1998uu}, 
to solve these nested sums in terms of the basis of harmonic sums. 
To speed up the calculation the results for a large number of 
basic integrals have been tabulated.

Let us emphasize that the method of recursion relations as described 
in section~\ref{sec:Method} allows for numerous checks at all stages 
of the calculation by means of the standard MINCER routine~\cite{Larin:1991fz}.
This is from a practical view point by far the most powerful feature 
of our approach, because the debugging of all our programs is extremely 
simplified.

A final remark is concerned with the analytic continuation of our $N$-space results. 
In the framework of the OPE we calculate either the even or the odd Mellin moments 
for a given structure function depending on the operators that contribute 
as detailed in section~\ref{sec:Formalism}. 
This is sufficient to determine all moments in the complex 
$N$ plane by analytic continuation.  
In order to do so we first perform the inverse Mellin transformation 
of our $N$-space result to obtain the complete expression in $x$-space. 
This mapping is unique provided we fix all factors of $(-1)^N$ 
according to whether we started from even or odd Mellin moments.
Subsequently, we may take this $x$-space result and execute 
another Mellin transformation back to $N$-space to obtain 
the final answer in terms of harmonic sums valid for all non-negative 
integer Mellin moments.
This step of restoring all those factors of $(-1)^N$ 
which are due to analyticity 
is particularly important for the determination of the flavour non-singlet 
anomalous dimensions and coefficient functions at two loops as will be detailed below.

\subsection{Renormalization and mass factorization}

Let us now concentrate on the issue of renormalization and mass 
factorization for the Mellin moments of the DIS structure functions. We 
wish to show explicitly how to extract the anomalous dimensions and 
coefficient functions from eqs.(\ref{TmunuPartonMomNS}), 
(\ref{TmunuPartonMomS}) and (\ref{TmunuPartonMomPhi}) if the partonic 
invariants  $T^N_{i,{\rm{p}}}$ and $T^N_{\f,{\rm{p}}}$ are expanded in 
powers of $\a_s$ and $\ve$. In particular, we will obtain the 2-loop 
anomalous dimensions
$\g^{(1)}_{\rm{gq}}$ and $\g^{(1)}_{\rm{gg}}$ from the invariants $T^N_{\f,
{\rm{p}}}$ 
of the scalar particle.

The calculation is performed in dimensional regularization 
$D=4-2\ve$,~\cite{'tHooft:1972fi,Bollini:1972ui}.
We briefly recall the procedure outlined in section~{\ref{sec:Formalism} and in  
refs.\cite{Zijlstra:1991qc,Larin:1993fv,Larin:1997wd}. 
The sum of all Feynman diagrams contributing to a given 
partonic invariant $T^N_{i,{\rm{p}}}$ or $T^N_{\f,{\rm{p}}}$ 
contains only ultraviolet and collinear divergences. 
The ultraviolet divergences need to be removed by coupling constant 
renormalization changing the bare $\a_s^{\rm{bare}}$ to the renormalized $\a_s$ 
as defined in eq.(\ref{alpha-s-renorm}).
If necessary, the additional renormalizations associated with 
the axial current, $\g_5$ and the interaction term $\f F_{\m\n}^{a}F^{\m\n}_a$  
of the scalar particle $\f$ have to be taken into account as described 
in section~\ref{sec:Formalism} and given 
in eqs.(\ref{overall-ren}) and (\ref{overall-ren-g5}).

Then one is left with the collinear divergences associated with the partonic 
initial states. Those need to be removed by mass factorization 
changing the bare densities of partons in the hadron to renormalized ones.
This is the same as renormalizing the operator matrix elements $A^j_{{\rm{p}},N}$ 
as done in eqs.(\ref{TmunuPartonMomNS}), (\ref{TmunuPartonMomS}) 
and (\ref{TmunuPartonMomPhi}). 

In the following, we list the results for the Mellin moments of the 
parton invariants order by order in $\a_s$. 
To that end it is useful to define the following 
expansions in powers of $\a_s$
for the coefficient functions and the anomalous dimensions,
\begin{eqnarray}
\g_{\rm{pp}}\left(\a_s\right) &=& \sum\limits_{n=0}^\infty\, 
  \left(\frac{\a_s}{4 \p}\right)^{n+1}\, \g^{(n)}_{\rm{pp}} \, ,\\
C_{i,\rm{p}}\left(\a_s\right) &=& \sum\limits_{n=0}^\infty\, 
  \left(\frac{\a_s}{4 \p}\right)^{n}\, c^{(n)}_{i,\rm{p}}\, ,
\end{eqnarray}
and, since the operator matrix elements $A_{{\rm{p}},N}^{\rm p, tree}$ factorize
after application of the projector $\cp_N$, also for the parton invariants,
\begin{eqnarray}
\label{partinv-exp}
T^{N}_{i,{\rm{p}}} &=& \left( 
T^{(0)}_{i,{\rm{p}}} + T^{(1)}_{i,{\rm{p}}} + T^{(2)}_{i,{\rm{p}}} + \dots 
\right) A^{{\rm p},{\rm{tree}}}_{{\rm p},N}\, ,
\end{eqnarray}
where all left-over singularities in the $T^{(l)}_{i,{\rm{p}}}$ 
are of collinear nature. 
The procedure of mass factorization works iteratively order by order, 
both in $\a_s$ and in the regularization parameter $\ve$, 
such that all physical quantities can be uniquely extracted. 
Notice in particular, that the gluon tree level operator matrix element 
$A^{{\rm g},{\rm{tree}}}_{{\rm g},N}$ is $(1-\ve) {\rm{const}}_N$.   
If factorized as in eq.(\ref{partinv-exp}), 
the factor $(1-\ve)$ due to the gluon polarization in $D$ dimensions 
is essential for a proper determination of the 1-loop coefficient function 
$c^{(1)}_{2,{\rm{g}}}$. 
As discussed in ref.\cite{Larin:1993fv}, the omission of the factor $(1-\ve)$ 
accounts for some of the differences with ref.\cite{Kazakov:1988jk}
for the structure function $F_L$.{\footnote{
The discrepancies were corrected in ref.\cite{Kazakov:1992xj}.}}

All results are given in dimensional regularization in the 
${\Bar{\rm{MS}}}$-scheme with $\a_s$ being the renormalized quantity 
according to eq.(\ref{alpha-s-renorm}). 
At leading order, we have normalized,
\begin{eqnarray}
\label{F-0}
T^{(0)}_{2,{\rm{q}}} = T^{(0)}_{3,{\rm{q}}} = 1\, , \quad\quad\quad\  
T^{(0)}_{2,{\rm{g}}} = T^{(0)}_{L,{\rm{q}}} = T^{(0)}_{L,{\rm{g}}} = 0\, .
\end{eqnarray}

At first order in $\a_s$, we have to expand up to order $\ve$ and find,
\begin{eqnarray}
\label{F2-NS-1}
T^{(1)}_{2,{\rm{q}}} &=& \frac{\a_s}{4 \p}\, S_\ve\, \left( \frac{\m^2}{Q^2} \right)^{\ve}
        \left[ \frac{1}{\ve} \g^{(0)}_{\rm{qq}} + 
                c^{(1)}_{2,{\rm{q}}}
                        + \ve a^{(1)}_{2,{\rm{q}}}
                \right] \, , \\[1ex]
\label{F3-NS-1}
T^{(1)}_{3,{\rm{q}}} &=& \frac{\a_s}{4 \p}\, S_\ve\, \left( \frac{\m^2}{Q^2} \right)^{\ve}
        \left[ \frac{1}{\ve} \g^{(0)}_{\rm{qq}} + 
        c^{(1)}_{3,{\rm{q}}}
                        + \ve a^{(1)}_{3,{\rm{q}}}
                \right] \, , \\[1ex]
\label{F2-G-1}
T^{(1)}_{2,{\rm{g}}} &=& n_f\, \frac{\a_s}{4 \p}\, S_\ve\, \left( \frac{\m^2}{Q^2} \right)^{\ve}
        \left[ \frac{1}{\ve} \g^{(0)}_{\rm{qg}} 
                + c^{(1)}_{2,{\rm{g}}}
                        + \ve a^{(1)}_{2,{\rm{g}}}
                \right] \, , \\[1ex]
\label{FL-NS-1}
T^{(1)}_{L,{\rm{q}}} &=& \frac{\a_s}{4 \p}\, S_\ve\, \left( \frac{\m^2}{Q^2} \right)^{\ve}
        \left[ c^{(1)}_{L,{\rm{q}}} + \ve a^{(1)}_{L,{\rm{q}}}
                \right] \, , \\[1ex]
\label{FL-G-1}
T^{(1)}_{L,{\rm{g}}} &=&  n_f\, \frac{\a_s}{4 \p}\, S_\ve\, \left( \frac{\m^2}{Q^2} \right)^{\ve}
        \left[ c^{(1)}_{L,{\rm{g}}} + \ve a^{(1)}_{L,{\rm{g}}}
                \right] \, ,
\end{eqnarray}
where the factor $S_\ve$ is defined by 
\begin{eqnarray}
S_\ve = \exp\left( \ve\{\ln(4\p) - \g_{\rm{E}}\} \right)\, .
\end{eqnarray}

At second order in $\a_s$, we need to split up the contributions into flavour 
non-singlet and singlet parts.
Allowing for electroweak interactions, one can consider 
in the non-singlet case the structure functions of four different physical processes,
$F_{2}^{\n{\rm{P}}\pm{\Bar{\n}{\rm{P}}}}$ and $F_{3}^{\n{\rm{P}}\pm{\Bar{\n}{\rm{P}}}}$.
This implies that we also need to distinguish even and odd moments. 
We have 
\begin{eqnarray}
\label{F2-NS-2}
T^{(2),\rm{ns},\pm}_{2,{\rm{q}}} &=& \left( \frac{\a_s}{4 \p} \right)^2 S_\ve^2\,
\left( \frac{\m^2}{Q^2} \right)^{2\ve}
        \left[ \frac{1}{\ve^2} \left\{ \frac{1}{2} \left(\g^{(0)}_{\rm{qq}}\right)^2 
                - \frac{1}{2} \b_0 \g^{(0)}_{\rm{qq}} \right\} \right. \\
& &
\left.
        + \frac{1}{\ve} \left\{ 
                \frac{1}{2} \g^{(1),\pm,\rm{V}}_{\rm{qq}} 
                + \g^{(0)}_{\rm{qq}} c^{(1)}_{2,{\rm{q}}} 
                \right\} 
+ c^{(2),\rm{ns},+}_{2,{\rm{q}}} \pm c^{(2),\rm{ns},-}_{2,{\rm{q}}}
+ \g^{(0)}_{\rm{qq}} a^{(1)}_{2,{\rm{q}}} 
                \right] , \nonumber\\[1ex]
\label{F3-NS-2}
T^{(2),\rm{ns},\pm}_{3,{\rm{q}}} &=& \left( \frac{\a_s}{4 \p} \right)^2 S_\ve^2\,
\left( \frac{\m^2}{Q^2} \right)^{2\ve}
        \left[ \frac{1}{\ve^2} \left\{ \frac{1}{2} \left(\g^{(0)}_{\rm{qq}}\right)^2 
                - \frac{1}{2} \b_0 \g^{(0)}_{\rm{qq}} \right\} \right. \\
& &
\left.
        + \frac{1}{\ve} \left\{ 
                \frac{1}{2} \g^{(1),\pm,\rm{V}}_{\rm{qq}} 
                + \g^{(0)}_{\rm{qq}} c^{(1)}_{3,{\rm{q}}} 
                \right\} 
+ c^{(2),\rm{ns},+}_{3,{\rm{q}}} \pm c^{(2),\rm{ns},-}_{3,{\rm{q}}}
+ \g^{(0)}_{\rm{qq}} a^{(1)}_{3,{\rm{q}}} 
                \right] ,\nonumber\\[1ex]
\label{FL-NS-2}
T^{(2),\rm{ns}}_{L,{\rm{q}}} &=& \left(\frac{\a_s}{4 \p} \right)^2 S_\ve^2\,
\left( \frac{\m^2}{Q^2} \right)^{2\ve}
        \left[ \frac{1}{\ve} \left\{ 
                \g^{(0)}_{\rm{qq}} c^{(1)}_{L,{\rm{q}}}
                \right\} 
+ c^{(2),\rm{ns}}_{L,{\rm{q}}}
+ \g^{(0)}_{\rm{qq}} a^{(1)}_{L,{\rm{q}}}
                \right] .
\end{eqnarray}
where the $\pm$-sign denotes even or odd moments 
in the expressions for $T^{(2),{\rm{ns}},\pm}_{2,{\rm{q}}}$ 
and $T^{(2),{\rm{ns}},\pm}_{3,{\rm{q}}}$.
We want to emphasize this distinction for even and odd moments as done 
in eqs.(\ref{F2-NS-2}) and (\ref{F3-NS-2}) is due to the fact that we 
deal with different scattering processes. 

It is relevant for the 2-loop anomalous dimensions entering 
in eqs.(\ref{F2-NS-2}) and (\ref{F3-NS-2}) 
because of flavour symmetry breaking in the anti-quark distributions.
We use the notation of eq.(\ref{splitting-functions-ns})
\begin{eqnarray}
\label{splitting-V-pm}
\g^{(1),\pm,\rm{V}}_{\rm{qq}} 
&=&
\g^{(1),\rm{V}}_{\rm{qq}} \pm \g^{(1),\rm{V}}_{\rm{q{\Bar{q}}}} \, .
\end{eqnarray}
It is also relevant for the 2-loop coefficient functions, which decompose 
for even and odd moments as 
\begin{eqnarray}
\label{c2-2loop-decomp}
c_{2,{\rm q}}^{(2),\rm ns} &=& c_{2,{\rm q}}^{(2),\rm ns,+} 
        + (-1)^N  c_{2,{\rm q}}^{(2),\rm ns,-}\, ,\\ 
\label{c3-2loop-decomp}
c_{3,{\rm q}}^{(2),\rm ns} &=& c_{2,{\rm q}}^{(2),\rm ns,+} 
        + (-1)^N  c_{3,{\rm q}}^{(2),\rm ns,-}\, .
\end{eqnarray}
After analytical continuation of our $N$-space results 
we can reconstruct these physical signs $(-1)^N$
in eqs.(\ref{splitting-V-pm})--(\ref{c3-2loop-decomp}).
That is to say, we can then determine $\g_{{\rm{q}}{\rm{q}}}^{\rm{V}}$ and 
$\g_{{\rm{q}}\bar{{\rm{q}}}}^{\rm{V}}$ by taking the sum or the difference of 
$\g_{{\rm{q}}{\rm{q}}}^{+,\rm{V}}$ and $\g_{{\rm{q}}{\rm{q}}}^{-,\rm{V}}$. 
Analogously, the individual contributions to the 
non-singlet coefficient functions at two loops, 
$c_{2,{\rm q}}^{\rm ns,+}$, $c_{2,{\rm q}}^{\rm ns,-}$, 
$c_{3,{\rm q}}^{\rm ns,+}$, and $c_{3,{\rm q}}^{\rm ns,-}$,
are obtained, but again 
the reconstruction works only after analytical continuation.

Note also, that in the equations above the partonic invariants $T^{(1)}_{3,{\rm{q}}}$ and 
$T^{(2),{\rm{ns}}}_{3,{\rm{q}}}$ are always 
understood to be renormalized according to eq.(\ref{overall-ren-g5}). 

We decompose the quark singlet invariant $T^{(2),\rm{s}}_{i,{\rm{q}}}$ 
into non-singlet and pure-singlet contributions,
\begin{eqnarray}
T^{(2),\rm{s}}_{2,{\rm{q}}} \,=\, T^{(2),\rm{ns},+}_{2,{\rm{q}}} 
                + T^{(2),\rm{ps}}_{2,{\rm{q}}}\, ,
\quad\quad\quad\quad\quad
T^{(2),\rm{s}}_{L,{\rm{q}}} \,=\, T^{(2),\rm{ns}}_{L,{\rm{q}}} 
                + T^{(2),\rm{ps}}_{L,{\rm{q}}}\, ,
\end{eqnarray}
and for the pure-singlet contributions we have at second order in $\a_s$, 
\begin{eqnarray}
\label{F2-PS-2}
T^{(2),\rm{ps}}_{2,{\rm{q}}} &=& n_f\, \left(\frac{\a_s}{4 \p} \right)^2 S_\ve^2\,
\left( \frac{\m^2}{Q^2} \right)^{2\ve}
        \left[ \frac{1}{\ve^2} 
\left\{ \frac{1}{2} \g^{(0)}_{\rm{qg}} \g^{(0)}_{\rm{gq}} \right\} \right. \\
& &\hspace*{30mm}
\left.
        + \frac{1}{\ve} \left\{ \frac{1}{2} \g^{(1),+,\rm{S}}_{\rm{qq}} 
                + \g^{(0)}_{\rm{gq}} c^{(1)}_{2,{\rm{g}}} \right\} 
+ c^{(2),\rm{ps}}_{2,{\rm{q}}} + \g^{(0)}_{\rm{gq}} a^{(1)}_{2,{\rm{g}}}
                \right] , \nonumber \\[1ex]
\label{FL-PS-2}
T^{(2),\rm{ps}}_{L,{\rm{q}}} &=& n_f\, \left( \frac{\a_s}{4 \p} \right)^2 S_\ve^2\,
\left( \frac{\m^2}{Q^2} \right)^{2\ve}
        \left[ \frac{1}{\ve} \left\{ 
                \g^{(0)}_{\rm{gq}} c^{(1)}_{L,{\rm{g}}} \right\} 
+ c^{(2),\rm{ps}}_{L,{\rm{q}}} + \g^{(0)}_{\rm{gq}} a^{(1)}_{L,{\rm{g}}}
                \right] .
\end{eqnarray}
In eq.(\ref{F2-PS-2}) we used again the notation for the anomalous 
dimensions of eqs.(\ref{splitting-functions-s-1}) and (\ref{splitting-functions-s-2})
\begin{eqnarray}
\g^{(1),\pm,\rm{S}}_{\rm{qq}} 
&=&\g^{(1),\rm{S}}_{\rm{qq}} \pm \g^{(1),\rm{S}}_{\rm{q{\Bar{q}}}} \, .
\end{eqnarray}
The pure singlet contribution to $T^{(2),{\rm{ps}}}_{3,\rm q}$ 
vanishes, implying that we have 
\begin{eqnarray}
 \g^{(1),-,\rm{S}}_{\rm{qq}} \,=&
\g^{(1),\rm{S}}_{\rm{qq}} - \g^{(1),\rm{S}}_{\rm{q{\Bar{q}}}} 
&=\, 0\, .
\end{eqnarray}

For the gluonic invariants, we find at second order in $\a_s$,
\begin{eqnarray}
\label{F2-G-2}
T^{(2)}_{2,{\rm{g}}} &=& n_f\, \left( \frac{\a_s}{4 \p} \right)^2 S_\ve^2\,
\left( \frac{\m^2}{Q^2} \right)^{2\ve}
        \left[ \frac{1}{\ve^2} \left\{ \frac{1}{2} \g^{(0)}_{\rm{qg}} 
                \left( \g^{(0)}_{\rm{qq}} + \g^{(0)}_{\rm{gg}}  \right) 
                - \frac{1}{2} \b_0 \g^{(0)}_{\rm{qg}} \right\} \right. \\
& &
\left.
        + \frac{1}{\ve} \left\{ \frac{1}{2} \g^{(1)}_{\rm{qg}} 
                + \g^{(0)}_{\rm{gg}} c^{(1)}_{2,{\rm{g}}} 
                + \g^{(0)}_{\rm{qg}} c^{(1)}_{2,{\rm{q}}} 
                \right\} 
+ c^{(2)}_{2,{\rm{g}}}
        + \g^{(0)}_{\rm{gg}} a^{(1)}_{2,{\rm{g}}}
        + \g^{(0)}_{\rm{qg}} a^{(1)}_{2,{\rm{q}}}
                \right] , \nonumber\\[1ex]
\label{FL-G-2}
T^{(2)}_{L,{\rm{g}}} &=& n_f\, \left( \frac{\a_s}{4 \p} \right)^2 S_\ve^2\,
\left( \frac{\m^2}{Q^2} \right)^{2\ve}
        \left[ \frac{1}{\ve} \left\{ 
                  \g^{(0)}_{\rm{gg}} c^{(1)}_{L,{\rm{g}}}
                + \g^{(0)}_{\rm{qg}} c^{(1)}_{L,{\rm{q}}} 
                \right\} 
+ c^{(2)}_{L,{\rm{g}}}
        + \g^{(0)}_{\rm{gg}} a^{(1)}_{L,{\rm{g}}}
        + \g^{(0)}_{\rm{qg}} a^{(1)}_{L,{\rm{q}}}
                \right] . 
\end{eqnarray}

Finally, we can give the expressions for the 
partonic invariants of the scalar particle $\f$. 
We find at leading order,
\begin{eqnarray}
T^{(0)}_{\f,{\rm{q}}} = 0\, , \quad\quad\quad
T^{(0)}_{\f,{\rm{g}}} = 1\, .
\end{eqnarray}
In the equations below the partonic invariants $T^{(1)}_{\f,{\rm{p}}}$ and 
$T^{(2)}_{\f,{\rm{p}}}$ are always 
understood to be renormalized according to eq.(\ref{overall-ren}).
Then we obtain at first order in $\a_s$,
\begin{eqnarray}
T^{(1)}_{\f,{\rm{q}}} &=& \frac{\a_s}{4 \p}\, S_\ve\, \left( \frac{\m^2}{Q^2} \right)^{\ve} 
        \left[ \frac{1}{\ve} \g^{(0)}_{\rm{gq}} + c^{(1)}_{\f,{\rm{q}}}
                \right] \, , \\[1ex]
T^{(1)}_{\f,{\rm{g}}} &=& \frac{\a_s}{4 \p}\, S_\ve\, \left( \frac{\m^2}{Q^2} \right)^{\ve} 
        \left[ \frac{1}{\ve}  \g^{(0)}_{\rm{gg}} + c^{(1)}_{\f,{\rm{g}}}
                \right] \, ,
\end{eqnarray}
where we only have to expand up the finte terms in $\ve$.

Finally, at second order in $\a_s$,
\begin{eqnarray}
\label{F-phiq-2}
T^{(2)}_{\f,{\rm{q}}} &=& \left( \frac{\a_s}{4 \p} \right)^2 S_\ve^2\,
\left( \frac{\m^2}{Q^2} \right)^{2\ve}
        \left[ \frac{1}{\ve^2} \left\{ \frac{1}{2} \g^{(0)}_{\rm{gq}} 
                \left( \g^{(0)}_{\rm{qq}} + \g^{(0)}_{\rm{gg}}  \right) 
                - \frac{1}{2} \b_0 \g^{(0)}_{\rm{gq}} \right\} \right. \\
& &
\left.
        + \frac{1}{\ve} \left\{ \frac{1}{2} \g^{(1)}_{\rm{gq}}
                + \g^{(0)}_{\rm{gq}} c^{(1)}_{\f,{\rm{g}}} 
                + \g^{(0)}_{\rm{qq}} c^{(1)}_{\f,{\rm{q}}} 
                \right\} 
                \right] , \nonumber\\[1ex]
\label{F-phig-2}
T^{(2)}_{\f,{\rm{g}}} &=& \left( \frac{\a_s}{4 \p} \right)^2 S_\ve^2\,
\left( \frac{\m^2}{Q^2} \right)^{2\ve}
        \left[ \frac{1}{\ve^2} \left\{ \frac{1}{2} \left( \g^{(0)}_{\rm{gg}} \right)^2 + 
                \frac{1}{2} \g^{(0)}_{\rm{qg}} \g^{(0)}_{\rm{gq}} 
                - \frac{1}{2} \b_0 \g^{(0)}_{\rm{gg}} \right\} \right. \\
& &
\left.
        + \frac{1}{\ve} \left\{ \frac{1}{2} \g^{(1)}_{\rm{gg}} 
                + \g^{(0)}_{\rm{gg}} c^{(1)}_{\f,{\rm{g}}} 
                + \g^{(0)}_{\rm{qg}} c^{(1)}_{\f,{\rm{q}}} 
                \right\}
                \right] . \nonumber 
\end{eqnarray}

This concludes our brief discussions on the extraction of the 
anomalous dimensions and coefficient functions from 
eqs.(\ref{TmunuPartonMomNS}), (\ref{TmunuPartonMomS}) and (\ref{TmunuPartonMomPhi})
by means of eqs.(\ref{F-0})--(\ref{F-phig-2}).

\subsection{Results in Mellin space}

We are now ready to present our results, which 
provide all necessary ingredients for the solution of the 
renormalization group equations (\ref{callannonsin}) and (\ref{callansinglet}) 
for the flavour singlet and non-singlet coefficient functions 
as detailed for example in ref.\cite{Furmanski:1982cw}.
However, as is well known, a complete solution to NNLO 
for the scale evolution of the coefficient functions
requires also the still unknown anomalous dimensions $\g^{(2)}_{\rm{pp}}$ 
at 3-loops.

To summarize, our results for the anomalous dimensions 
agree with the ones published in refs.\cite{Gross:1973rr,Floratos:1977au,Gonzalez-Arroyo:1979df,
  Curci:1980uw,Furmanski:1980cm,Hamberg:1992qt,Ellis:1996nn}
while  our results for the coefficient functions agree with 
those of refs.\cite{Bardeen:1978yd,vanNeerven:1991nn,Zijlstra:1991qc,Zijlstra:1992kj,Larin:1993fv}.
As far as the earlier calculations of the longitudinal structure function $F_{L}$ 
at two loops of 
refs.\cite{Duke:1982ga,Kazakov:1988jk,Kazakov:1990fu,SanchezGuillen:1991iq} 
are concerned, let us mention that there is only complete agreement 
for $c_{L,{\rm{q}}}^{\rm{NS}}$ and $c_{L,{\rm{q}}}^{\rm{ps}}$ 
with ref.\cite{SanchezGuillen:1991iq}. 
Part of the discrepancies with ref.\cite{Kazakov:1988jk}
have been explained above.

For the anomalous dimensions, we find at 1-loop,
\begin{eqnarray}
\g^{(0)}_{\rm{qq}}(N) &=&  
          C_F   \left\{
          - 3
          + 2 S_{1}(N\!-\!1)
          + 2 S_{1}(N\!+\!1)
          \right\}\, , \\
\g^{(0)}_{\rm{qg}}(N) &=&  
          n_f   \left\{
            2  S_{1}(N\!-\!1)
          + 8 S_{1}(N\!+\!1)
          - 4 S_{1}(N\!+\!2)
          - 6 S_{1}(N) \right\}\, , \\
\g^{(0)}_{\rm{gq}}(N) &=&  
          C_F   \left\{
            4 S_{1}(N\!-\!2)
          - 8 S_{1}(N\!-\!1)
          - 2 S_{1}(N\!+\!1)
          + 6  S_{1}(N)
          \right\}\, ,\\
\g^{(0)}_{\rm{gg}}(N) &=& 
          C_A   \left\{
            4 S_{1}(N\!-\!2)
          - 8 S_{1}(N\!-\!1)
          - 8 S_{1}(N\!+\!1)
          + 4 S_{1}(N\!+\!2) 
          + 12 S_{1}(N)
          \right\}
          - \b_0\, . 
\end{eqnarray}

At two loops, we obtain,
\begin{eqnarray}
\lefteqn{
\g^{(1),{\rm{V}}}_{\rm{qq}}(N)  \,=\, (-1)^N\, \times} \\
& &
\!\!\!\Biggl[
         C_F C_A \Biggl\{  
       (-1)^N\,\Bigl(
          - \frac{17}{6}
          + 4 \z_3
          + \Bigl( \frac{268}{9} 
                 - 8 \z_2 \Bigr) S_{1}(N\!-\!1) 
          - \frac{44}{3} S_{2}(N\!-\!1)
          + 8 S_{3}(N\!-\!1)
          \Bigr)
\nonumber\\
& &\hspace*{10mm}
          - 4 S_{-3}(N\!-\!1)
          - 4 S_{-3}(N\!+\!1)
          + 8 S_{-3}(N)
          + \frac{10}{3} S_{-2}(N\!-\!1)
          + \frac{10}{3} S_{-2}(N\!+\!1)
\nonumber\\
& &\hspace*{10mm}
          - \frac{20}{3} S_{-2}(N)
          + \Bigl( \frac{106}{9} 
                 + 4 \z_2 \Bigr) S_{-1}(N\!-\!1) 
          - \Bigl( \frac{374}{9} 
                 - 4 \z_2 \Bigr) S_{-1}(N\!+\!1) 
\nonumber\\
& &\hspace*{10mm}
          + \Bigl( \frac{268}{9} 
                 - 8 \z_2 \Bigr) S_{-1}(N)
          \Biggr\}
\nonumber\\
&+\!\!&
         C_F n_f   \Biggl\{  
       (-1)^N\,\Bigl(
            \frac{1}{3}
          - \frac{40}{9} S_{1}(N\!-\!1)
          + \frac{8}{3} S_{2}(N\!-\!1)
          \Bigr)
          - \frac{4}{3} S_{-2}(N\!-\!1)
          - \frac{4}{3} S_{-2}(N\!+\!1)
          + \frac{8}{3} S_{-2}(N)
\nonumber\\
& &\hspace*{10mm}
          - \frac{4}{9} S_{-1}(N\!-\!1)
          + \frac{44}{9} S_{-1}(N\!+\!1)
          - \frac{40}{9} S_{-1}(N)
          \Biggr\}
\nonumber\\
&+\!\!&
         C_F^2   \Biggl\{  
       (-1)^N\,\Bigl(
          - \frac{3}{2}
          - 8 \z_3
          + 16 S_{1}(N\!-\!1) \z_2
          - 16 S_{1,2}(N\!-\!1)
          + 12 S_{2}(N\!-\!1)
          - 16 S_{2,1}(N\!-\!1)
          \Bigr)
\nonumber\\
& &\hspace*{10mm}
          - 4 S_{-3}(N\!-\!1)
          - 4 S_{-3}(N\!+\!1)
          + 8 S_{-3}(N)
          + 8 S_{-2}(N\!+\!1)
          - 8 S_{-2}(N)
          + 8 S_{-2,1}(N\!-\!1)
\nonumber\\
& &\hspace*{10mm}
          + 8 S_{-2,1}(N\!+\!1)
          - 16 S_{-2,1}(N)
          - \Bigl( 20 
                 + 8 \z_2 \Bigr) S_{-1}(N\!-\!1) 
          + \Bigl( 20 
                 - 8 \z_2 \Bigr) S_{-1}(N\!+\!1) 
\nonumber\\
& &\hspace*{10mm}
          + 16 S_{-1}(N) \z_2
          + 8 S_{-1,2}(N\!-\!1)
          + 8 S_{-1,2}(N\!+\!1)
          - 16 S_{-1,2}(N)
          \Biggr\}\Biggr]
\, ,
\nonumber\\[2ex]
\lefteqn{
\g^{(1),{\rm{V}}}_{\rm{q{\Bar{q}}}}(N) \,=\, (-1)^N\, \times} \\
& & 
C_F \Bigl(C_F-\frac{C_A}{2}\Bigr)   
       \Biggl\{
            8 \z_3
          + 8 S_{-3}(N\!-\!1)
          + 8 S_{-3}(N\!+\!1)
          - 8 S_{-2}(N\!-\!1)
          - 8 S_{-2}(N\!+\!1)
\nonumber\\
& &\hspace*{10mm}
          + 16 S_{-2}(N)
          + 16 S_{-1}(N\!-\!1)
          - 16 S_{-1}(N\!+\!1)
          - 8 S_{1}(N\!-\!1) \z_2
          - 8 S_{1}(N\!+\!1) \z_2
\nonumber\\
& &\hspace*{10mm}
          - 16 S_{1,-2}(N\!-\!1)
          - 16 S_{1,-2}(N\!+\!1)
          \Biggr\}
\, ,
\nonumber\\[2ex]
\lefteqn{
\g^{(1),{\rm{S}}}_{\rm{qq}}(N) \,=\, 
\g^{(1),{\rm{S}}}_{\rm{q{\Bar{q}}}}(N) \, =\, 
(-1)^N\, \times } \\
& &
   C_F n_f   \Biggl\{
          - 4 S_{-3}(N\!-\!1)
          - 4 S_{-3}(N\!+\!1)
          + 8 S_{-3}(N)
          - 2 S_{-2}(N\!-\!1)
          - \frac{46}{3} S_{-2}(N\!+\!1)
\nonumber\\
& &\hspace*{10mm}
          + \frac{16}{3} S_{-2}(N\!+\!2)
          + 12 S_{-2}(N)
          - \frac{40}{9} S_{-1}(N\!-\!2)
          + \frac{4}{9} S_{-1}(N\!-\!1)
          - \frac{4}{9} S_{-1}(N\!+\!1)
\nonumber\\
& &\hspace*{10mm}
          + \frac{112}{9} S_{-1}(N\!+\!2)
          - 8 S_{-1}(N)
          \Biggr\}     
\, ,
\nonumber\\[2ex]
\lefteqn{
\g^{(1)}_{\rm{qg}}(N) \,=\, (-1)^N\, \times} \\
& &
\!\!\!\Biggl[
       C_F n_f   \Biggl\{
            4 S_{-3}(N\!-\!1)
          + 8 S_{-3}(N\!+\!1)
          - 16 S_{-3}(N\!+\!2)
          + 4 S_{-3}(N)
          - 6 S_{-2}(N\!-\!1)
\nonumber\\
& &\hspace*{10mm}
          - 8 S_{-2}(N\!+\!1)
          + 16 S_{-2}(N\!+\!2)
          - 2 S_{-2}(N)
          - 8 S_{-2,1}(N\!-\!1)
          + 16 S_{-2,1}(N\!+\!2)
\nonumber\\
& &\hspace*{10mm}
          - 8 S_{-2,1}(N)
          + 28 S_{-1}(N\!-\!1)
          - 18 S_{-1}(N\!+\!1)
          - 40 S_{-1}(N\!+\!2)
          + 30 S_{-1}(N)
\nonumber\\
& &\hspace*{10mm}
          - 16 S_{-1,1}(N\!+\!2)
          + 16 S_{-1,1}(N)
          + 8 S_{-1,1,1}(N\!-\!1)
          - 16 S_{-1,1,1}(N\!+\!2)
          + 8 S_{-1,1,1}(N)
\nonumber\\
& &\hspace*{10mm}
          - 8 S_{-1,2}(N\!-\!1)
          + 16 S_{-1,2}(N\!+\!2)
          - 8 S_{-1,2}(N)
          \Biggr\}
\nonumber\\
&+\!\!&
        C_A n_f   \Biggl\{
          - 8 S_{-3}(N\!-\!1)
          - 16 S_{-3}(N\!+\!1)
          + 24 S_{-3}(N)
          - 4 S_{-2}(N\!-\!1)
          - \frac{272}{3} S_{-2}(N\!+\!1)
\nonumber\\
& &\hspace*{10mm}
          + \frac{176}{3} S_{-2}(N\!+\!2)
          + 36 S_{-2}(N)
          - \frac{80}{9} S_{-1}(N\!-\!2)
          + \Bigl( \frac{8}{9} 
                + 4 \z_2 \Bigr) S_{-1}(N\!-\!1) 
\nonumber\\
& &\hspace*{10mm}
          + \frac{28}{9} S_{-1}(N\!+\!1)
          + \Bigl( \frac{872}{9} 
                 - 8 \z_2 \Bigr) S_{-1}(N\!+\!2) 
          - \Bigl( 92 
                 - 4 \z_2 \Bigr) S_{-1}(N)
          + 16 S_{-1,1}(N\!+\!2)
\nonumber\\
& &\hspace*{10mm}
          - 16 S_{-1,1}(N)
          - 8 S_{-1,1,1}(N\!-\!1)
          + 16 S_{-1,1,1}(N\!+\!2)
          - 8 S_{-1,1,1}(N)
          - 4 S_{1}(N\!-\!1) \z_2
\nonumber\\
& &\hspace*{10mm}
          - 16 S_{1}(N\!+\!1) \z_2
          + 8 S_{1}(N\!+\!2) \z_2
          + 12 S_{1}(N) \z_2
          - 8 S_{1,-2}(N\!-\!1)
          - 32 S_{1,-2}(N\!+\!1)
\nonumber\\
& &\hspace*{10mm}
          + 16 S_{1,-2}(N\!+\!2)
          + 24 S_{1,-2}(N)
          \Biggr\} \Biggr]
\, ,
\nonumber\\[2ex]
\lefteqn{
\g^{(1)}_{\rm{gq}}(N) \,=\, (-1)^N\, \times} \\
& &
\!\!\!\Biggl[
       C_F C_A   \Biggl\{
            16 S_{-3}(N\!-\!1)
          + 8 S_{-3}(N\!+\!1)
          - 24 S_{-3}(N)
          + 48 S_{-2}(N\!-\!1)
          + \frac{92}{3} S_{-2}(N\!+\!1)
\nonumber\\
& &\hspace*{10mm}
          - \frac{32}{3} S_{-2}(N\!+\!2)
          - 68 S_{-2}(N)
          + 16 S_{-2,1}(N\!-\!2)
          - 8 S_{-2,1}(N\!+\!1)
          - 8 S_{-2,1}(N)
\nonumber\\
& &\hspace*{10mm}
          - \Bigl( 4 
                 + 8 \z_2 \Bigr) S_{-1}(N\!-\!2) 
          + \frac{112}{9} S_{-1}(N\!-\!1)
          + \Bigl( 36 
                 + 4 \z_2 \Bigr) S_{-1}(N\!+\!1)
          - \frac{176}{9} S_{-1}(N\!+\!2)
\nonumber\\
& &\hspace*{10mm}
          - \Bigl( \frac{224}{9} 
                 - 4 \z_2 \Bigr) S_{-1}(N) 
          + \frac{88}{3} S_{-1,1}(N\!-\!2)
          - \frac{68}{3} S_{-1,1}(N\!+\!1)
          - \frac{20}{3} S_{-1,1}(N)
\nonumber\\
& &\hspace*{10mm}
          - 16 S_{-1,1,1}(N\!-\!2)
          + 8 S_{-1,1,1}(N\!+\!1)
          + 8 S_{-1,1,1}(N)
          + 16 S_{-1,2}(N\!-\!2)
          - 8 S_{-1,2}(N\!+\!1)
\nonumber\\
& &\hspace*{10mm}
          - 8 S_{-1,2}(N)
          - 8 S_{1}(N\!-\!2) \z_2
          + 16 S_{1}(N\!-\!1) \z_2
          + 4 S_{1}(N\!+\!1) \z_2
          - 12 S_{1}(N) \z_2
\nonumber\\
& &\hspace*{10mm}
          - 16 S_{1,-2}(N\!-\!2)
          + 32 S_{1,-2}(N\!-\!1)
          + 8 S_{1,-2}(N\!+\!1)
          - 24 S_{1,-2}(N)
          \Biggr\}
\nonumber\\
&+\!\!&
       C_F n_f   \Biggl\{
            \frac{80}{9} S_{-1}(N\!-\!2)
          - \frac{64}{9} S_{-1}(N\!+\!1)
          - \frac{16}{9} S_{-1}(N)
          - \frac{16}{3} S_{-1,1}(N\!-\!2)
          + \frac{8}{3} S_{-1,1}(N\!+\!1)
\nonumber\\
& &\hspace*{10mm}
          + \frac{8}{3} S_{-1,1}(N)
          \Biggr\}
\nonumber\\
&+\!\!&
       C_F^2   \Biggl\{
          - 8 S_{-3}(N\!-\!1)
          + 4 S_{-3}(N\!+\!1)
          + 4 S_{-3}(N)
          - 8 S_{-2}(N\!-\!1)
          - 14 S_{-2}(N\!+\!1)
          + 22 S_{-2}(N)
\nonumber\\
& &\hspace*{10mm}
          - 10 S_{-1}(N\!-\!1)
          - 14 S_{-1}(N\!+\!1)
          + 24 S_{-1}(N)
          - 24 S_{-1,1}(N\!-\!2)
          + 20 S_{-1,1}(N\!+\!1)
\nonumber\\
& &\hspace*{10mm}
          + 4 S_{-1,1}(N)
          + 16 S_{-1,1,1}(N\!-\!2)
          - 8 S_{-1,1,1}(N\!+\!1)
          - 8 S_{-1,1,1}(N)
          \Biggr\}\Biggr]
\, ,
\nonumber\\[2ex]
\lefteqn{
\g^{(1)}_{\rm{gg}}(N) \,=\, (-1)^N\, \times} \\
& &
\!\!\!\Biggl[
         C_F n_f  \Biggl\{
            2 (-1)^N  
          - 8 S_{-3}(N\!-\!1)
          - 8 S_{-3}(N\!+\!1)
          + 16 S_{-3}(N)
          + 12 S_{-2}(N\!-\!1)
          + 20 S_{-2}(N\!+\!1)
\nonumber\\
& &\hspace*{10mm}
          - 32 S_{-2}(N)
          - \frac{8}{3} S_{-1}(N\!-\!2)
          - \frac{88}{3} S_{-1}(N\!-\!1)
          + \frac{88}{3} S_{-1}(N\!+\!1)
          - \frac{40}{3} S_{-1}(N\!+\!2)
\nonumber\\
& &\hspace*{10mm}
          + 16 S_{-1}(N)
          \Biggr\}
\nonumber\\
&+\!\!&
         C_A n_f  \Biggl\{
            (-1)^N \Bigl(  
            \frac{8}{3}
          - \frac{40}{9} S_{1}(N\!-\!1)
          \Bigr)
          + \frac{8}{3} S_{-2}(N\!-\!1)
          + \frac{8}{3} S_{-2}(N\!+\!1)
          - \frac{16}{3} S_{-2}(N)
\nonumber\\
& &\hspace*{10mm}
          + \frac{92}{9} S_{-1}(N\!-\!2)
          + \frac{8}{3} S_{-1}(N\!-\!1)
          + \frac{16}{9} S_{-1}(N\!+\!1)
          - \frac{92}{9} S_{-1}(N\!+\!2)
          - \frac{40}{9} S_{-1}(N)
          \Biggr\}
\nonumber\\
&+\!\!&
         C_A^2   \Biggl\{
            (-1)^N \Bigl(
          - \frac{32}{3}
          - 4 \z_3
          + \Bigl( \frac{268}{9} 
                 + 8 \z_2 \Bigr) S_{1}(N\!-\!1) 
          - 16 S_{1,2}(N\!-\!1)
          - 16 S_{2,1}(N\!-\!1)
\nonumber\\
& &\hspace*{10mm}
          + 8 S_{3}(N\!-\!1)
          \Bigr)
          + 4 \z_3
          + 8 S_{-3}(N\!-\!1)
          + 16 S_{-3}(N\!+\!1)
          + 16 S_{-3}(N\!+\!2)
          - 32 S_{-3}(N)
\nonumber\\
& &\hspace*{10mm}
          + \frac{100}{3} S_{-2}(N\!-\!1)
          + 44 S_{-2}(N\!+\!1)
          - \frac{176}{3} S_{-2}(N\!+\!2)
          - \frac{56}{3} S_{-2}(N)
          + 16 S_{-2,1}(N\!-\!2)
\nonumber\\
& &\hspace*{10mm}
          + 16 S_{-2,1}(N\!-\!1)
          - 16 S_{-2,1}(N\!+\!2)
          - 16 S_{-2,1}(N)
          - 8 S_{-1}(N\!-\!2) \z_2
\nonumber\\
& &\hspace*{10mm}
          - \Bigl( \frac{50}{9} 
                 + 8 \z_2 \Bigr) S_{-1}(N\!-\!1) 
          - \frac{218}{9} S_{-1}(N\!+\!1)
          + 8 S_{-1}(N\!+\!2) \z_2
          + \Bigl( \frac{268}{9} 
                 + 8 \z_2 \Bigr) S_{-1}(N)
\nonumber\\
& &\hspace*{10mm}
          + 16 S_{-1,2}(N\!-\!2)
          + 16 S_{-1,2}(N\!-\!1)
          - 16 S_{-1,2}(N\!+\!2)
          - 16 S_{-1,2}(N)
          - 8 S_{1}(N\!-\!2) \z_2
\nonumber\\
& &\hspace*{10mm}
          + 16 S_{1}(N\!-\!1) \z_2
          + 16 S_{1}(N\!+\!1) \z_2
          - 8 S_{1}(N\!+\!2) \z_2
          - 24 S_{1}(N) \z_2
          - 16 S_{1,-2}(N\!-\!2)
\nonumber\\
& &\hspace*{10mm}
          + 32 S_{1,-2}(N\!-\!1)
          + 32 S_{1,-2}(N\!+\!1)
          - 16 S_{1,-2}(N\!+\!2)
          - 48 S_{1,-2}(N)
          \Biggr\}\Biggr]
\, .
\nonumber
\end{eqnarray}

For the coefficient functions, we obtain 
at tree level,
\begin{eqnarray}
c^{(0)}_{2,{\rm{q}}}(N) = c^{(0)}_{3,{\rm{q}}}(N) = 1\, ,\quad\quad\quad
c^{(0)}_{2,{\rm{g}}}(N) = c^{(0)}_{L,{\rm{q}}}(N) = c^{(0)}_{L,{\rm{g}}}(N) = 0\, ,
\end{eqnarray}

At 1-loop we find,
\begin{eqnarray}
c^{(1)}_{2,{\rm{q}}}(N)  &=& 
          C_F   \left\{
          - 9
          - 3 S_{1}(N\!-\!1)
          + 4 S_{1}(N\!+\!1)
          + 2 S_{1}(N)
          + 2 S_{1,1}(N\!-\!1)
\right.\\
& &\left.\hspace*{3.0cm}
          + 2 S_{1,1}(N\!+\!1)
          - 2 S_{2}(N\!-\!1)
          - 2 S_{2}(N\!+\!1)
          \right\} \, ,\nonumber\\
c^{(1)}_{3,{\rm{q}}}(N)  &=&  
c^{(1)}_{2,{\rm{q}}}(N)  + 
          C_F \left\{
            2 S_{1}(N\!-\!1)
          - 2 S_{1}(N\!+\!1)
          \right\} \, ,\\
c^{(1)}_{2,{\rm{g}}}(N)  &=&  
           n_f   \left\{
            2 S_{1}(N\!-\!1)
          + 32 S_{1}(N\!+\!1)
          - 16 S_{1}(N\!+\!2)
          - 18 S_{1}(N)
          + 2 S_{1,1}(N\!-\!1)
\right.\\
& &\hspace*{1.0cm}
          + 8 S_{1,1}(N\!+\!1)
          - 4 S_{1,1}(N\!+\!2)
          - 6 S_{1,1}(N)
          - 2 S_{2}(N\!-\!1)
          - 8 S_{2}(N\!+\!1) \nonumber\\
& &\hspace*{1.0cm}\left.
          + 4 S_{2}(N\!+\!2)
          + 6 S_{2}(N)
          \right\}\, ,\nonumber\\
c^{(1)}_{L,{\rm{q}}}(N)  &=&  
          4 C_F   \left\{
            S_{1}(N\!+\!1)
          - S_{1}(N)
          \right\}\, ,\\
c^{(1)}_{L,{\rm{g}}}(N)  &=&  
          8  n_f   \left\{
            2 S_{1}(N\!+\!1)
          - S_{1}(N\!+\!2)
          - S_{1}(N)
          \right\}\, .
\end{eqnarray}

Our results for the Mellin moments of 2-loop coefficient functions 
$c^{(2),\rm{ns}}_{2,{\rm{q}}}$,$c^{(2),\rm{ns}}_{3,{\rm{q}}}$, 
$c^{(2),\rm{ps}}_{2,{\rm{q}}}$, $c^{(2)}_{2,{\rm{g}}}$
$c^{(2),\rm{ns}}_{L,{\rm{q}}}$, $c^{(2),\rm{ps}}_{L,{\rm{q}}}$ and $c^{(2)}_{L,{\rm{g}}}$
can be found in appendix B.

\subsection{Results in $x$-space}
Let us now present our results in momentum space.  
The scale evolution of the DIS structure 
functions in $x$-space has been discussed in 
ref.\cite{Gribov:1972ri}. 
It is governed by the Altarelli-Parisi splitting functions, defined as 
\begin{eqnarray}
\label{splitting-def}
\g_{ij}(N) = - \int\limits_0^1 dx\, x^{N-1}\, P_{ij}(x)\, ,
\end{eqnarray}
where the conventional relation between the anomalous dimensions and 
the splitting functions in eq.(\ref{splitting-def}) 
involves a relative sign. 
In the following, all expressions which diverge for $x \to 1$ 
are understood in the sense of $+$-distributions as defined 
in eq.(\ref{eq:mellinplus}).

For the splitting functions, we find at 1-loop,
\begin{eqnarray}
P^{(0)}_{\rm{qq}}(x) &=& 
        C_F   \left\{
          2 p_{\rm{qq}}(x) 
       + 3 \d(1 - x)
          \right\}\, , \\
P^{(0)}_{\rm{qg}}(x) &=&    
        2 n_f\, p_{\rm{qg}}(x) \, , \\
P^{(0)}_{\rm{gq}}(x) &=&    
        2 C_F\, p_{\rm{gg}}(x) \, , \\
P^{(0)}_{\rm{gg}}(x) &=&       
        4 C_A\,  p_{\rm{gg}}(x) + \b_0 \d(1 - x)\, ,
\end{eqnarray}
where we have introduced the following polynomials,
\begin{eqnarray}
p_{\rm{qq}}(x) &=& 
            \frac{2}{1- x}
          - 1
          - x\, , \\
p_{\rm{qg}}(x) &=&    
          1
          - 2 x
          + 2 x^2\, , \\
p_{\rm{gq}}(x) &=&    
            \frac{2}{x}
          - 2
          + x, \\
p_{\rm{gg}}(x) &=&       
            \frac{1}{1- x}
          + \frac{1}{x}
          - 2
          + x
          - x^2\, .
\end{eqnarray}

At two loops, we obtain,
{\footnote{The definition of $S_2$ in ref.\cite{Furmanski:1980cm} contains 
a typographical error. 
The lower integration boundary $(1+x)/x$ should read  $x/(1+x)$, 
see for instance eq.(61) in ref.\cite{Ellis:1996nn}.
Also the timelike quark-quark splitting function in eq.(12) 
of ref.\cite{Furmanski:1980cm} contains a typographical mistake. 
The term $( 10 - 18x - 16/3 x^2 ) \ln(x)$ should read 
$(-10 - 18x - 16/3 x^2 ) \ln(x)$.}}
\begin{eqnarray}
\lefteqn{
P^{(1),{\rm{V}}}_{\rm{qq}}(x)  \,=}\\
& & 
         n_f C_F   \Biggl\{
          - \frac{8}{3}(1-x)
          - \frac{4}{3} \Bigl(\frac{5}{3} + H_{0}(x)\Bigr) p_{\rm{qq}}(x)
          - \Bigl( \frac{1}{3} + \frac{8}{3} \z_2\Bigr) \d(1 - x)
          \Biggr\}
\nonumber\\
&+\!\!&
         C_F C_A   \Biggl\{
            \frac{80}{3}(1-x)
          + 4 (1 + x) H_{0}(x)
          + 4 \Bigl(\frac{67}{18} - \z_2 
                + \frac{11}{6} H_{0}(x) + H_{0,0}(x)\Bigr) p_{\rm{qq}}(x)
\nonumber\\
& &\hspace*{10mm}
          + \Bigl(\frac{17}{6} + \frac{44}{3} \z_2 - 12 \z_3\Bigr) \d(1 - x)
          \Biggr\}
\nonumber\\
&+\!\!&
         C_F^2   \Biggl\{
          - 20(1-x)
          - 2 (3 + 7 x) H_{0}(x)
          - 8 \Bigl( \frac{3}{4} H_{0}(x) - H_{1,0}(x) 
                  - H_{2}(x)\Bigr) p_{\rm{qq}}(x)
\nonumber\\
& &\hspace*{10mm}
          - 4 (1 + x) H_{0,0}(x)
          + \Bigl(\frac{3}{2} - 12 \z_2 + 24 \z_3\Bigr) \d(1 - x) 
          \Biggr\}
\, ,
\nonumber\\[2ex]
\lefteqn{
P^{(1),{\rm{V}}}_{\rm{q{\Bar{q}}}}(x)  \,=}\\
& & 
         C_F \Bigl(C_F-\frac{C_A}{2}\Bigr)   \Biggl\{
            16 (1 - x)
          + 8 (1 + x) H_{0}(x) 
          + 8 \Bigl( H_{0,0}(x) - 2 H_{-1,0}(x) - \z_2 \Bigr) p_{\rm{qq}}( - x)
          \Biggr\}
\, ,
\nonumber\\[2ex]
\lefteqn{
P^{(1),{\rm{S}}}_{\rm{qq}}(x)  \,=\, P^{(1),{\rm{S}}}_{\rm{q{\Bar{q}}}}(x) \,= }\\
& &
         n_f C_F \Biggl\{
          - 4
          + \frac{40}{9 x}
          + 12 x
          - \frac{112}{9} x^2
          - 4 (1 + x) H_{0,0}(x)
          + 2 \Bigl(1 + 5 x + \frac{8}{3} x^2 \Bigr) H_{0}(x)
          \Biggr\}
\, ,
\nonumber
\end{eqnarray}
%
%
%
\begin{eqnarray}
\lefteqn{
P^{(1)}_{\rm{qg}}(x)  \,=}\\
& &
         n_f C_F   \Biggl\{
            8
          - 18 x
          - 2 (1 - 4 x) H_{0}(x)
          - 8 H_{1}(x)
          - 4 (1 - 2 x) H_{0,0}(x)
\nonumber\\
& &\hspace*{10mm}
          + 8 \Bigl(\frac{5}{2} - \z_2 + H_{0}(x) + H_{0,0}(x) 
                + H_{1}(x) + H_{1,0}(x) 
                + H_{1,1}(x) + H_{2}(x)\Bigr) p_{\rm{qg}}(x)
          \Biggr\}
\nonumber\\
&+\!\!&
         n_f C_A   \Biggl\{
            \frac{364}{9}
          + \frac{80}{9 x}
          + \frac{28}{9} x
          - 8 \Bigl( \frac{109}{18} - \frac{11}{3} 
                H_{0}(x) + H_{1}(x) + H_{1,1}(x)\Bigr) p_{\rm{qg}}(x)
\nonumber\\
& &\hspace*{10mm}
          - \frac{4}{3} (19 - 68 x) H_{0}(x)
          + 8 H_{1}(x)
          - 8 (1 + 2 x) H_{0,0}(x)
          - 8 H_{-1,0}(x) p_{\rm{qg}}( - x)
          - 16 x \z_2
          \Biggr\}
\, ,
\nonumber\\[2ex]
\lefteqn{
P^{(1)}_{\rm{gq}}(x)  \,=}\\
& & 
         n_f C_F   \Biggl\{
          - \frac{8}{3} x
          - \frac{8}{3} \Bigl( \frac{5}{3} - H_{1}(x)\Bigr) p_{\rm{gq}}(x)
          \Biggr\}
\nonumber\\
&+\!\!&
         C_F C_A   \Biggl\{
            \frac{112}{9}
          + \frac{130}{9} x
          + \frac{176}{9} x^2
          + 8 \Bigl(\frac{1}{4} - \frac{11}{6} H_{1}(x) 
                + H_{1,0}(x) + H_{1,1}(x) + H_{2}(x)\Bigr) p_{\rm{gq}}(x)
\nonumber\\
& &\hspace*{10mm}
          - 4 \Bigl(12 + 5 x + \frac{8}{3} x^2\Bigr) H_{0}(x)
          - 8 x H_{1}(x)
          + 8 (2 + x) H_{0,0}(x)
          - 8 H_{-1,0}(x) p_{\rm{gq}}( - x)
          + 16 \z_2
          \Biggr\}
\nonumber\\
&+\!\!&
         C_F^2   \Biggl\{
          - 10
          - 14 x
          - 4 (2 - x) H_{0,0}(x)
          + 2 (4 + 7 x) H_{0}(x)
          + 8 x H_{1}(x)
\nonumber\\
& &\hspace*{10mm}
          + 8 \Bigl(\frac{3}{2} H_{1}(x) - H_{1,1}(x)\Bigr) p_{\rm{gq}}(x)
          \Biggr\}
\, ,
\nonumber\\[2ex]
\lefteqn{
P^{(1)}_{\rm{gg}}(x)  \,=} \\
& & 
         n_f C_F   \Biggl\{
          - 32
          + \frac{8}{3 x}
          + 16 x
          + \frac{40}{3} x^2
          - 8 (1 + x) H_{0,0}(x)
          - 4 (3 + 5 x) H_{0}(x)
          - 2 \d(1 - x)
          \Biggr\}
\nonumber\\
&+\!\!&
         n_f C_A   \Biggl\{
            4(1 - x)
          - \frac{52}{9} \Bigl( \frac{1}{x} - x^2 \Bigr)
          - \frac{40}{9} p_{\rm{gg}}(x)
          - \frac{8}{3} (1 + x) H_{0}(x)
          - \frac{8}{3} \d(1 - x) 
          \Biggr\}
\nonumber\\
&+\!\!&
         C_A^2   \Biggl\{
            54(1 - x)
          - \frac{268}{9} \Bigl( \frac{1}{x} - x^2 \Bigr)
          - \frac{4}{3} (25 - 11 x + 44 x^2) H_{0}(x)
          + 32 (1 + x) H_{0,0}(x)
\nonumber\\
& &\hspace*{10mm}
          + 8 \Bigl(\frac{67}{18} - \z_2 + H_{0,0}(x) + 2 H_{1,0}(x) 
            + 2 H_{2}(x)\Bigr) p_{\rm{gg}}(x)
\nonumber\\
& &\hspace*{10mm}
          + 8 \Bigl( H_{0,0}(x) - 2 H_{-1,0}(x) - \z_2 \Bigr) p_{\rm{gg}}( - x)
          + \Bigl(\frac{32}{3} + 12 \z_3\Bigr) \d(1 - x)
          \Biggr\}
\nonumber
\, .
\end{eqnarray}

The $x$-space expressions for the coefficient functions 
are commonly defined by, 
\begin{eqnarray}
\label{coefficient-def}
c_{i,{\rm{p}}}(N) = \int\limits_0^1 dx\, x^{N-1}\, c_{i,{\rm{p}}}(x)\, .
\end{eqnarray}

We obtain at tree level,
\begin{eqnarray}
c^{(0)}_{2,{\rm{q}}}(x) = c^{(0)}_{3,{\rm{q}}}(x)  = \d(1 - x)\, ,\quad\quad\quad
c^{(0)}_{2,{\rm{g}}}(x)  = c^{(0)}_{L,{\rm{q}}}(x)  = c^{(0)}_{L,{\rm{g}}}(x)  = 0\, ,
\end{eqnarray}

At 1-loop we find,
\begin{eqnarray}
c^{(1)}_{2,{\rm{q}}}(x) &=& 
        C_F   \Biggl\{
           \frac{9}{2} + \frac{5}{2} x
- 2 \Bigl( \frac{3}{4} + H_{0}(x) + H_{1}(x) \Bigr) p_{\rm{qq}}(x)   
       -  \Bigl(
          9 + 4 \z_2 \Bigr) \d(1 - x) \Biggr\}\, , \\
c^{(1)}_{3,{\rm{q}}}(x) &=& c^{(1)}_{2,{\rm{q}}}(x)
       - 2 C_F \,  (1 + x) \, , \\
c^{(1)}_{2,{\rm{g}}}(x) &=& 
         n_f \Biggl\{
          6 - 2  \Bigl( 4 + H_{0}(x) + H_{1}(x) 
          \Bigr) p_{\rm{qg}}(x)   
          \Biggr\} \, ,\\
c^{(1)}_{L,{\rm{q}}}(x)  &=& 4 C_F x\, , \\
c^{(1)}_{L,{\rm{g}}}(x)  &=& 8 n_f x (1 - x) \, .
\end{eqnarray}

The results for the 2-loop coefficient functions 
$c^{(2),\rm{ns}}_{2,{\rm{q}}}$,$c^{(2),\rm{ns}}_{3,{\rm{q}}}$, 
$c^{(2),\rm{ps}}_{2,{\rm{q}}}$, $c^{(2)}_{2,{\rm{g}}}$
$c^{(2),\rm{ns}}_{L,{\rm{q}}}$, $c^{(2),\rm{ps}}_{L,{\rm{q}}}$ and $c^{(2)}_{L,{\rm{g}}}$
have been deferred to appendix C. 

\section{Conclusions
\label{sec:Conclusions}}

In the present paper we have calculated 
the Mellin moments of the perturbative QCD corrections 
up to second order for the DIS structure functions 
$F_{2}$, $F_{3}$ and $F_{L}$
in leading twist approximation 
using the method of projection~\cite{Gorishnii:1983su}.
We have presented the analytic results for the 1- and 2-loop contributions 
to the anomalous dimensions of the singlet and non-singlet operator matrix 
elements and the 1- and 2-loop coefficient functions of 
$F_{2}$, $F_{3}$ and $F_{L}$.
Our results are in agreement with the literature as has been 
detailed in section~\ref{sec:CalculationResults}.

Our choice to work in Mellin space enabled us to solve the 2-loop integrals 
in dimensional regularization, $D=4-2\ve$, 
by means of recursion relations in the Mellin moment $N$. 
This approach has systematically extended previous work in this 
direction~\cite{Kazakov:1988jk} and relied on the improved understanding of harmonic 
sums~\cite{Gonzalez-Arroyo:1979df,Vermaseren:1998uu,Blumlein:1998if}.
Progress was possible in particular due to the development of new algorithms 
for a large class of series that involve harmonic sums. 

The method as we applied it turned out to be very flexible 
and in the expansion in $\ve$ it is in principle not limited 
to a certain order. This allowed us to calculate anomalous dimensions 
and coefficient functions at the same time. 
As a byproduct, we have also performed the calculation of all $\co(\ve^2)$-terms 
at 1-loops and all $\co(\ve)$-terms at two loops, 
for all unpolarized structure functions, as needed to extract 
the coefficient functions at 3-loops after mass factorization. 
However these results will be published elsewhere~\cite{MVinprep}.

We have been able to perform the inverse Mellin transformation 
of our $N$-space results analytically 
and we have given the corresponding expressions in $x$-space as well.
Our $x$-space results are expressed in terms of harmonic polylogarithms, 
which we believe to be the natural class of functions for 
calculations of DIS structure functions. 
Again, this has been achieved due to deeper insight gained into 
the subtle interplay between harmonic sums and harmonic polylogarithms~\cite{Remiddi:1999ew}.

We are very confident, that the program presented in this work 
can be successfully applied to the ultimate goal, the 
calculation of the anomalous dimensions and the coefficient functions 
of the DIS structure functions at 3-loops.
Finally we would like to note that a calculation
of the perturbative QCD corrections to the polarized  structure function $g_1$ 
up to second order and in leading twist approximation 
is in progress~\cite{MVinprep}.

\subsection*{Acknowledgments}
We would like to thank D.A. Broadhurst, E. Laenen, S.A. Larin, 
A. Retey, T. van Ritbergen and F.J. Yndur\'ain for fruitful discussions 
during the various phases of this project. 

This work is part of the research program of the
Foundation for Fundamental Research of Matter (FOM) and
the National Organization for Scientific Research (NWO).

\section*{Appendix A}

Here we give the results for the harmonic polylogarithms 
of weight two and three in terms of standard polylogarithms,
\begin{eqnarray}
   \Li_2(x) = - \int_0^x \frac{dz}{z}\ln(1-z)\, ,\quad\quad\quad
   \Li_3(x) =   \int_0^x \frac{dz}{z}\Li_2(z)\, ,
\label{Li2} 
\end{eqnarray}
where $\Li_2(x)$ is Euler's dilogarithm and $\Li_3(x)$ the usual trilogarithm, 
see ref.\cite{lewin:book}.
For 
It is obvious, that we only need to consider a limited subset at weight 
two and three. We find at weight two,

\begin{eqnarray}
\label{eq:weight2_1}
   H_{-1,-1}(x)  &=&
         \frac{1}{2} \ln^2(1+x)
      \, ,\\
\label{eq:weight2_2}
   H_{-1,0}(x)  &=&  
         \ln{x} \ln(1+x) + \Li_2(-x) 
      \, ,\\
\label{eq:weight2_3}
   H_{-1,1}(x)  &=&
       - \frac{1}{2} \z_2
       + \frac{1}{2} \ln^2{2}
       - \ln(1+x) \ln{2}
       + \Li_2\!\! \left(\frac{1\!+\!x}{2}\right)
      \, ,\\
\label{eq:weight2_4}
   H_{0,-1}(x)  &=&
       - \Li_2(-x);
      \, ,\\
\label{eq:weight2_5}
   H_{0,0}(x)  &=&  
         \frac{1}{2}\ln^2{x}
      \, ,\\
\label{eq:weight2_6}
   H_{0,1}(x)  &=&  
         \Li_2(x) 
      \ , \\
\label{eq:weight2_7}
   H_{1,-1}(x)  &=& 
         \frac{1}{2} \z_2
       - \frac{1}{2} \ln^2{2}
       - \ln(1-x) \ln(1+x)
       + \ln(1+x) \ln{2}
       - \Li_2\!\! \left(\frac{1\!+\!x}{2}\right)
      \, ,\\
\label{eq:weight2_8}
   H_{1,0}(x)  &=&  
       - \ln{x} \ln(1-x) 
       - \Li_2(x) 
      \, ,\\
\label{eq:weight2_9}
   H_{1,1}(x)  &=&  
         \frac{1}{2} \ln^2(1-x) 
      \, .
\end{eqnarray}
At weight three we obtain,
\begin{eqnarray}
\label{eq:weight3_1}
   H_{-1,-1,-1}(x) &=&
         \frac{1}{6} \ln^3(1\!+\!x)
      \, ,\\
\label{eq:weight3_2}
   H_{-1,-1,0}(x) &=&
         \z_3
       - \ln(1\!+\!x) \z_2
       + \frac{1}{6} \ln^3(1\!+\!x)
       - \Li_3\!\! \left(\frac{1}{1\!+\!x}\right)
      \, ,\\
\label{eq:weight3_3}
   H_{-1,-1,1}(x) &=&
         \frac{1}{2} \z_2 \ln{2}
       - \frac{7}{8} \z_3
       - \frac{1}{6} \ln^3{2}
       - \frac{1}{2} \ln(1\!+\!x) \z_2
       + \frac{1}{2} \ln(1\!+\!x) \ln^2{2}
\\
& &
       - \frac{1}{2} \ln^2(1\!+\!x) \ln{2}
       + \Li_3\!\! \left(\frac{1\!+\!x}{2}\right)
      \, ,\nonumber\\
\label{eq:weight3_4}
   H_{-1,0,-1}(x) &=&
       - 2 \z_3
       + 2 \ln(1\!+\!x) \z_2
       - \frac{1}{3} \ln^3(1\!+\!x)
       + \ln(1\!+\!x) \Li_2( - x)
       + \ln{x} \ln^2(1\!+\!x)
\\
& &
       + 2 \Li_3\!\! \left(\frac{1}{1\!+\!x}\right)
      \, ,\nonumber\\
\label{eq:weight3_5}
   H_{-1,0,0}(x) &=&
         \frac{1}{2} \ln^2{x} \ln(1\!+\!x)
       + \ln{x} \Li_2( - x)
       - \Li_3( - x)
      \, ,\\
\label{eq:weight3_6}
   H_{-1,0,1}(x) &=&
         \frac{1}{4} \z_3
       - \frac{1}{2} \ln(1\!-\!x) \z_2
       - \ln(1\!-\!x) \Li_2( - x)
       + \frac{1}{2} \ln(1\!+\!x) \z_2
       + \frac{1}{6} \ln^3(1\!+\!x)
\\
& &
       - \ln{x} \ln(1\!-\!x) \ln(1\!+\!x)
       - \Li_3(1\!-\!x)
       - \Li_3\!\! \left( -\frac{1\!-\!x}{1\!+\!x}\right)
       - \Li_3\!\! \left(\frac{1}{1\!+\!x}\right)
\nonumber\\
& &
       + \Li_3\!\! \left(\frac{1\!-\!x}{1\!+\!x}\right)
      \, ,\nonumber\\
\label{eq:weight3_7}
   H_{-1,1,-1}(x) &=&
       - \z_2 \ln{2}
       + \frac{7}{4} \z_3
       + \frac{1}{3} \ln^3{2}
       + \frac{1}{2} \ln(1\!+\!x) \z_2
       - \frac{1}{2} \ln(1\!+\!x) \ln^2{2}
\\
& &
       + \ln(1\!+\!x) \Li_2\!\! \left(\frac{1\!+\!x}{2}\right)
       - 2 \Li_3\!\! \left(\frac{1\!+\!x}{2}\right)
      \, ,\nonumber\\
\label{eq:weight3_8}
   H_{-1,1,0}(x) &=&
       - \frac{1}{2} \z_2 \ln{2}
       - \frac{1}{8} \z_3
       + \frac{1}{6} \ln^3{2}
       + \frac{3}{2} \ln(1\!+\!x) \z_2
       - \frac{1}{2} \ln(1\!+\!x) \ln^2{2}
\\
& &
       + \frac{1}{2} \ln^2(1\!+\!x) \ln{2}
       - \frac{1}{3} \ln^3(1\!+\!x)
       - \frac{1}{2} \ln{x} \z_2
       + \frac{1}{2} \ln{x} \ln^2{2}
       - \ln{x} \ln(1\!+\!x) \ln{2}
\nonumber\\
& &
       + \frac{1}{2} \ln{x} \ln^2(1\!+\!x)
       + \ln{x} \Li_2\!\! \left(\frac{1\!+\!x}{2}\right)
       - \Li_3\!\! \left(\frac{1\!+\!x}{2}\right)
       + \Li_3( - x)
       + \Li_3\!\! \left(\frac{1}{1\!+\!x}\right)
\nonumber\\
& &
       + \Li_3\!\! \left(\frac{2x}{1\!+\!x}\right)
       - \Li_3(x)
      \, ,\nonumber\\
\label{eq:weight3_9}
   H_{-1,1,1}(x) &=&
         \frac{1}{2} \z_2 \ln{2}
       - \frac{1}{8} \z_3
       - \frac{1}{6} \ln^3{2}
       + \ln(1\!-\!x) \z_2
       - \frac{1}{2} \ln(1\!-\!x) \ln^2{2}
\\
& &
       + \frac{1}{2} \ln^2(1\!-\!x) \ln{2}
       - \frac{1}{2} \ln^2(1\!-\!x) \ln(1\!+\!x)
       + \frac{1}{2} \ln(1\!-\!x) \ln^2(1\!+\!x)
\nonumber\\
& &
       - \ln(1\!-\!x) \Li_2\!\! \left(\frac{1\!+\!x}{2}\right)
       - \ln(1\!+\!x) \z_2
       + \frac{1}{2} \ln(1\!+\!x) \ln^2{2}
       - \frac{1}{6} \ln^3(1\!+\!x)
\nonumber\\
& &
       + \Li_3\!\! \left(\frac{1\!+\!x}{2}\right)
       + \Li_3\!\! \left( -\frac{1\!-\!x}{1\!+\!x}\right)
      \, ,\nonumber\\
\label{eq:weight3_10}
   H_{0,-1,-1}(x) &=&
         \z_3
       - \ln(1\!+\!x) \z_2
       + \frac{1}{6} \ln^3(1\!+\!x)
       - \ln(1\!+\!x) \Li_2( - x)
       - \frac{1}{2} \ln{x} \ln^2(1\!+\!x)
\\
& &
       - \Li_3\!\! \left(\frac{1}{1\!+\!x}\right)
      \nonumber\\
&=&
         {\rm{S}}_{n,p}(-x) 
      \, ,\nonumber\\
\label{eq:weight3_11}
   H_{0,-1,0}(x) &=&
       - \ln{x} \Li_2( - x)
       + 2 \Li_3( - x)
      \, ,\\
\label{eq:weight3_12}
   H_{0,-1,1}(x) &=&
         \frac{1}{2} \z_2 \ln{2}
       - \frac{1}{8} \z_3
       - \frac{1}{6} \ln^3{2}
       + \frac{1}{2} \ln(1\!-\!x) \z_2
       + \ln(1\!-\!x) \Li_2( - x)
\\
& &
       - 2 \ln(1\!+\!x) \z_2
       + \frac{1}{2} \ln(1\!+\!x) \ln^2{2}
       - \frac{1}{2} \ln^2(1\!+\!x) \ln{2}
       + \frac{1}{6} \ln^3(1\!+\!x)
\nonumber\\
& &
       + \ln{x} \ln(1\!-\!x) \ln(1\!+\!x)
       - \frac{1}{2} \ln{x} \ln^2(1\!+\!x)
       + \Li_3\!\! \left(\frac{1\!+\!x}{2}\right)
       + \Li_3(1\!-\!x)
\nonumber\\
& &
       - \Li_3( - x)
       + \Li_3\!\! \left( -\frac{1\!-\!x}{1\!+\!x}\right)
       - \Li_3\!\! \left(\frac{1\!-\!x}{1\!+\!x}\right)
       - \Li_3\!\! \left(\frac{2x}{1\!+\!x}\right)
       + \Li_3(x)
      \, ,\nonumber\\
\label{eq:weight3_13}
   H_{0,0,-1}(x) &=&
       - \Li_3( - x)
      \, ,\\
\label{eq:weight3_14}
   H_{0,0,0}(x) &=&
         \frac{1}{6} \ln^3{x}
      \, ,\\
\label{eq:weight3_15}
   H_{0,0,1}(x) &=&
         \Li_3(x)
      \, ,\\
\label{eq:weight3_16}
   H_{0,1,-1}(x) &=&
       - \frac{1}{2} \z_2 \ln{2}
       - \frac{1}{8} \z_3
       + \frac{1}{6} \ln^3{2}
       + \frac{3}{2} \ln(1\!+\!x) \z_2
       - \frac{1}{2} \ln(1\!+\!x) \ln^2{2}
\\
& &
       + \frac{1}{2} \ln^2(1\!+\!x) \ln{2}
       - \frac{1}{3} \ln^3(1\!+\!x)
       + \ln(1\!+\!x) \Li_2(x)
       + \frac{1}{2} \ln{x} \ln^2(1\!+\!x)
\nonumber\\
& &
       - \Li_3\!\! \left(\frac{1\!+\!x}{2}\right)
       + \Li_3( - x)
       + \Li_3\!\! \left(\frac{1}{1\!+\!x}\right)
       + \Li_3\!\! \left(\frac{2x}{1\!+\!x}\right)
       - \Li_3(x)
      \, ,\nonumber\\
\label{eq:weight3_17}
   H_{0,1,0}(x) &=&
         \ln{x} \Li_2(x)
       - 2 \Li_3(x)
      \, ,\\
\label{eq:weight3_18}
   H_{0,1,1}(x) &=&
         \z_3
       + \ln(1\!-\!x) \z_2
       - \ln(1\!-\!x) \Li_2(x)
       - \frac{1}{2} \ln{x} \ln^2(1\!-\!x)
       - \Li_3(1\!-\!x)
      \, ,\\
\label{eq:weight3_19}
   H_{1,-1,-1}(x) &=&
         \frac{1}{2} \z_2 \ln{2}
       - \frac{7}{8} \z_3
       - \frac{1}{6} \ln^3{2}
       - \frac{1}{2} \ln(1\!-\!x) \ln^2(1\!+\!x)
       + \frac{1}{2} \ln^2(1\!+\!x) \ln{2}
\\
& &
       - \ln(1\!+\!x) \Li_2\!\! \left(\frac{1\!+\!x}{2}\right)
       + \Li_3\!\! \left(\frac{1\!+\!x}{2}\right)
      \, ,\nonumber\\
\label{eq:weight3_20}
   H_{1,-1,0}(x) &=&
         \frac{1}{2} \z_2 \ln{2}
       - \frac{1}{8} \z_3
       - \frac{1}{6} \ln^3{2}
       + \frac{1}{2} \ln(1\!-\!x) \z_2
       - 2 \ln(1\!+\!x) \z_2
\\
& &
       + \frac{1}{2} \ln(1\!+\!x) \ln^2{2}
       - \frac{1}{2} \ln^2(1\!+\!x) \ln{2}
       + \frac{1}{6} \ln^3(1\!+\!x)
       + \frac{1}{2} \ln{x} \z_2
       - \frac{1}{2} \ln{x} \ln^2{2}
\nonumber\\
& &
       + \ln{x} \ln(1\!+\!x) \ln{2}
       - \frac{1}{2} \ln{x} \ln^2(1\!+\!x)
       - \ln{x} \Li_2\!\! \left(\frac{1\!+\!x}{2}\right)
       + \Li_3\!\! \left(\frac{1\!+\!x}{2}\right)
\nonumber\\
& &
       + \Li_3(1\!-\!x)
       - \Li_3( - x)
       + \Li_3\!\! \left( -\frac{1\!-\!x}{1\!+\!x}\right)
       - \Li_3\!\! \left(\frac{1\!-\!x}{1\!+\!x}\right)
       - \Li_3\!\! \left(\frac{2x}{1\!+\!x}\right)
       + \Li_3(x)
      \, ,\nonumber\\
\label{eq:weight3_21}
   H_{1,-1,1}(x) &=&
       - \z_2 \ln{2}
       + \frac{1}{4} \z_3
       + \frac{1}{3} \ln^3{2}
       - \frac{3}{2} \ln(1\!-\!x) \z_2
       + \frac{1}{2} \ln(1\!-\!x) \ln^2{2}
\\
& &
       - \ln^2(1\!-\!x) \ln{2}
       + \ln^2(1\!-\!x) \ln(1\!+\!x)
       + \ln(1\!-\!x) \ln(1\!+\!x) \ln{2}
\nonumber\\
& &
       - \ln(1\!-\!x) \ln^2(1\!+\!x)
       + \ln(1\!-\!x) \Li_2\!\! \left(\frac{1\!+\!x}{2}\right)
       + 2 \ln(1\!+\!x) \z_2
       - \ln(1\!+\!x) \ln^2{2}
\nonumber\\
& &
       + \frac{1}{3} \ln^3(1\!+\!x)
       - 2 \Li_3\!\! \left(\frac{1\!+\!x}{2}\right)
       - 2 \Li_3\!\! \left( -\frac{1\!-\!x}{1\!+\!x}\right)
      \, ,\nonumber\\
\label{eq:weight3_22}
   H_{1,0,-1}(x) &=&
         \frac{1}{4} \z_3
       - \frac{1}{2} \ln(1\!-\!x) \z_2
       + \frac{1}{2} \ln(1\!+\!x) \z_2
       + \frac{1}{6} \ln^3(1\!+\!x)
       - \ln(1\!+\!x) \Li_2(x)
\\
& &
       - \ln{x} \ln(1\!-\!x) \ln(1\!+\!x)
       - \Li_3(1\!-\!x)
       - \Li_3\!\! \left( -\frac{1\!-\!x}{1\!+\!x}\right)
       - \Li_3\!\! \left(\frac{1}{1\!+\!x}\right)
\nonumber\\
& &
       + \Li_3\!\! \left(\frac{1\!-\!x}{1\!+\!x}\right)
      \, ,\nonumber\\
\label{eq:weight3_23}
   H_{1,0,0}(x) &=&
       - \frac{1}{2} \ln^2x \ln(1\!-\!x)
       - \ln{x} \Li_2(x)
       + \Li_3(x)  
      \\
&=&
         \z_3
       - {\rm{S}}_{n,p}(1-x) 
      \, ,\nonumber\\
\label{eq:weight3_24}
   H_{1,0,1}(x) &=&
       - 2 \z_3
       - 2 \ln(1\!-\!x) \z_2
       + \ln(1\!-\!x) \Li_2(x)
       + \ln{x} \ln^2(1\!-\!x)
       + 2 \Li_3(1\!-\!x)
      \, ,\\
\label{eq:weight3_25}
   H_{1,1,-1}(x) &=&
         \frac{1}{2} \z_2 \ln{2}
       - \frac{1}{8} \z_3
       - \frac{1}{6} \ln^3{2}
       + \frac{1}{2} \ln(1\!-\!x) \z_2
       + \frac{1}{2} \ln^2(1\!-\!x) \ln{2}
\\
& &
       - \ln(1\!-\!x) \ln(1\!+\!x) \ln{2}
       + \frac{1}{2} \ln(1\!-\!x) \ln^2(1\!+\!x)
       - \ln(1\!+\!x) \z_2
       + \frac{1}{2} \ln(1\!+\!x) \ln^2{2}
\nonumber\\
& &
       - \frac{1}{6} \ln^3(1\!+\!x)
       + \Li_3\!\! \left(\frac{1\!+\!x}{2}\right)
       + \Li_3\!\! \left( -\frac{1\!-\!x}{1\!+\!x}\right)
      \, ,\nonumber\\
\label{eq:weight3_26}
   H_{1,1,0}(x) &=&
         \z_3
       + \ln(1\!-\!x) \z_2
       - \Li_3(1\!-\!x)
      \, ,\\
\label{eq:weight3_27}
   H_{1,1,1}(x) &=&
       - \frac{1}{6} \ln^3(1\!-\!x)
      \, .
\end{eqnarray}

The function ${\rm{S}}_{n,p}$ in eqs.(\ref{eq:weight3_10}) 
and (\ref{eq:weight3_23}) denote 
the Nielsen functions~\cite{lewin:book}, defined as 
\begin{eqnarray}
{\rm{S}}_{n,p}(x) &=& \frac{(-1)^{p+n-1}}{p! \, (n-1)!}
   \int\limits_0^1 \frac{dz}{z}\, \ln^{n-1}(z)\, \ln^p(1-xz)\, .
\end{eqnarray}

\section*{Appendix B}

Here we present the formulae for the Mellin moments of the 2-loop 
coefficient functions. We obtain,
\begin{eqnarray}
\lefteqn{
c^{(2),+{\rm{ns}}}_{2,{\rm{q}}}(N)  \,=\, 
\d(N\!-\!2) \Biggl\{
         \Bigl(
            \frac{11581}{360}
          - \frac{86}{9} \z_2
          - \frac{32}{5} \z_2^2
          - \frac{9}{5} \z_3
          \Bigr) C_F C_A  
       - 4 C_F n_f }\\
& &
\hspace*{20mm}
       - \Bigl(
            \frac{24359}{1620}
          - \frac{172}{9} \z_2
          - \frac{64}{5} \z_2^2
          + \frac{142}{5} \z_3
          \Bigr) C_F^2
          \Biggr\}
\,+\, \q(N\!-\!3)\, (-1)^N\, \times  \nonumber\\
& &
\!\!\!\Biggl[
       C_F C_A   \Biggl\{ (-1)^N \Bigl(
          - \frac{5465}{72}
          - 4 \z_2
          - \frac{32}{5} \z_2^2
          + 54 \z_3
          - 12 S_{-2}(N\!-\!1) \z_2
          - 24 S_{-2,-2}(N\!-\!1)
\nonumber\\
& &\hspace*{10mm}
          + \Bigl( \frac{3155}{54} 
                 - 32 \z_3 \Bigr) S_{1}(N\!-\!1) 
          + \Bigl( \frac{367}{9} 
                 - 16 \z_2 \Bigr) S_{1,1}(N\!-\!1) 
          + \frac{44}{3} S_{1,1,1}(N\!-\!1)
\nonumber\\
& &\hspace*{10mm}
          + 8 S_{1,1,2}(N\!-\!1)
          - \frac{44}{3} S_{1,2}(N\!-\!1)
          - 8 S_{1,2,1}(N\!-\!1)
          + 16 S_{1,3}(N\!-\!1)
\nonumber\\
& &\hspace*{10mm}
          - \Bigl( \frac{239}{3} 
                 - 12 \z_2 \Bigr) S_{2}(N\!-\!1) 
          - \frac{88}{3} S_{2,1}(N\!-\!1)
          + \frac{110}{3} S_{3}(N\!-\!1)
          + 8 S_{3,1}(N\!-\!1)
\nonumber\\
& &\hspace*{10mm}
          - 12 S_{4}(N\!-\!1)
          \Bigr)
          + 6 S_{-4}(N\!-\!1)
          + 6 S_{-4}(N\!+\!1)
          - 12 S_{-4}(N)
          - \frac{43}{3} S_{-3}(N\!-\!1)
\nonumber\\
& &\hspace*{10mm}
          + \frac{5}{3} S_{-3}(N\!+\!1)
          - \frac{156}{5} S_{-3}(N\!+\!2)
          + \frac{36}{5} S_{-3}(N\!+\!3)
          + \frac{110}{3} S_{-3}(N)
          - 4 S_{-3,1}(N\!-\!1)
\nonumber\\
& &\hspace*{10mm}
          - 4 S_{-3,1}(N\!+\!1)
          + 8 S_{-3,1}(N)
          - \frac{4}{5} S_{-2}(N\!-\!2)
          + \Bigl( \frac{616}{15} 
                 - 4 \z_2 \Bigr) S_{-2}(N\!-\!1)
\nonumber\\
& &\hspace*{10mm}
          + \Bigl( \frac{1366}{15} 
                 - 16 \z_2 \Bigr) S_{-2}(N\!+\!1)
          + \frac{36}{5} S_{-2}(N\!+\!2)
          - \Bigl( \frac{2078}{15} 
                 - 20\z_2  \Bigr) S_{-2}(N) 
\nonumber\\
& &\hspace*{10mm}
          + \frac{32}{3} S_{-2,1}(N\!-\!1)
          + \frac{32}{3} S_{-2,1}(N\!+\!1)
          - \frac{64}{3} S_{-2,1}(N)
          - \frac{2}{5} S_{-1}(N\!-\!3) \z_2
\nonumber\\
& &\hspace*{10mm}
          - \Bigl( \frac{4}{5} 
                 - \frac{2}{5} \z_2 \Bigr) S_{-1}(N\!-\!2)
          - \Bigl( \frac{1414}{135} 
                 + 8 \z_2
                 - 6 \z_3 \Bigr) S_{-1}(N\!-\!1)
\nonumber\\
& &\hspace*{10mm}
          - \Bigl( \frac{17761}{135} 
                 + 4 \z_2
                 - 66 \z_3 \Bigr) S_{-1}(N\!+\!1) 
          - \Bigl( \frac{36}{5} 
                 - \frac{78}{5} \z_2 \Bigr) S_{-1}(N\!+\!2) 
          - \frac{18}{5} S_{-1}(N\!+\!3) \z_2
\nonumber\\
& &\hspace*{10mm}
          + \Bigl( \frac{4051}{27} 
                 - 72 \z_3 \Bigr) S_{-1}(N)
          + \Bigl( \frac{16}{9} 
                 + 4 \z_2 \Bigr) S_{-1,1}(N\!-\!1) 
          - \Bigl( \frac{740}{9} 
                 - 28 \z_2 \Bigr) S_{-1,1}(N\!+\!1) 
\nonumber\\
& &\hspace*{10mm}
          + \Bigl( \frac{724}{9} 
                 - 32 \z_2 \Bigr) S_{-1,1}(N) 
          - \frac{22}{3} S_{-1,1,1}(N\!-\!1)
          - \frac{22}{3} S_{-1,1,1}(N\!+\!1)
          + \frac{44}{3} S_{-1,1,1}(N)
\nonumber\\
& &\hspace*{10mm}
          - 4 S_{-1,1,2}(N\!-\!1)
          - 4 S_{-1,1,2}(N\!+\!1)
          + 8 S_{-1,1,2}(N)
          + \frac{22}{3} S_{-1,2}(N\!-\!1)
          + \frac{22}{3} S_{-1,2}(N\!+\!1)
\nonumber\\
& &\hspace*{10mm}
          - \frac{44}{3} S_{-1,2}(N)
          + 4 S_{-1,2,1}(N\!-\!1)
          + 4 S_{-1,2,1}(N\!+\!1)
          - 8 S_{-1,2,1}(N)
          - 4 S_{-1,3}(N\!-\!1)
\nonumber\\
& &\hspace*{10mm}
          - 28 S_{-1,3}(N\!+\!1)
          + 32 S_{-1,3}(N)
          + \frac{2}{5} S_{1}(N\!-\!3) \z_2
          - \frac{2}{5} S_{1}(N\!-\!2) \z_2
          + 10 S_{1}(N\!-\!1) \z_2
\nonumber\\
& &\hspace*{10mm}
          - 10 S_{1}(N\!+\!1) \z_2
          + \frac{78}{5} S_{1}(N\!+\!2) \z_2
          - \frac{18}{5} S_{1}(N\!+\!3) \z_2
          - 12 S_{1}(N) \z_2
          + \frac{4}{5} S_{1,-2}(N\!-\!3)
\nonumber\\
& &\hspace*{10mm}
          - \frac{4}{5} S_{1,-2}(N\!-\!2)
          + 20 S_{1,-2}(N\!-\!1)
          - 20 S_{1,-2}(N\!+\!1)
          + \frac{156}{5} S_{1,-2}(N\!+\!2)
\nonumber\\
& &\hspace*{10mm}
          - \frac{36}{5} S_{1,-2}(N\!+\!3)
          - 24 S_{1,-2}(N)
          + 4 S_{2}(N\!-\!1) \z_2
          + 16 S_{2}(N\!+\!1) \z_2
          - 20 S_{2}(N) \z_2
\nonumber\\
& &\hspace*{10mm}
          + 8 S_{2,-2}(N\!-\!1)
          + 32 S_{2,-2}(N\!+\!1)
          - 40 S_{2,-2}(N)
          \Biggr\}
\nonumber\\
&+\!\!&
       C_F n_f   \Biggl\{ (-1)^N \Bigl(
            \frac{457}{36}
          - \frac{247}{27} S_{1}(N\!-\!1)
          - \frac{58}{9} S_{1,1}(N\!-\!1)
          - \frac{8}{3} S_{1,1,1}(N\!-\!1)
          + \frac{8}{3} S_{1,2}(N\!-\!1)
\nonumber\\
& &\hspace*{10mm}
          + \frac{38}{3} S_{2}(N\!-\!1)
          + \frac{16}{3} S_{2,1}(N\!-\!1)
          - \frac{20}{3} S_{3}(N\!-\!1)
          \Bigr)
          + \frac{10}{3} S_{-3}(N\!-\!1)
          + \frac{10}{3} S_{-3}(N\!+\!1)
\nonumber\\
& &\hspace*{10mm}
          - \frac{20}{3} S_{-3}(N)
          - \frac{26}{3} S_{-2}(N\!-\!1)
          - \frac{38}{3} S_{-2}(N\!+\!1)                  
          + \frac{64}{3} S_{-2}(N)
          - \frac{8}{3} S_{-2,1}(N\!-\!1)
\nonumber\\
& &\hspace*{10mm} 
          - \frac{8}{3} S_{-2,1}(N\!+\!1)
          + \frac{16}{3} S_{-2,1}(N)
          + \frac{158}{27} S_{-1}(N\!-\!1)
          + \frac{488}{27} S_{-1}(N\!+\!1)
          - \frac{646}{27} S_{-1}(N)
\nonumber\\
& &\hspace*{10mm}
          + \frac{32}{9} S_{-1,1}(N\!-\!1)
          + \frac{68}{9} S_{-1,1}(N\!+\!1)
          - \frac{100}{9} S_{-1,1}(N)
          + \frac{4}{3} S_{-1,1,1}(N\!-\!1)
\nonumber\\
& &\hspace*{10mm}
          + \frac{4}{3} S_{-1,1,1}(N\!+\!1)
          - \frac{8}{3} S_{-1,1,1}(N)
          - \frac{4}{3} S_{-1,2}(N\!-\!1)
          - \frac{4}{3} S_{-1,2}(N\!+\!1)
          + \frac{8}{3} S_{-1,2}(N)
          \Biggr\}
\nonumber\\
&+\!\!&
       C_F^2   \Biggl\{ (-1)^N \Bigl(
            \frac{331}{8}
          + 8 \z_2
          + \frac{64}{5} \z_2^2
          - 72 \z_3
          + 24 S_{-2}(N\!-\!1) \z_2
          + 48 S_{-2,-2}(N\!-\!1)
\nonumber\\
& &\hspace*{10mm}
          - \Bigl( \frac{51}{2} 
                 - 16 \z_3 \Bigr) S_{1}(N\!-\!1) 
          - \Bigl( 27 
                 - 32 \z_2 \Bigr) S_{1,1}(N\!-\!1) 
          + 36 S_{1,1,1}(N\!-\!1)
\nonumber\\
& &\hspace*{10mm}
          + 48 S_{1,1,1,1}(N\!-\!1)
          - 64 S_{1,1,2}(N\!-\!1)
          - 36 S_{1,2}(N\!-\!1)
          - 48 S_{1,2,1}(N\!-\!1)
          + 24 S_{1,3}(N\!-\!1)
\nonumber\\
& &\hspace*{10mm}
          + \Bigl( 61 
                 - 24 \z_2 \Bigr) S_{2}(N\!-\!1) 
          - 24 S_{2,1}(N\!-\!1)
          - 56 S_{2,1,1}(N\!-\!1)
          + 48 S_{2,2}(N\!-\!1)
\nonumber\\
& &\hspace*{10mm}
          + 6 S_{3}(N\!-\!1)
          + 48 S_{3,1}(N\!-\!1)
          - 16 S_{4}(N\!-\!1)
          \Bigr)
          + 18 S_{-4}(N\!-\!1)
          + 18 S_{-4}(N\!+\!1)
\nonumber\\
& &\hspace*{10mm}
          - 36 S_{-4}(N)
          - 32 S_{-3}(N\!-\!1)
          - 76 S_{-3}(N\!+\!1)
          + \frac{312}{5} S_{-3}(N\!+\!2)
          - \frac{72}{5} S_{-3}(N\!+\!3)
\nonumber\\
& &\hspace*{10mm}
          + 60 S_{-3}(N)
          - 32 S_{-3,1}(N\!-\!1)
          - 32 S_{-3,1}(N\!+\!1)
          + 64 S_{-3,1}(N)
          + \frac{8}{5} S_{-2}(N\!-\!2)
\nonumber\\
& &\hspace*{10mm}
          - \Bigl( \frac{154}{5} 
                 - 8 \z_2 \Bigr) S_{-2}(N\!-\!1) 
          - \Bigl( \frac{284}{5} 
                 - 32 \z_2 \Bigr) S_{-2}(N\!+\!1)
          - \frac{72}{5} S_{-2}(N\!+\!2)
\nonumber\\
& &\hspace*{10mm}
          + \Bigl( \frac{502}{5} 
                 - 40 \z_2 \Bigr) S_{-2}(N) 
          + 40 S_{-2,1}(N\!-\!1)
          + 56 S_{-2,1}(N\!+\!1)
          - 96 S_{-2,1}(N)
\nonumber\\
& &\hspace*{10mm}
          + 32 S_{-2,1,1}(N\!-\!1)
          + 32 S_{-2,1,1}(N\!+\!1)
          - 64 S_{-2,1,1}(N)
          - 28 S_{-2,2}(N\!-\!1)
\nonumber\\
& &\hspace*{10mm}
          - 28 S_{-2,2}(N\!+\!1)
          + 56 S_{-2,2}(N)
          + \frac{4}{5} S_{-1}(N\!-\!3) \z_2
          + \Bigl( \frac{8}{5} 
                 - \frac{4}{5} \z_2 \Bigr) S_{-1}(N\!-\!2)
\nonumber\\
& &\hspace*{10mm}
          - \Bigl( \frac{46}{5} 
                 - 16 \z_2
                 - 12 \z_3 \Bigr) S_{-1}(N\!-\!1) 
          + \Bigl( \frac{471}{5} 
                 + 8 \z_2
                 - 108 \z_3 \Bigr) S_{-1}(N\!+\!1)
\nonumber\\
& &\hspace*{10mm}
          + \Bigl( \frac{72}{5} 
                 - \frac{156}{5} \z_2 \Bigr) S_{-1}(N\!+\!2) 
          + \frac{36}{5} S_{-1}(N\!+\!3) \z_2
          - \Bigl( 101 
                 - 96 \z_3 \Bigr) S_{-1}(N) 
\nonumber\\
& &\hspace*{10mm}
          - \Bigl( 32 
                 + 8 \z_2 \Bigr) S_{-1,1}(N\!-\!1) 
          + \Bigl( 84 
                 - 56 \z_2 \Bigr) S_{-1,1}(N\!+\!1) 
          - \Bigl( 52 
                 - 64 \z_2 \Bigr) S_{-1,1}(N)
\nonumber\\
& &\hspace*{10mm}
          - 28 S_{-1,1,1}(N\!-\!1)
          - 36 S_{-1,1,1}(N\!+\!1)
          + 64 S_{-1,1,1}(N)
          - 24 S_{-1,1,1,1}(N\!-\!1)
\nonumber\\
& &\hspace*{10mm}
          - 24 S_{-1,1,1,1}(N\!+\!1)
          + 48 S_{-1,1,1,1}(N)
          + 32 S_{-1,1,2}(N\!-\!1)
          + 32 S_{-1,1,2}(N\!+\!1)
\nonumber\\
& &\hspace*{10mm}
          - 64 S_{-1,1,2}(N)
          + 32 S_{-1,2}(N\!-\!1)
          + 32 S_{-1,2}(N\!+\!1)
          - 64 S_{-1,2}(N)
          + 24 S_{-1,2,1}(N\!-\!1)
\nonumber\\
& &\hspace*{10mm}
          + 24 S_{-1,2,1}(N\!+\!1)
          - 48 S_{-1,2,1}(N)
          - 20 S_{-1,3}(N\!-\!1)
          + 28 S_{-1,3}(N\!+\!1)
          - 8 S_{-1,3}(N)
\nonumber\\
& &\hspace*{10mm}
          - \frac{4}{5} S_{1}(N\!-\!3) \z_2
          + \frac{4}{5} S_{1}(N\!-\!2) \z_2
          - 20 S_{1}(N\!-\!1) \z_2
          + 20 S_{1}(N\!+\!1) \z_2
          - \frac{156}{5} S_{1}(N\!+\!2) \z_2
\nonumber\\
& &\hspace*{10mm}
          + \frac{36}{5} S_{1}(N\!+\!3) \z_2
          + 24 S_{1}(N) \z_2
          - \frac{8}{5} S_{1,-2}(N\!-\!3)
          + \frac{8}{5} S_{1,-2}(N\!-\!2)
          - 40 S_{1,-2}(N\!-\!1)
\nonumber\\
& &\hspace*{10mm}
          + 40 S_{1,-2}(N\!+\!1)
          - \frac{312}{5} S_{1,-2}(N\!+\!2)
          + \frac{72}{5} S_{1,-2}(N\!+\!3)
          + 48 S_{1,-2}(N)
          - 8 S_{2}(N\!-\!1) \z_2
\nonumber\\
& &\hspace*{10mm}
          - 32 S_{2}(N\!+\!1) \z_2
          + 40 S_{2}(N) \z_2
          - 16 S_{2,-2}(N\!-\!1)
          - 64 S_{2,-2}(N\!+\!1)
          + 80 S_{2,-2}(N)
          \Biggr\}
\Biggr]
\, ,
\nonumber\\[2ex]
\lefteqn{
c^{(2),-{\rm{ns}}}_{2,{\rm{q}}}(N)  \,=}\\
& &  \d(N\!-\!2) \Biggl\{
          \Bigl(
            \frac{5327}{540}
          - \frac{172}{9} \z_2
          - \frac{64}{5} \z_2^2
          + \frac{238}{5} \z_3
          \Bigr)  C_F \Bigl(C_F-\frac{C_A}{2}\Bigr)
          \Biggr\}
\,+\, \q(N\!-\!3)\, (-1)^N\, \times  
\nonumber\\
& &
\!\!\!\Biggl[
       C_F \Bigl(C_F-\frac{C_A}{2}\Bigr)   \Biggl\{
          - 8 \z_2
          - \frac{64}{5} \z_2^2
          - 12 S_{-4}(N\!-\!1)
          - 12 S_{-4}(N\!+\!1)
          + 8 S_{-3}(N\!-\!1)
\nonumber\\
& &\hspace*{10mm}
          + 80 S_{-3}(N\!+\!1)
          - \frac{168}{5} S_{-3}(N\!+\!2)
          - \frac{72}{5} S_{-3}(N\!+\!3)
          - 40 S_{-3}(N)
          + 8 S_{-3,1}(N\!-\!1)
\nonumber\\
& &\hspace*{10mm}
          + 8 S_{-3,1}(N\!+\!1)
          + \frac{8}{5} S_{-2}(N\!-\!2)
          - \Bigl( \frac{74}{5} 
                 + 16 \z_2 \Bigr) S_{-2}(N\!-\!1)
          - \Bigl( \frac{74}{5} 
                 + 32 \z_2 \Bigr) S_{-2}(N\!+\!1) 
\nonumber\\
& &\hspace*{10mm}
          - \frac{72}{5} S_{-2}(N\!+\!2)
          + \Bigl( \frac{132}{5} 
                 + 24 \z_2 \Bigr) S_{-2}(N)
          - 8 S_{-2,1}(N\!-\!1)
          - 8 S_{-2,1}(N\!+\!1)
\nonumber\\
& &\hspace*{10mm}
          + 16 S_{-2,1}(N)
          + \frac{4}{5} S_{-1}(N\!-\!3) \z_2
          + \Bigl( \frac{8}{5} 
                 - \frac{4}{5} \z_2 \Bigr) S_{-1}(N\!-\!2) 
          + \Bigl( \frac{154}{5} 
                 + 20 \z_2 \Bigr) S_{-1}(N\!-\!1)
\nonumber\\
& &\hspace*{10mm}
          - \Bigl( \frac{154}{5} 
                 + 28 \z_2 \Bigr) S_{-1}(N\!+\!1) 
          + \Bigl( \frac{72}{5} 
                 + \frac{84}{5} \z_2 \Bigr) S_{-1}(N\!+\!2) 
          + \frac{36}{5} S_{-1}(N\!+\!3) \z_2
\nonumber\\
& &\hspace*{10mm}
          - \Bigl( 16 
                 + 16 \z_2 \Bigr) S_{-1}(N)
          + 16 S_{-1,1}(N\!-\!1)
          - 16 S_{-1,1}(N\!+\!1)
          - \frac{4}{5} S_{1}(N\!-\!3) \z_2
\nonumber\\
& &\hspace*{10mm}
          + \frac{4}{5} S_{1}(N\!-\!2) \z_2
          - \Bigl( 16 \z_2
                 - 20 \z_3 \Bigr) S_{1}(N\!-\!1)
          - \Bigl( 40 \z_2
                 - 36 \z_3 \Bigr) S_{1}(N\!+\!1)
          + \frac{84}{5} S_{1}(N\!+\!2) \z_2
\nonumber\\
& &\hspace*{10mm}
          + \frac{36}{5} S_{1}(N\!+\!3) \z_2
          + \Bigl( 32 \z_2
                 - 24 \z_3 \Bigr) S_{1}(N)
          + 48 S_{1,-3}(N\!-\!1)
          + 80 S_{1,-3}(N\!+\!1)
\nonumber\\
& &\hspace*{10mm}
          - 48 S_{1,-3}(N)
          - \frac{8}{5} S_{1,-2}(N\!-\!3)
          + \frac{8}{5} S_{1,-2}(N\!-\!2)
          - 32 S_{1,-2}(N\!-\!1)
          - 80 S_{1,-2}(N\!+\!1)
\nonumber\\
& &\hspace*{10mm}
          + \frac{168}{5} S_{1,-2}(N\!+\!2)
          + \frac{72}{5} S_{1,-2}(N\!+\!3)
          + 64 S_{1,-2}(N)
          - 16 S_{1,-2,1}(N\!-\!1)
\nonumber\\
& &\hspace*{10mm}
          - 16 S_{1,-2,1}(N\!+\!1)
          - 24 S_{1,1}(N\!-\!1) \z_2
          - 56 S_{1,1}(N\!+\!1) \z_2
          + 48 S_{1,1}(N) \z_2
\nonumber\\
& &\hspace*{10mm}
          - 48 S_{1,1,-2}(N\!-\!1)
          - 112 S_{1,1,-2}(N\!+\!1)
          + 96 S_{1,1,-2}(N)
          + 16 S_{2}(N\!-\!1) \z_2
\nonumber\\
& &\hspace*{10mm}
          + 32 S_{2}(N\!+\!1) \z_2
          - 24 S_{2}(N) \z_2
          + 32 S_{2,-2}(N\!-\!1)
          + 64 S_{2,-2}(N\!+\!1)
          - 48 S_{2,-2}(N)
          \Biggr\}
\Biggr]
\, ,
\nonumber\\[2ex]
\lefteqn{
c^{(2),+{\rm{ns}}}_{3,{\rm{q}}}(N)  \,=\, c^{(2),+{\rm{ns}}}_{2,{\rm{q}}}(N) 
+  \d(N\!-\!1) \Biggl\{
          \Bigl(
          - \frac{175}{8}
          - 4 \z_2
          - \frac{32}{5} \z_2^2
          + 19 \z_3
          \Bigr) C_F C_A 
       + 4 C_F n_f }\\
& &\hspace*{20mm}
       +  \Bigl(
            \frac{33}{4}
          + 8 \z_2
          + \frac{64}{5} \z_2^2
          - 38 \z_3
          \Bigr) C_F^2  
          \Biggr\}
+  \d(N\!-\!2) \Biggl\{
         \Bigl(
          - \frac{13669}{540}
          + \frac{1}{3} \z_2
          + \frac{44}{5} \z_3
          \Bigr) C_F C_A 
\nonumber\\
& &\hspace*{20mm}
       + \frac{106}{27} C_F n_f  
       + \Bigl(
            \frac{794}{45}
          - \frac{2}{3} \z_2
          - \frac{88}{5} \z_3
          \Bigr) C_F^2  
          \Biggr\}
\,+\, \q(N\!-\!3)\, (-1)^N\, \times \nonumber\\
& &
\!\!\!\Biggl[
       C_F C_A   \Biggl\{
            4 S_{-3}(N\!-\!1)
          - 8 S_{-3}(N\!+\!1)
          + \frac{136}{5} S_{-3}(N\!+\!2)
          - \frac{36}{5} S_{-3}(N\!+\!3)
          - 16 S_{-3}(N)
\nonumber\\
& &\hspace*{10mm}
          + \frac{4}{5} S_{-2}(N\!-\!2)
          - \Bigl( \frac{136}{15} 
                 + 4 \z_2 \Bigr) S_{-2}(N\!-\!1) 
          - \Bigl( \frac{376}{15} 
                 - 12 \z_2 \Bigr) S_{-2}(N\!+\!1) 
          - \frac{36}{5} S_{-2}(N\!+\!2)
\nonumber\\
& &\hspace*{10mm}
          + \Bigl( \frac{608}{15} 
                 - 8 \z_2 \Bigr) S_{-2}(N)
          + \frac{2}{5} S_{-1}(N\!-\!3) \z_2
          + \Bigl( \frac{4}{5} 
                 - \frac{12}{5} \z_2 \Bigr) S_{-1}(N\!-\!2) 
\nonumber\\
& &\hspace*{10mm}
          - \Bigl( \frac{737}{45} 
                 - 10 \z_2
                 - 20 \z_3 \Bigr) S_{-1}(N\!-\!1)
          + \Bigl( \frac{2887}{45} 
                 + 2 \z_2
                 - 60 \z_3 \Bigr) S_{-1}(N\!+\!1) 
\nonumber\\
& &\hspace*{10mm}
          + \Bigl( \frac{36}{5} 
                 - \frac{68}{5} \z_2 \Bigr) S_{-1}(N\!+\!2) 
          + \frac{18}{5} S_{-1}(N\!+\!3) \z_2
          - \Bigl( \frac{502}{9} 
                 - 40  \z_3 \Bigr) S_{-1}(N)
\nonumber\\
& &\hspace*{10mm}
          - \Bigl( \frac{50}{3}
               - 8 \z_2 \Bigr) S_{-1,1}(N\!-\!1) 
          + \Bigl( \frac{142}{3} 
                 - 24 \z_2 \Bigr) S_{-1,1}(N\!+\!1) 
          - \Bigl( \frac{92}{3} 
                 - 16 \z_2 \Bigr) S_{-1,1}(N) 
\nonumber\\
& &\hspace*{10mm}
          - 8 S_{-1,3}(N\!-\!1)
          + 24 S_{-1,3}(N\!+\!1)
          - 16 S_{-1,3}(N)
          - \frac{2}{5} S_{1}(N\!-\!3) \z_2
          + \frac{12}{5} S_{1}(N\!-\!2) \z_2
\nonumber\\
& &\hspace*{10mm}
          - 14 S_{1}(N\!-\!1) \z_2
          + 6 S_{1}(N\!+\!1) \z_2
          - \frac{68}{5} S_{1}(N\!+\!2) \z_2
          + \frac{18}{5} S_{1}(N\!+\!3) \z_2
          + 16 S_{1}(N) \z_2
\nonumber\\
& &\hspace*{10mm}
          - \frac{4}{5} S_{1,-2}(N\!-\!3)
          + \frac{24}{5} S_{1,-2}(N\!-\!2)
          - 28 S_{1,-2}(N\!-\!1)
          + 12 S_{1,-2}(N\!+\!1)
\nonumber\\
& &\hspace*{10mm}
          - \frac{136}{5} S_{1,-2}(N\!+\!2)
          + \frac{36}{5} S_{1,-2}(N\!+\!3)
          + 32 S_{1,-2}(N)
          + 4 S_{2}(N\!-\!1) \z_2
          - 12 S_{2}(N\!+\!1) \z_2
\nonumber\\
& &\hspace*{10mm}
          + 8 S_{2}(N) \z_2
          + 8 S_{2,-2}(N\!-\!1)
          - 24 S_{2,-2}(N\!+\!1)
          + 16 S_{2,-2}(N)
          \Biggr\}
\nonumber\\
&+\!\!&
       C_F n_f   \Biggl\{
            \frac{8}{3} S_{-2}(N\!-\!1)
          + \frac{8}{3} S_{-2}(N\!+\!1)
          - \frac{16}{3} S_{-2}(N)
          - \frac{14}{9} S_{-1}(N\!-\!1)
          - \frac{62}{9} S_{-1}(N\!+\!1)
\nonumber\\
& &\hspace*{10mm}
          + \frac{76}{9} S_{-1}(N)
          - \frac{4}{3} S_{-1,1}(N\!-\!1)
          - \frac{4}{3} S_{-1,1}(N\!+\!1)
          + \frac{8}{3} S_{-1,1}(N)
          \Biggr\}
\nonumber\\
&+\!\!&
       C_F^2   \Biggl\{
            24 S_{-3}(N\!+\!1)
          - \frac{272}{5} S_{-3}(N\!+\!2)
          + \frac{72}{5} S_{-3}(N\!+\!3)
          + 16 S_{-3}(N)
          - \frac{8}{5} S_{-2}(N\!-\!2)
\nonumber\\
& &\hspace*{10mm}
          + \Bigl( \frac{24}{5} 
                 + 8 \z_2 \Bigr) S_{-2}(N\!-\!1) 
          + \Bigl( \frac{124}{5} 
                 - 24 \z_2 \Bigr) S_{-2}(N\!+\!1) 
          + \frac{72}{5} S_{-2}(N\!+\!2)
\nonumber\\
& &\hspace*{10mm}
          - \Bigl( \frac{212}{5} 
                 - 16  \z_2 \Bigr) S_{-2}(N) 
          - 12 S_{-2,1}(N\!-\!1)
          - 12 S_{-2,1}(N\!+\!1)
          + 24 S_{-2,1}(N)
\nonumber\\
& &\hspace*{10mm}
          - \frac{4}{5} S_{-1}(N\!-\!3) \z_2
          - \Bigl( \frac{8}{5} 
                 - \frac{24}{5} \z_2 \Bigr) S_{-1}(N\!-\!2) 
          + \Bigl( \frac{251}{5} 
                 - 20 \z_2
                 - 40 \z_3 \Bigr) S_{-1}(N\!-\!1) 
\nonumber\\
& &\hspace*{10mm}
          - \Bigl( \frac{421}{5} 
                 + 4 \z_2
                 - 120 \z_3 \Bigr) S_{-1}(N\!+\!1) 
          - \Bigl( \frac{72}{5} 
                 - \frac{136}{5} \z_2 \Bigr) S_{-1}(N\!+\!2) 
          - \frac{36}{5} S_{-1}(N\!+\!3) \z_2
\nonumber\\
& &\hspace*{10mm}
          + \Bigl( 50 
                 - 80 \z_3 \Bigr) S_{-1}(N)
          + \Bigl( 42 
                 - 16 \z_2 \Bigr) S_{-1,1}(N\!-\!1)
          - \Bigl( 78 
                 - 48 \z_2 \Bigr) S_{-1,1}(N\!+\!1)
\nonumber\\
& &\hspace*{10mm}
          + \Bigl( 36 
                 - 32 \z_2 \Bigr) S_{-1,1}(N) 
          + 8 S_{-1,1,1}(N\!-\!1)
          + 8 S_{-1,1,1}(N\!+\!1)
          - 16 S_{-1,1,1}(N)
\nonumber\\
& &\hspace*{10mm}
          - 8 S_{-1,2}(N\!-\!1)
          - 8 S_{-1,2}(N\!+\!1)
          + 16 S_{-1,2}(N)
          + 16 S_{-1,3}(N\!-\!1)
          - 48 S_{-1,3}(N\!+\!1)
\nonumber\\
& &\hspace*{10mm}
          + 32 S_{-1,3}(N)
          + \frac{4}{5} S_{1}(N\!-\!3) \z_2
          - \frac{24}{5} S_{1}(N\!-\!2) \z_2
          + 28 S_{1}(N\!-\!1) \z_2
          - 12 S_{1}(N\!+\!1) \z_2
\nonumber\\
& &\hspace*{10mm}
          + \frac{136}{5} S_{1}(N\!+\!2) \z_2
          - \frac{36}{5} S_{1}(N\!+\!3) \z_2
          - 32 S_{1}(N) \z_2
          + \frac{8}{5} S_{1,-2}(N\!-\!3)
          - \frac{48}{5} S_{1,-2}(N\!-\!2)
\nonumber\\
& &\hspace*{10mm}
          + 56 S_{1,-2}(N\!-\!1)
          - 24 S_{1,-2}(N\!+\!1)
          + \frac{272}{5} S_{1,-2}(N\!+\!2)
          - \frac{72}{5} S_{1,-2}(N\!+\!3)
          - 64 S_{1,-2}(N)
\nonumber\\
& &\hspace*{10mm}
          - 8 S_{2}(N\!-\!1) \z_2
          + 24 S_{2}(N\!+\!1) \z_2
          - 16 S_{2}(N) \z_2
          - 16 S_{2,-2}(N\!-\!1)
          + 48 S_{2,-2}(N\!+\!1)
\nonumber\\
& &\hspace*{10mm}
          - 32 S_{2,-2}(N)
          \Biggr\}
\Biggr]
\, ,
\nonumber\\[2ex]
\lefteqn{
c^{(2),-{\rm{ns}}}_{3,{\rm{q}}}(N)  \,=\, c^{(2),-{\rm{ns}}}_{2,{\rm{q}}}(N) 
+  \d(N\!-\!1) \Biggl\{
        \Bigl(
          - \frac{9}{4}
          + 8 \z_2
          + \frac{64}{5} \z_2^2
          - 38 \z_3
          \Bigr) C_F \Bigl(C_F-\frac{C_A}{2}\Bigr) 
\Biggr\} }\\
& &\hspace*{20mm}
+ \d(N\!-\!2) \Biggl\{
        \Bigl(
            \frac{19}{270}
          + \frac{2}{3} \z_2
          - \frac{8}{5} \z_3
          \Bigr) C_F \Bigl(C_F-\frac{C_A}{2}\Bigr) 
          \Biggr\}
\,+\, \q(N\!-\!3)\, (-1)^N\, \times \nonumber \\
& &
\!\!\!\Biggl[
       C_F \Bigl(C_F-\frac{C_A}{2}\Bigr)   \Biggl\{
            8 S_{-3}(N\!-\!1)
          - 64 S_{-3}(N\!+\!1)
          + \frac{128}{5} S_{-3}(N\!+\!2)
          + \frac{72}{5} S_{-3}(N\!+\!3)
\nonumber\\
& &\hspace*{10mm}
          + 16 S_{-3}(N)
          - \frac{8}{5} S_{-2}(N\!-\!2)
          + \Bigl( \frac{64}{5} 
                 + 8 \z_2 \Bigr) S_{-2}(N\!-\!1)
          + \Bigl( \frac{224}{5} 
                 + 24 \z_2 \Bigr) S_{-2}(N\!+\!1)
\nonumber\\
& &\hspace*{10mm}
          + \frac{72}{5} S_{-2}(N\!+\!2)
          - \Bigl( \frac{352}{5} 
                 + 32 \z_2 \Bigr) S_{-2}(N) 
          - \frac{4}{5} S_{-1}(N\!-\!3) \z_2
          - \Bigl( \frac{8}{5} 
                 + \frac{16}{5} \z_2 \Bigr) S_{-1}(N\!-\!2)
\nonumber\\
& &\hspace*{10mm}
          - \Bigl( \frac{304}{5} 
                 + 20 \z_2 \Bigr) S_{-1}(N\!-\!1) 
          + \Bigl( \frac{304}{5}
                 + 28 \z_2 \Bigr) S_{-1}(N\!+\!1)
          - \Bigl( \frac{72}{5} 
                 + \frac{64}{5} \z_2 \Bigr) S_{-1}(N\!+\!2) 
\nonumber\\
& &\hspace*{10mm}
          - \frac{36}{5} S_{-1}(N\!+\!3) \z_2
          + \Bigl( 16 
                 + 16  \z_2 \Bigr) S_{-1}(N) 
          + \frac{4}{5} S_{1}(N\!-\!3) \z_2
          + \frac{16}{5} S_{1}(N\!-\!2) \z_2
\nonumber\\
& &\hspace*{10mm}
          + \Bigl( 12 \z_2
                 - 8 \z_3 \Bigr) S_{1}(N\!-\!1)
          + \Bigl( 36 \z_2
                 - 24 \z_3 \Bigr) S_{1}(N\!+\!1) 
          - \frac{64}{5} S_{1}(N\!+\!2) \z_2
\nonumber\\
& &\hspace*{10mm}
          - \frac{36}{5} S_{1}(N\!+\!3) \z_2
          - \Bigl( 32 \z_2
                 - 32 \z_3 \Bigr) S_{1}(N)
          - 16 S_{1,-3}(N\!-\!1)
          - 48 S_{1,-3}(N\!+\!1)
\nonumber\\
& &\hspace*{10mm}
          + 64 S_{1,-3}(N)
          + \frac{8}{5} S_{1,-2}(N\!-\!3)
          + \frac{32}{5} S_{1,-2}(N\!-\!2)
          + 24 S_{1,-2}(N\!-\!1)
          + 72 S_{1,-2}(N\!+\!1)
\nonumber\\
& &\hspace*{10mm}
          - \frac{128}{5} S_{1,-2}(N\!+\!2)
          - \frac{72}{5} S_{1,-2}(N\!+\!3)
          - 64 S_{1,-2}(N)
          + 16 S_{1,1}(N\!-\!1) \z_2
\nonumber\\
& &\hspace*{10mm}
          + 48 S_{1,1}(N\!+\!1) \z_2
          - 64 S_{1,1}(N) \z_2
          + 32 S_{1,1,-2}(N\!-\!1)
          + 96 S_{1,1,-2}(N\!+\!1)
\nonumber\\
& &\hspace*{10mm}
          - 128 S_{1,1,-2}(N)
          - 8 S_{2}(N\!-\!1) \z_2
          - 24 S_{2}(N\!+\!1) \z_2
          + 32 S_{2}(N) \z_2
          - 16 S_{2,-2}(N\!-\!1)
\nonumber\\
& &\hspace*{10mm}
          - 48 S_{2,-2}(N\!+\!1)
          + 64 S_{2,-2}(N)
          \Biggr\}
\Biggr]
\, ,
\nonumber\\[2ex]
\lefteqn{
c^{(2),{\rm{ps}}}_{2,{\rm{q}}}(N)  \,=\, \d(N\!-\!2) \Biggl\{
          - \frac{133}{81} C_F n_f
          \Biggr\}
\,+\, \q(N\!-\!3)\, (-1)^N\, \times } \\
& &
\!\!\!\Biggl[
       C_F n_f   \Biggl\{
            20 S_{-4}(N\!-\!1)
          + 20 S_{-4}(N\!+\!1)
          - 40 S_{-4}(N)
          + 2 S_{-3}(N\!-\!1)
          - \frac{26}{3} S_{-3}(N\!+\!1)
\nonumber\\
& &\hspace*{10mm}
          - \frac{64}{3} S_{-3}(N\!+\!2)
          + 28 S_{-3}(N)
          - 16 S_{-3,1}(N\!-\!1)
          - 16 S_{-3,1}(N\!+\!1)
          + 32 S_{-3,1}(N)
\nonumber\\
& &\hspace*{10mm}
          + 56 S_{-2}(N\!-\!1)
          - \frac{392}{9} S_{-2}(N\!+\!1)
          + \frac{128}{9} S_{-2}(N\!+\!2)
          - \frac{80}{3} S_{-2}(N)
          - 16 S_{-2,1}(N\!+\!1)
\nonumber\\
& &\hspace*{10mm}
          + 16 S_{-2,1}(N\!+\!2)
          + 8 S_{-2,1,1}(N\!-\!1)
          + 8 S_{-2,1,1}(N\!+\!1)
          - 16 S_{-2,1,1}(N)
          - 8 S_{-2,2}(N\!-\!1)
\nonumber\\
& &\hspace*{10mm}
          - 8 S_{-2,2}(N\!+\!1)
          + 16 S_{-2,2}(N)
          + \Bigl( \frac{344}{27} 
                 - \frac{8}{3} \z_2 \Bigr) S_{-1}(N\!-\!2) 
          - \Bigl( \frac{818}{27} 
                 + \frac{16}{3} \z_2 \Bigr) S_{-1}(N\!-\!1) 
\nonumber\\
& &\hspace*{10mm}
          + \Bigl( \frac{818}{27} 
                 + \frac{16}{3} \z_2 \Bigr) S_{-1}(N\!+\!1) 
          + \Bigl( \frac{448}{27} 
                 + \frac{8}{3} \z_2 \Bigr) S_{-1}(N\!+\!2) 
          - \frac{88}{3} S_{-1}(N)
\nonumber\\
& &\hspace*{10mm}
          - \frac{104}{9} S_{-1,1}(N\!-\!2)
          - \frac{208}{9} S_{-1,1}(N\!-\!1)
          + \frac{208}{9} S_{-1,1}(N\!+\!1)
          + \frac{32}{9} S_{-1,1}(N\!+\!2)
\nonumber\\
& &\hspace*{10mm}
          + 8 S_{-1,1}(N)
          + \frac{16}{3} S_{-1,1,1}(N\!-\!2)
          - \frac{28}{3} S_{-1,1,1}(N\!-\!1)
          + \frac{28}{3} S_{-1,1,1}(N\!+\!1)
\nonumber\\
& &\hspace*{10mm}
          - \frac{16}{3} S_{-1,1,1}(N\!+\!2)
          - \frac{16}{3} S_{-1,2}(N\!-\!2)
          + \frac{28}{3} S_{-1,2}(N\!-\!1)
          - \frac{28}{3} S_{-1,2}(N\!+\!1)
\nonumber\\
& &\hspace*{10mm}
          + \frac{16}{3} S_{-1,2}(N\!+\!2)
          - \frac{8}{3} S_{1}(N\!-\!2) \z_2
          + \frac{32}{3} S_{1}(N\!-\!1) \z_2
          + \frac{32}{3} S_{1}(N\!+\!1) \z_2
\nonumber\\
& &\hspace*{10mm}
          - \frac{8}{3} S_{1}(N\!+\!2) \z_2
          - 16 S_{1}(N) \z_2
          - \frac{16}{3} S_{1,-2}(N\!-\!2)
          + \frac{64}{3} S_{1,-2}(N\!-\!1)
          + \frac{64}{3} S_{1,-2}(N\!+\!1)
\nonumber\\
& &\hspace*{10mm}
          - \frac{16}{3} S_{1,-2}(N\!+\!2)
          - 32 S_{1,-2}(N)
          \Biggr\}
\Biggr]
\, ,
\nonumber\\[2ex]
\lefteqn{
c^{(2)}_{2,{\rm{g}}}(N)  \,=}           \\
& &
\d(N\!-\!2) \Biggl\{
       - \Bigl(\frac{4799}{810}
          - \frac{16}{5} \z_3
          \Bigr) C_F n_f
       + \Bigl(
           \frac{115}{324}
          - 2 \z_3
          \Bigr) C_A n_f   
          \Biggr\}
\,+\, \q(N\!-\!3)\, (-1)^N\, \times
\nonumber\\
& &
\!\!\!\Biggl[
       C_F n_f   \Biggl\{
          - 10 S_{-4}(N\!-\!1)
          - 20 S_{-4}(N\!+\!1)
          + 40 S_{-4}(N\!+\!2)
          - 10 S_{-4}(N)
          + 3 S_{-3}(N\!-\!1)
\nonumber\\
& &\hspace*{10mm}
          + \frac{172}{3} S_{-3}(N\!+\!1)
          - \frac{456}{5} S_{-3}(N\!+\!2)
          + \frac{96}{5} S_{-3}(N\!+\!3)
          + \frac{35}{3} S_{-3}(N)
          + 16 S_{-3,1}(N\!-\!1)
\nonumber\\
& &\hspace*{10mm}
          + 16 S_{-3,1}(N\!+\!1)
          - 48 S_{-3,1}(N\!+\!2)
          + 16 S_{-3,1}(N)
          + \frac{8}{15} S_{-2}(N\!-\!2)
\nonumber\\
& &\hspace*{10mm}
          - \Bigl( \frac{244}{15}
                 + 16 \z_2 \Bigr) S_{-2}(N\!-\!1)
          - \Bigl( \frac{103}{5} 
                 + 16 \z_2 \Bigr) S_{-2}(N\!+\!1)
          + \Bigl( \frac{216}{5} 
                 + 16 \z_2 \Bigr) S_{-2}(N\!+\!2) 
\nonumber\\
& &\hspace*{10mm}
          - \Bigl( \frac{103}{15} 
                 - 16 \z_2 \Bigr) S_{-2}(N) 
          - 16 S_{-2,1}(N\!-\!1)
          - 16 S_{-2,1}(N\!+\!1)
          + 72 S_{-2,1}(N\!+\!2)
\nonumber\\
& &\hspace*{10mm}
          - 40 S_{-2,1}(N)
          - 16 S_{-2,1,1}(N\!-\!1)
          - 8 S_{-2,1,1}(N\!+\!1)
          + 40 S_{-2,1,1}(N\!+\!2)
          - 16 S_{-2,1,1}(N)
\nonumber\\
& &\hspace*{10mm}
          + 12 S_{-2,2}(N\!-\!1)
          + 8 S_{-2,2}(N\!+\!1)
          - 32 S_{-2,2}(N\!+\!2)
          + 12 S_{-2,2}(N)
          + \frac{4}{15} S_{-1}(N\!-\!3) \z_2
\nonumber\\
& &\hspace*{10mm}
          + \Bigl( \frac{8}{15} 
                 - \frac{4}{15} \z_2 \Bigr) S_{-1}(N\!-\!2) 
          + \Bigl( \frac{213}{5} 
                 + 24 \z_2
                 + 16 \z_3 \Bigr) S_{-1}(N\!-\!1) 
\nonumber\\
& &\hspace*{10mm}
          - \Bigl( \frac{203}{5} 
                 + \frac{32}{3} \z_2
                 + 40 \z_3 \Bigr) S_{-1}(N\!+\!1) 
          - \Bigl( \frac{36}{5} 
                 - \frac{48}{5} \z_2
                 - 8 \z_3 \Bigr) S_{-1}(N\!+\!2) 
          - \frac{48}{5} S_{-1}(N\!+\!3) \z_2
\nonumber\\
& &\hspace*{10mm}
          + \Bigl( \frac{14}{3} 
                 - \frac{40}{3} \z_2
                 + 16 \z_3 \Bigr) S_{-1}(N) 
          + \Bigl( 14 
                 + 16 \z_2 \Bigr) S_{-1,1}(N\!-\!1)
          - \Bigl( 16 
                 + 16 \z_2 \Bigr) S_{-1,1}(N\!+\!1) 
\nonumber\\
& &\hspace*{10mm}
          - \Bigl( 24 
                 + 16 \z_2 \Bigr) S_{-1,1}(N\!+\!2)
          + \Bigl( 26 
                 + 16 \z_2 \Bigr) S_{-1,1}(N)
          + 26 S_{-1,1,1}(N\!-\!1)
\nonumber\\
& &\hspace*{10mm}
          - 8 S_{-1,1,1}(N\!+\!1)
          - 72 S_{-1,1,1}(N\!+\!2)
          + 54 S_{-1,1,1}(N)
          + 20 S_{-1,1,1,1}(N\!-\!1)
\nonumber\\
& &\hspace*{10mm}
          - 40 S_{-1,1,1,1}(N\!+\!2)
          + 20 S_{-1,1,1,1}(N)
          - 16 S_{-1,1,2}(N\!-\!1)
          + 32 S_{-1,1,2}(N\!+\!2)
\nonumber\\
& &\hspace*{10mm}
          - 16 S_{-1,1,2}(N)
          - 26 S_{-1,2}(N\!-\!1)
          + 8 S_{-1,2}(N\!+\!1)
          + 72 S_{-1,2}(N\!+\!2)
          - 54 S_{-1,2}(N)
\nonumber\\
& &\hspace*{10mm}
          - 24 S_{-1,2,1}(N\!-\!1)
          + 48 S_{-1,2,1}(N\!+\!2)
          - 24 S_{-1,2,1}(N)
          + 4 S_{-1,3}(N\!-\!1)
\nonumber\\
& &\hspace*{10mm}
          + 16 S_{-1,3}(N\!+\!1)
          - 24 S_{-1,3}(N\!+\!2)
          + 4 S_{-1,3}(N)
          - \frac{4}{15} S_{1}(N\!-\!3) \z_2
          + \frac{4}{15} S_{1}(N\!-\!2) \z_2
\nonumber\\
& &\hspace*{10mm}
          - \Bigl( 24 \z_2
                  - 8 \z_3 \Bigr) S_{1}(N\!-\!1) 
          - \Bigl( \frac{32}{3} \z_2
                  - 24 \z_3 \Bigr) S_{1}(N\!+\!1)
          + \Bigl( \frac{48}{5} \z_2
                 - 8 \z_3 \Bigr) S_{1}(N\!+\!2) 
\nonumber\\
& &\hspace*{10mm}
          - \frac{48}{5} S_{1}(N\!+\!3) \z_2
          + \Bigl( \frac{104}{3} \z_2
                 - 24 \z_3 \Bigr) S_{1}(N)
          + 16 S_{1,-3}(N\!-\!1)
          + 48 S_{1,-3}(N\!+\!1)
\nonumber\\
& &\hspace*{10mm}
          - 16 S_{1,-3}(N\!+\!2)
          - 48 S_{1,-3}(N)
          - \frac{8}{15} S_{1,-2}(N\!-\!3)
          + \frac{8}{15} S_{1,-2}(N\!-\!2)
          - 48 S_{1,-2}(N\!-\!1)
\nonumber\\
& &\hspace*{10mm}
          - \frac{64}{3} S_{1,-2}(N\!+\!1)
          + \frac{96}{5} S_{1,-2}(N\!+\!2)
          - \frac{96}{5} S_{1,-2}(N\!+\!3)
          + \frac{208}{3} S_{1,-2}(N)
\nonumber\\
& &\hspace*{10mm}
          - 16 S_{1,1}(N\!-\!1) \z_2
          - 48 S_{1,1}(N\!+\!1) \z_2
          + 16 S_{1,1}(N\!+\!2) \z_2
          + 48 S_{1,1}(N) \z_2
\nonumber\\
& &\hspace*{10mm}
          - 32 S_{1,1,-2}(N\!-\!1)
          - 96 S_{1,1,-2}(N\!+\!1)
          + 32 S_{1,1,-2}(N\!+\!2)
          + 96 S_{1,1,-2}(N)
\nonumber\\
& &\hspace*{10mm}
          + 16 S_{2}(N\!-\!1) \z_2
          + 16 S_{2}(N\!+\!1) \z_2
          - 16 S_{2}(N\!+\!2) \z_2
          - 16 S_{2}(N) \z_2
          + 32 S_{2,-2}(N\!-\!1)
\nonumber\\
& &\hspace*{10mm}
          + 32 S_{2,-2}(N\!+\!1)
          - 32 S_{2,-2}(N\!+\!2)
          - 32 S_{2,-2}(N)
          \Biggr\}
\nonumber\\
&+\!\!&
       C_A n_f   \Biggl\{
            20 S_{-4}(N\!-\!1)
          + 56 S_{-4}(N\!+\!1)
          - 76 S_{-4}(N)
          + 2 S_{-3}(N\!-\!1)
          - \frac{140}{3} S_{-3}(N\!+\!1)
\nonumber\\
& &\hspace*{10mm}
          - \frac{388}{3} S_{-3}(N\!+\!2)
          + 174 S_{-3}(N)
          - 8 S_{-3,1}(N\!-\!1)
          - 48 S_{-3,1}(N\!+\!1)
          - 16 S_{-3,1}(N\!+\!2)
\nonumber\\
& &\hspace*{10mm}
          + 72 S_{-3,1}(N)
          + 58 S_{-2}(N\!-\!1)
          - \Bigl( \frac{338}{9} 
                 + 8 \z_2 \Bigr) S_{-2}(N\!+\!1)
          + \Bigl( \frac{2090}{9} 
                 + 8 \z_2 \Bigr) S_{-2}(N\!+\!2) 
\nonumber\\
& &\hspace*{10mm}
          - \frac{758}{3} S_{-2}(N)
          - 8 S_{-2,1}(N\!-\!1)
          - 4 S_{-2,1}(N\!+\!1)
          + 148 S_{-2,1}(N\!+\!2)
          - 136 S_{-2,1}(N)
\nonumber\\
& &\hspace*{10mm}
          + 32 S_{-2,1,1}(N\!+\!1)
          + 16 S_{-2,1,1}(N\!+\!2)
          - 48 S_{-2,1,1}(N)
          - 32 S_{-2,2}(N\!+\!1)
\nonumber\\
& &\hspace*{10mm}
          - 16 S_{-2,2}(N\!+\!2)
          + 48 S_{-2,2}(N)
          + \Bigl( \frac{344}{27} 
                 - \frac{8}{3} \z_2 \Bigr) S_{-1}(N\!-\!2) 
\nonumber\\
& &\hspace*{10mm}
          - \Bigl( \frac{1061}{27} 
                 + \frac{28}{3} \z_2
                 + 14 \z_3 \Bigr) S_{-1}(N\!-\!1) 
          + \Bigl( \frac{1277}{27} 
                 + \frac{40}{3} \z_2
                 + 20 \z_3 \Bigr) S_{-1}(N\!+\!1) 
\nonumber\\
& &\hspace*{10mm}
          - \Bigl( \frac{4493}{27} 
                 + \frac{40}{3} \z_2
                 - 8 \z_3 \Bigr) S_{-1}(N\!+\!2) 
          + \Bigl( \frac{437}{3} 
                 + 12 \z_2
                 - 14 \z_3 \Bigr) S_{-1}(N) 
          - \frac{104}{9} S_{-1,1}(N\!-\!2)
\nonumber\\
& &\hspace*{10mm}
          - \Bigl( \frac{82}{9} 
                 + 4 \z_2 \Bigr) S_{-1,1}(N\!-\!1) 
          + \Bigl( \frac{208}{9} 
                 + 8 \z_2 \Bigr) S_{-1,1}(N\!+\!1) 
          - \frac{1570}{9} S_{-1,1}(N\!+\!2)
\nonumber\\
& &\hspace*{10mm}
          + \Bigl( 172 
                 - 4 \z_2 \Bigr) S_{-1,1}(N)
          + \frac{16}{3} S_{-1,1,1}(N\!-\!2)
          - \frac{4}{3} S_{-1,1,1}(N\!-\!1)
          + \frac{28}{3} S_{-1,1,1}(N\!+\!1)
\nonumber\\
& &\hspace*{10mm}
          - \frac{244}{3} S_{-1,1,1}(N\!+\!2)
          + 68 S_{-1,1,1}(N)
          + 4 S_{-1,1,1,1}(N\!-\!1)
          - 8 S_{-1,1,1,1}(N\!+\!2)
\nonumber\\
& &\hspace*{10mm}
\hspace*{-0.3pt}
          + 4 S_{-1,1,1,1}(N)
          - 12 S_{-1,1,2}(N\!-\!1)
          + 24 S_{-1,1,2}(N\!+\!2)
          - 12 S_{-1,1,2}(N)
          - \frac{16}{3} S_{-1,2}(N\!-\!2)
\nonumber\\
& &\hspace*{10mm}
          + \frac{4}{3} S_{-1,2}(N\!-\!1)
          - \frac{28}{3} S_{-1,2}(N\!+\!1)
          + \frac{268}{3} S_{-1,2}(N\!+\!2)
          - 76 S_{-1,2}(N)
          - 4 S_{-1,2,1}(N\!-\!1)
\nonumber\\
& &\hspace*{10mm}
          + 8 S_{-1,2,1}(N\!+\!2)
          - 4 S_{-1,2,1}(N)
          + 12 S_{-1,3}(N\!-\!1)
          - 8 S_{-1,3}(N\!+\!1)
          - 16 S_{-1,3}(N\!+\!2)
\nonumber\\
& &\hspace*{10mm}
          + 12 S_{-1,3}(N)
          - \frac{8}{3} S_{1}(N\!-\!2) \z_2
          + \Bigl( \frac{44}{3} \z_2
                 + 2 \z_3 \Bigr) S_{1}(N\!-\!1) 
          - \Bigl( \frac{40}{3} \z_2
                 - 12 \z_3 \Bigr) S_{1}(N\!+\!1) 
\nonumber\\
& &\hspace*{10mm}
          + \Bigl( \frac{40}{3} \z_2
                 - 8 \z_3 \Bigr) S_{1}(N\!+\!2) 
          - \Bigl( 12 \z_2
                 + 6 \z_3 \Bigr) S_{1}(N) 
          + 8 S_{1,-3}(N\!-\!1)
          + 40 S_{1,-3}(N\!+\!1)
\nonumber\\
& &\hspace*{10mm}
          - 24 S_{1,-3}(N\!+\!2)
          - 24 S_{1,-3}(N)
          - \frac{16}{3} S_{1,-2}(N\!-\!2)
          + \frac{88}{3} S_{1,-2}(N\!-\!1)
          - \frac{80}{3} S_{1,-2}(N\!+\!1)
\nonumber\\
& &\hspace*{10mm}
          + \frac{80}{3} S_{1,-2}(N\!+\!2)
          - 24 S_{1,-2}(N)
          - 8 S_{1,-2,1}(N\!-\!1)
          - 32 S_{1,-2,1}(N\!+\!1)
\nonumber\\
& &\hspace*{10mm}
          + 16 S_{1,-2,1}(N\!+\!2)
          + 24 S_{1,-2,1}(N)
          + 4 S_{1,1}(N\!-\!1) \z_2
          + 8 S_{1,1}(N\!+\!1) \z_2
          - 12 S_{1,1}(N) \z_2
\nonumber\\
& &\hspace*{10mm}
          + 8 S_{1,1,-2}(N\!-\!1)
          + 16 S_{1,1,-2}(N\!+\!1)
          - 24 S_{1,1,-2}(N)
          + 8 S_{2}(N\!+\!1) \z_2
          - 8 S_{2}(N\!+\!2) \z_2
\nonumber\\
& &\hspace*{10mm}
          + 16 S_{2,-2}(N\!+\!1)
          - 16 S_{2,-2}(N\!+\!2)
          \Biggr\}
\Biggr]
\, ,
\nonumber\\[2ex]
\lefteqn{
c^{(2),{\rm{ns}}}_{L,{\rm{q}}}(N)  \,=}\\
& & 
\d(N\!-\!2) \Biggl\{
          - \frac{92}{27} C_F n_f 
-  \Bigl(
            \frac{5756}{135}
          - \frac{64}{5} \z_3
          \Bigr) C_F \Bigl(C_F-\frac{C_A}{2}\Bigr)
          + \frac{770}{27} C_F^2
          \Biggr\}
\,+\, \q(N\!-\!3)\, (-1)^N\, \times
\nonumber\\
& &
\!\!\!\Biggl[
       C_F n_f   \Biggl\{
          - \frac{16}{3} S_{-2}(N\!+\!1)
          + \frac{16}{3} S_{-2}(N)
          - \frac{8}{3} S_{-1}(N\!-\!1)
          + \frac{100}{9} S_{-1}(N\!+\!1)
          - \frac{76}{9} S_{-1}(N)
\nonumber\\
& &\hspace*{10mm}
          + \frac{8}{3} S_{-1,1}(N\!+\!1)
          - \frac{8}{3} S_{-1,1}(N)
          \Biggr\}
\nonumber\\
&+\!\!&
       C_F \Bigl(C_F-\frac{C_A}{2}\Bigr)   \Biggl\{
            32 S_{-3}(N\!+\!1)
          + \frac{96}{5} S_{-3}(N\!+\!2)
          - \frac{96}{5} S_{-3}(N\!+\!3)
          - 32 S_{-3}(N)
\nonumber\\
& &\hspace*{10mm}
          + \frac{64}{5} S_{-2}(N\!-\!2)
          - \frac{32}{5} S_{-2}(N\!-\!1)
          - \frac{1216}{15} S_{-2}(N\!+\!1)
          - \frac{96}{5} S_{-2}(N\!+\!2)
          + \frac{1408}{15} S_{-2}(N)
\nonumber\\
& &\hspace*{10mm}
          + \frac{32}{5} S_{-1}(N\!-\!3) \z_2
          + \Bigl( \frac{64}{5} 
                 - \frac{32}{5} \z_2 \Bigr) S_{-1}(N\!-\!2) 
          - \Bigl( \frac{584}{15} 
                  - 32 \z_2 \Bigr) S_{-1}(N\!-\!1) 
\nonumber\\
& &\hspace*{10mm}
          + \Bigl( \frac{6052}{45} 
                 - 16 \z_2
                 - 80 \z_3 \Bigr) S_{-1}(N\!+\!1) 
          + \Bigl( \frac{96}{5} 
                 - \frac{48}{5} \z_2 \Bigr) S_{-1}(N\!+\!2)
          + \frac{48}{5} S_{-1}(N\!+\!3) \z_2
\nonumber\\
& &\hspace*{10mm}
          - \Bigl( \frac{1148}{9} 
                 + 16 \z_2
                 - 80 \z_3 \Bigr) S_{-1}(N) 
          + \Bigl( \frac{184}{3} 
                 - 32 \z_2 \Bigr) S_{-1,1}(N\!+\!1) 
          - \Bigl( \frac{184}{3} 
                 - 32 \z_2 \Bigr) S_{-1,1}(N) 
\nonumber\\
& &\hspace*{10mm}
          + 32 S_{-1,3}(N\!+\!1)
          - 32 S_{-1,3}(N)
          - \frac{32}{5} S_{1}(N\!-\!3) \z_2
          + \frac{32}{5} S_{1}(N\!-\!2) \z_2
          - 32 S_{1}(N\!-\!1) \z_2
\nonumber\\
& &\hspace*{10mm}
          - \Bigl( 16 \z_2
                 - 16 \z_3 \Bigr) S_{1}(N\!+\!1)
          - \frac{48}{5} S_{1}(N\!+\!2) \z_2
          + \frac{48}{5} S_{1}(N\!+\!3) \z_2
          + \Bigl( 48 \z_2
                 - 16 \z_3 \Bigr) S_{1}(N) 
\nonumber\\
& &\hspace*{10mm}
          + 32 S_{1,-3}(N\!+\!1)
          - 32 S_{1,-3}(N)
          - \frac{64}{5} S_{1,-2}(N\!-\!3)
          + \frac{64}{5} S_{1,-2}(N\!-\!2)
          - 64 S_{1,-2}(N\!-\!1)
\nonumber\\
& &\hspace*{10mm}
          - 32 S_{1,-2}(N\!+\!1)
          - \frac{96}{5} S_{1,-2}(N\!+\!2)
          + \frac{96}{5} S_{1,-2}(N\!+\!3)
          + 96 S_{1,-2}(N)
          - 32 S_{1,1}(N\!+\!1) \z_2
\nonumber\\
& &\hspace*{10mm}
          + 32 S_{1,1}(N) \z_2
          - 64 S_{1,1,-2}(N\!+\!1)
          + 64 S_{1,1,-2}(N)
          \Biggr\}
\nonumber\\
&+\!\!&
       C_F^2   \Biggl\{
          - 16 S_{-3}(N\!+\!1)
          + 16 S_{-3}(N)
          - 8 S_{-2}(N\!-\!1)
          + \frac{176}{3} S_{-2}(N\!+\!1)
          - \frac{152}{3} S_{-2}(N)
\nonumber\\
& &\hspace*{10mm}
          + 24 S_{-2,1}(N\!+\!1)
          - 24 S_{-2,1}(N)
          + \frac{52}{3} S_{-1}(N\!-\!1)
          - \frac{710}{9} S_{-1}(N\!+\!1)
          + \frac{554}{9} S_{-1}(N)
\nonumber\\
& &\hspace*{10mm}
          + 8 S_{-1,1}(N\!-\!1)
          - \frac{100}{3} S_{-1,1}(N\!+\!1)
          + \frac{76}{3} S_{-1,1}(N)
          - 16 S_{-1,1,1}(N\!+\!1)
          + 16 S_{-1,1,1}(N)
\nonumber\\
& &\hspace*{10mm}
          + 16 S_{-1,2}(N\!+\!1)
          - 16 S_{-1,2}(N)
          \Biggr\}
\Biggl]
\, ,
\nonumber
\\[2ex]
\lefteqn{
c^{(2),{\rm{ps}}}_{L,{\rm{q}}}(N) \,=\, \d(N\!-\!2) \Biggl\{
- \frac{80}{27} C_F n_f 
          \Biggr\}
\,+\, \q(N\!-\!3)\, (-1)^N\, \times
}\\
& &
\!\!\!\Biggl[ 
       C_F n_f   \Biggl\{
          - 32 S_{-3}(N\!+\!1)
          + 32 S_{-3}(N)
          + 16 S_{-2}(N\!-\!1)
          - 48 S_{-2}(N\!+\!1)
          + 32 S_{-2}(N\!+\!2)
\nonumber\\
& &\hspace*{10mm}
          + 16 S_{-2,1}(N\!+\!1)
          - 16 S_{-2,1}(N)
          - \frac{16}{9} S_{-1}(N\!-\!2)
          - \frac{32}{9} S_{-1}(N\!-\!1)
          + \frac{32}{9} S_{-1}(N\!+\!1)
\nonumber\\
& &\hspace*{10mm}
          + \frac{160}{9} S_{-1}(N\!+\!2)
          - 16 S_{-1}(N)
          - \frac{16}{3} S_{-1,1}(N\!-\!2)
          - \frac{32}{3} S_{-1,1}(N\!-\!1)
          + \frac{32}{3} S_{-1,1}(N\!+\!1)
\nonumber\\
& &\hspace*{10mm}
          - \frac{32}{3} S_{-1,1}(N\!+\!2)
          + 16 S_{-1,1}(N)
          \Biggr\}
\Biggl]
\, ,
\nonumber\\[2ex]
\lefteqn{
c^{(2)}_{L,{\rm{g}}}(N) \,=\, \d(N\!-\!2) \Biggl\{
-  \Bigl(
            \frac{116}{135}
          + \frac{16}{5} \z_3
          \Bigr) C_F n_f 
+ \frac{173}{27} C_A n_f 
          \Biggr\}
\,+\, \q(N\!-\!3)\, (-1)^N\, \times
}\\
& &
\!\!\!\!\Biggl[ 
          C_F n_f   \Biggl\{
            \frac{64}{3} S_{-3}(N\!+\!1)
          - \frac{64}{5} S_{-3}(N\!+\!2)
          + \frac{64}{5} S_{-3}(N\!+\!3)
          - \frac{64}{3} S_{-3}(N)
          + \frac{32}{15} S_{-2}(N\!-\!2)
\nonumber\\
& &\hspace*{10mm}
          - \frac{136}{15} S_{-2}(N\!-\!1)
          - \frac{112}{5} S_{-2}(N\!+\!1)
          - \frac{96}{5} S_{-2}(N\!+\!2)
          + \frac{728}{15} S_{-2}(N)
\nonumber\\
& &\hspace*{10mm}
          - 16 S_{-2,1}(N\!+\!1)
          + 16 S_{-2,1}(N)
          + \frac{16}{15} S_{-1}(N\!-\!3) \z_2
          + \Bigl( \frac{32}{15} 
                 - \frac{16}{15} \z_2 \Bigr) S_{-1}(N\!-\!2) 
\nonumber\\
& &\hspace*{10mm}
          + \frac{32}{5} S_{-1}(N\!-\!1)
          - \Bigl( \frac{32}{5} 
                 - \frac{16}{3} \z_2 \Bigr) S_{-1}(N\!+\!1)
          + \Bigl( \frac{336}{5} 
                 + \frac{32}{5} \z_2 \Bigr) S_{-1}(N\!+\!2)
\nonumber\\
& &\hspace*{10mm}
          - \frac{32}{5} S_{-1}(N\!+\!3) \z_2
          - \Bigl( \frac{208}{3} 
                 + \frac{16}{3} \z_2 \Bigr) S_{-1}(N) 
          + 8 S_{-1,1}(N\!-\!1)
          - 8 S_{-1,1}(N\!+\!1)
\nonumber\\
& &\hspace*{10mm}
          + 32 S_{-1,1}(N\!+\!2)
          - 32 S_{-1,1}(N)
          - \frac{16}{15} S_{1}(N\!-\!3) \z_2
          + \frac{16}{15} S_{1}(N\!-\!2) \z_2
          + \frac{16}{3} S_{1}(N\!+\!1) \z_2
\nonumber\\
& &\hspace*{10mm}
          + \frac{32}{5} S_{1}(N\!+\!2) \z_2
          - \frac{32}{5} S_{1}(N\!+\!3) \z_2
          - \frac{16}{3} S_{1}(N) \z_2
          - \frac{32}{15} S_{1,-2}(N\!-\!3)
          + \frac{32}{15} S_{1,-2}(N\!-\!2)
\nonumber\\
& &\hspace*{10mm}
          + \frac{32}{3} S_{1,-2}(N\!+\!1)
          + \frac{64}{5} S_{1,-2}(N\!+\!2)
          - \frac{64}{5} S_{1,-2}(N\!+\!3)
          - \frac{32}{3} S_{1,-2}(N)
          \Biggr\}
\nonumber\\
&+\!\!&
       C_A n_f   \Biggl\{
          - 96 S_{-3}(N\!+\!1)
          + 96 S_{-3}(N)
          + 16 S_{-2}(N\!-\!1)
          - 80 S_{-2}(N\!+\!1)
          + 208 S_{-2}(N\!+\!2)
\nonumber\\
& &\hspace*{10mm}
          - 144 S_{-2}(N)
          + 64 S_{-2,1}(N\!+\!1)
          + 32 S_{-2,1}(N\!+\!2)
          - 96 S_{-2,1}(N)
          - \frac{16}{9} S_{-1}(N\!-\!2)
\nonumber\\
& &\hspace*{10mm}
          - \frac{32}{9} S_{-1}(N\!-\!1)
          + \frac{32}{9} S_{-1}(N\!+\!1)
          - \Bigl( \frac{848}{9} 
                 + 16 \z_2 \Bigr) S_{-1}(N\!+\!2) 
          + \Bigl( 96 
                 + 16 \z_2 \Bigr) S_{-1}(N) 
\nonumber\\
& &\hspace*{10mm}
          - \frac{16}{3} S_{-1,1}(N\!-\!2)
          - \frac{32}{3} S_{-1,1}(N\!-\!1)      
          + \frac{32}{3} S_{-1,1}(N\!+\!1)
          - \frac{464}{3} S_{-1,1}(N\!+\!2)
\nonumber\\
& &\hspace*{10mm}
          + 160 S_{-1,1}(N)
          - 32 S_{-1,1,1}(N\!+\!2)
          + 32 S_{-1,1,1}(N)
          + 32 S_{-1,2}(N\!+\!2)
          - 32 S_{-1,2}(N)
\nonumber\\
& &\hspace*{10mm}
          - 32 S_{1}(N\!+\!1) \z_2
          + 16 S_{1}(N\!+\!2) \z_2
          + 16 S_{1}(N) \z_2 
          - 64 S_{1,-2}(N\!+\!1)
          + 32 S_{1,-2}(N\!+\!2)
\nonumber\\
& &\hspace*{10mm}
          + 32 S_{1,-2}(N)
          \Biggr\}
\Biggl] 
\nonumber\, .
\end{eqnarray}

\section*{Appendix C}

Here we present the $x$-space expressions of the 2-loop 
coefficient functions. We find,{\footnote{
The expression for the coefficient function $c^{(2),{\rm{ps}}}_{2,{\rm{q}}}$ 
in eq.(13) of ref.\cite{vanNeerven:1991nn}
contains a typographical error. The term $+488/27 x^2$ should read $+448/27 x^2$.}}

\begin{eqnarray}
\lefteqn{
c^{(2),+{\rm{ns}}}_{2,{\rm{q}}}(x)  \,=} \\
& &
         C_F C_A   \Biggl\{
          - \frac{3229}{180}
          - \frac{4}{5 x}
          + \frac{2191}{20} x
          - \frac{36}{5} x^2
          + 4 p_{\rm{qq}}(x) \Bigl( - \frac{3155}{432} + \frac{11}{3} \z_2 + \frac{1}{2} \z_3 
                - 3 H_{-2,0}(x) 
\nonumber\\
& &\hspace*{10mm}
                - \frac{239}{24} H_{0}(x) + H_{0}(x) \z_2 - \frac{55}{12} H_{0,0}(x) 
                - \frac{3}{2} H_{0,0,0}(x) 
                - \frac{367}{72} H_{1}(x) + 3 H_{1}(x) \z_2 
\nonumber\\
& &\hspace*{10mm}
                - \frac{11}{6} H_{1,0}(x) 
                - 2 H_{1,0,0}(x) 
                - \frac{11}{6} H_{1,1}(x) + H_{1,1,0}(x) - H_{1,2}(x) - \frac{11}{3} H_{2}(x) 
                - H_{3}(x)\Bigr)
\nonumber\\
& &\hspace*{10mm}
          - \frac{1}{6} (133 - 371 x) H_{1}(x)
          - 4 \Bigl(5 + \frac{1}{5 x^2} + x - 6 x^2 - \frac{9}{5} x^3\Bigr) H_{-1,0}(x)
          + \frac{1}{30} (13 + \frac{24}{x} 
\nonumber\\
& &\hspace*{10mm}
+ 1753 x - 216 x^2\Bigr) H_{0}(x)
          + 4 (1 - 5 x) \Bigl(\z_3 
                        + H_{1}(x) \z_2
                - H_{-2,0}(x) - H_{1,0,0}(x)\Bigr)
\nonumber\\
& &\hspace*{10mm}
          + 2 \Bigl(1 - 3 x + 12 x^2 + \frac{18}{5} x^3\Bigr) \z_2
          - 4 \Bigl(1 - x + 6 x^2 + \frac{9}{5} x^3\Bigr) H_{0,0}(x)
          - 4 (1 + x) H_{2}(x)
\nonumber\\
& &\hspace*{10mm}
          - \Bigl( \frac{5465}{72} + \frac{251}{3} \z_2 - \frac{71}{5} \z_2^2 
                - \frac{140}{3} \z_3\Bigr) \d(1 - x)
          \Biggr\}
\nonumber\\
&+\!\!&
         C_F n_f   \Biggl\{
          - \frac{23}{18}
          - \frac{27}{2} x
          + \frac{4}{3} p_{\rm{qq}}(x) \Bigl(\frac{247}{72} - 2 \z_2 + \frac{19}{4} H_{0}(x) 
                + \frac{5}{2} H_{0,0}(x)
          + \frac{29}{12} H_{1}(x) + H_{1,0}(x) 
\nonumber\\
& &\hspace*{10mm}
                + H_{1,1}(x) + 2 H_{2}(x)\Bigr)
          - \frac{1}{3} (7 + 19 x) H_{0}(x)
          - \frac{1}{3} (1 + 13 x) H_{1}(x)
\nonumber\\
& &\hspace*{10mm}
          + \Bigl(\frac{457}{36} + \frac{38}{3} \z_2 + \frac{4}{3} \z_3\Bigr) \d(1 - x)
          \Biggr\}
\nonumber\\
&+\!\!&
         C_F^2   \Biggl\{
            \frac{407}{20}
          + \frac{8}{5 x}
          - \frac{1917}{20} x
          + \frac{72}{5} x^2
          + 4 p_{\rm{qq}}(x) \Bigl(\frac{51}{16} + 3 \z_2 + 8 \z_3 + 6 H_{-2,0}(x) 
                        + \frac{61}{8} H_{0}(x) 
\nonumber\\
& &\hspace*{10mm}
                + 6 H_{0}(x) \z_2 - \frac{3}{4} H_{0,0}(x) - 2 H_{0,0,0}(x) 
                + \frac{27}{8} H_{1}(x) + 2 H_{1}(x) \z_2 - \frac{9}{2} H_{1,0}(x) 
\nonumber\\
& &\hspace*{10mm}
                - 3 H_{1,0,0}(x) 
                - \frac{9}{2} H_{1,1}(x) - 8 H_{1,1,0}(x) - 6 H_{1,1,1}(x) - 6 H_{1,2}(x) 
                - 3 H_{2}(x) - 6 H_{2,0}(x) 
\nonumber\\
& &\hspace*{10mm}
                - 7 H_{2,1}(x) - 6 H_{3}(x)\Bigr)
          + \Bigl(\frac{13}{10} - \frac{8}{5 x} - \frac{407}{10} x 
                + \frac{72}{5} x^2\Bigr) H_{0}(x)
          + \Bigl(\frac{91}{2} - \frac{141}{2} x\Bigr) H_{1}(x)
\nonumber\\
& &\hspace*{10mm}
          + 8 \Bigl(5 + \frac{1}{5 x^2} + x - 6 x^2 - \frac{9}{5} x^3\Bigr) H_{-1,0}(x)
          + \Bigl(29 + 25 x + 48 x^2 + \frac{72}{5} x^3\Bigr) H_{0,0}(x)
\nonumber\\
& &\hspace*{10mm}
          - 12 (1 - 3 x) \z_3
          + 8 (1 - 5 x) \Bigl(H_{-2,0}(x) + H_{1,0,0}(x)
                        -  H_{1}(x) \z_2\Bigr)
\nonumber\\
& &\hspace*{10mm}
          + 2 (1 + x) \Bigl(5 H_{0,0,0}(x) 
                        - 4 H_{0}(x) \z_2 
                + 7 H_{1,0}(x) + 2 H_{2,0}(x) 
                + 2 H_{2,1}(x) + 4 H_{3}(x)\Bigr)
\nonumber\\
& &\hspace*{10mm}
          - 24 \Bigl(1 + x + 2 x^2 + \frac{3}{5} x^3\Bigr) \z_2
          + 4 (7 + 11 x) H_{2}(x)
          + 2 (5 + 9 x) H_{1,1}(x)
\nonumber\\
& &\hspace*{10mm}
          + \Bigl(\frac{331}{8} + 69 \z_2 + 6 \z_2^2 - 78 \z_3\Bigr) \d(1 - x)
          \Biggr\}
\, ,
\nonumber\\[2ex]
\lefteqn{
c^{(2),-{\rm{ns}}}_{2,{\rm{q}}}(x)  \,=} \\
& &
         C_F \Bigl(C_F-\frac{C_A}{2}\Bigr)   \Biggl\{
          - \frac{162}{5}
          + \frac{8}{5 x}
          + \frac{82}{5} x
          + \frac{72}{5} x^2
          + 4 p_{\rm{qq}}( - x) \Bigl(7 \z_3 + 6 H_{-2,0}(x) - 8 H_{-1}(x) \z_2 
\nonumber\\
& &\hspace*{10mm}
                - 8 H_{-1,-1,0}(x) + 10 H_{-1,0,0}(x) + 4 H_{-1,2}(x) - 2 H_{0}(x) 
                + 2 H_{0}(x) \z_2 - 3 H_{0,0,0}(x) 
\nonumber\\
& &\hspace*{10mm}
- 2 H_{3}(x)\Bigr)
          + 32 \Bigl(1 + \frac{1}{20 x^2} + x + \frac{3}{2} x^2 
                - \frac{9}{20} x^3\Bigr) H_{-1,0}(x)
          - 16 (1 - x) H_{1}(x)
\nonumber\\
& &\hspace*{10mm}
          - \frac{2}{5} (13 + \frac{4}{x} + 53 x - 36 x^2\Bigr) H_{0}(x)
          - 8 (1 + x) H_{2}(x)
          - 8 \Bigl(1 + 4 x + 6 x^2 - \frac{9}{5} x^3\Bigr) H_{0,0}(x)
\nonumber\\
& &\hspace*{10mm}
          + 8 (1 + 5 x) \Bigl(\z_3 + H_{-2,0}(x) - 2 H_{-1,-1,0}(x) + H_{-1,0,0}(x)
                        - H_{-1}(x) \z_2 \Bigr)
\nonumber\\
& &\hspace*{10mm}
          + 4 \Bigl(1 + 7 x + 12 x^2 - \frac{18}{5} x^3\Bigr) \z_2
          \Biggr\}
\, ,
\nonumber\\[2ex]
\lefteqn{
c^{(2),+{\rm{ns}}}_{3,{\rm{q}}}(x)  \,=\, c^{(2),+{\rm{ns}}}_{2,{\rm{q}}}(x) \, +} \\
& &          C_F C_A   \Biggl\{
            \frac{701}{45}
          + \frac{4}{5 x}
          - \frac{3211}{45} x
          + \frac{36}{5} x^2
          + \frac{2}{3} (25 - 71 x) H_{1}(x)
          - 8 (1 - 3 x) \Bigl(\z_3 
                       + H_{1}(x) \z_2
\nonumber\\
& &\hspace*{10mm}
                - H_{-2,0}(x) - H_{1,0,0}(x)\Bigr)
          + 4 \Bigl(6 + \frac{1}{5 x^2} + \frac{1}{x} + 2 x - 5 x^2 
                - \frac{9}{5} x^3\Bigr) H_{-1,0}(x)
\nonumber\\
& &\hspace*{10mm}
          - \frac{4}{15} \Bigl(31 + \frac{3}{x} + 121 x - 27 x^2\Bigr) H_{0}(x)
          + 4 \Bigl(1 + 3 x - 5 x^2 - \frac{9}{5} x^3\Bigr) 
                \Bigl(\z_2 - H_{0,0}(x)\Bigr)
          \Biggr\}
\nonumber\\
&+\!\!&
         C_F n_f   \Biggl\{
            \frac{14}{9}
          + \frac{62}{9} x
          + \frac{4}{3} (1 + x) \Bigl(2 H_{0}(x) + H_{1}(x)\Bigr)
          \Biggr\}
\nonumber\\
&+\!\!&
         C_F^2   \Biggl\{
          - \frac{243}{5}
          - \frac{8}{5 x}
          + \frac{493}{5} x
          - \frac{72}{5} x^2
          - 2 (21 - 39 x) H_{1}(x)
          + \frac{8}{5} \Bigl(2 + \frac{1}{x} + \frac{49}{2} x - 9 x^2\Bigr) H_{0}(x)
\nonumber\\
& &\hspace*{10mm}
          - 8 \Bigl(6 + \frac{1}{5 x^2} + \frac{1}{x} + 2 x - 5 x^2 
                - \frac{9}{5} x^3\Bigr) H_{-1,0}(x)
          + 4 \Bigl(1 - 3 x + 10 x^2 + \frac{18}{5} x^3\Bigr) \z_2
\nonumber\\
& &\hspace*{10mm}
          + 16 (1 - 3 x) \Bigl(\z_3 
                        + H_{1}(x) \z_2
                - H_{-2,0}(x) - H_{1,0,0}(x)\Bigr)
\nonumber\\
& &\hspace*{10mm}
          - 8 (1 + x) \Bigl(H_{1,0}(x) + H_{1,1}(x) + \frac{3}{2} H_{2}(x)\Bigr)
          + 8 \Bigl(2 x - 5 x^2 - \frac{9}{5} x^3\Bigr) H_{0,0}(x)
          \Biggr\}
\, ,
\nonumber\\[2ex]
\lefteqn{
c^{(2),-{\rm{ns}}}_{3,{\rm{q}}}(x)  \,=\, c^{(2),-{\rm{ns}}}_{2,{\rm{q}}}(x) \, +} \\
& &         C_F \Bigl(C_F-\frac{C_A}{2}\Bigr)   \Biggl\{
            \frac{312}{5}
          - \frac{8}{5 x}
          - \frac{232}{5} x
          - \frac{72}{5} x^2
          - 8 \Bigl(4 + \frac{1}{5 x^2} - \frac{1}{x} + 4 x + 5 x^2 
                - \frac{9}{5} x^3) H_{-1,0}(x)
\nonumber\\
& &\hspace*{10mm}
          + \frac{8}{5} \Bigl(7 + \frac{1}{x} + 37 x - 9 x^2\Bigr) H_{0}(x)
          + 8 \Bigl(1 - 3 x - 5 x^2 + \frac{9}{5} x^3\Bigr) 
                        \Bigl(\z_2 - H_{0,0}(x)\Bigr)
\nonumber\\
& &\hspace*{10mm}
          - 16 (1 + 3 x) \Bigl(\z_3 
                        - H_{-1}(x) \z_2
                + H_{-2,0}(x) - 2 H_{-1,-1,0}(x) 
                + H_{-1,0,0}(x)\Bigr)
          \Biggr\}
\, ,
\nonumber\\[2ex]
\lefteqn{
c^{(2),{\rm{ps}}}_{2,{\rm{q}}}(x)  \,=}\\
& & 
         n_f C_F   \Biggl\{
            \frac{158}{9}
          + \frac{344}{27 x}
          - \frac{422}{9} x
          + \frac{448}{27} x^2
          + \frac{8}{3} \Bigl(13 - \frac{13}{3 x} - 10 x + \frac{4}{3} x^2\Bigr) H_{1}(x)
\nonumber\\
& &\hspace*{10mm}
          - 16 \Bigl(1 + \frac{1}{3 x} + x + \frac{1}{3} x^2\Bigr) H_{-1,0}(x)
          + 4 \Bigl(1 + \frac{4}{3 x} - x - \frac{4}{3} x^2\Bigr) 
                \Bigl(H_{1,0}(x) + H_{1,1}(x)\Bigr)
\nonumber\\
& &\hspace*{10mm}
          - 16 x^2 H_{2}(x)
          - 2 \Bigl(1 - 15 x + \frac{32}{3} x^2\Bigr) H_{0,0}(x)
          + 56 \Bigl(1 - \frac{11}{21} x - \frac{16}{63} x^2\Bigr) H_{0}(x)
\nonumber\\
& &\hspace*{10mm}
          - 8 (1 + x) \Bigl(\z_3 + 2 H_{0}(x) \z_2 - \frac{5}{2} H_{0,0,0}(x) - H_{2,0}(x) 
                - H_{2,1}(x) - 2 H_{3}(x)\Bigr)
\nonumber\\
& &\hspace*{10mm}
          - \frac{16}{3} \Bigl(\frac{1}{x} + 3 x - 3 x^2\Bigr) \z_2
          \Biggr\}
\, ,
\nonumber\\[2ex]
\lefteqn{
c^{(2)}_{2,{\rm{g}}}(x)  \,=}\\
& & 
         n_f C_F   \Biggl\{
          - \frac{647}{15}
          + \frac{8}{15 x}
          + \frac{239}{5} x
          - \frac{36}{5} x^2
          + 48 \Bigl(1 + \frac{1}{90 x^2} + \frac{4}{9} x + \frac{2}{5} x^3\Bigr) H_{-1,0}(x)
\nonumber\\
& &\hspace*{10mm}
          - \frac{4}{15} \Bigl(59 + \frac{2}{x} - \frac{339}{4} x + 162 x^2\Bigr) H_{0}(x)
          - 3 \Bigl(1 - \frac{44}{9} x + 24 x^2 + \frac{32}{5} x^3\Bigr) H_{0,0}(x)
\nonumber\\
& &\hspace*{10mm}
          - 8 \Bigl(2 - 7 x + 9 x^2\Bigr) H_{2}(x)
          - 2 \Bigl(13 - 40 x + 36 x^2\Bigr) \Bigl(H_{1,0}(x) + H_{1,1}(x)\Bigr)
\nonumber\\
& &\hspace*{10mm}
          - 2 \Bigl(7 - 20 x + 12 x^2\Bigr) H_{1}(x)
          + 16 \Bigl(1 - \frac{13}{6} x + \frac{9}{2} x^2 + \frac{6}{5} x^3\Bigr) \z_2
          + 8 (4 + 9 x^2) \z_3
\nonumber\\
& &\hspace*{10mm}
          + 32 (1 + x^2) H_{-2,0}(x)
          + 16 x^2 \Bigl(H_{-1}(x) \z_2 + 2 H_{-1,-1,0}(x) - H_{-1,0,0}(x) 
                + H_{0}(x) \z_2 
\nonumber\\
& &\hspace*{10mm}
          - \frac{5}{4} H_{0,0,0}(x) + H_{1}(x) \z_2 - H_{1,0,0}(x)
          - \frac{1}{2} H_{2,0}(x) - \frac{1}{2} H_{2,1}(x) - H_{3}(x)\Bigr)
\nonumber\\
& &\hspace*{10mm}
          - 16 \Bigl(H_{-1}(x) \z_2 + 2 H_{-1,-1,0}(x) - H_{-1,0,0}(x)\Bigr) p_{\rm{qg}}(- x)
          + 16 \Bigl(H_{0}(x) \z_2 - \frac{5}{8} H_{0,0,0}(x) 
\nonumber\\
& &\hspace*{10mm}
+ \frac{1}{2} H_{1}(x) \z_2 
                - \frac{1}{4} H_{1,0,0}(x) - H_{1,1,0}(x) - \frac{5}{4} H_{1,1,1}(x) 
                - \frac{3}{2} H_{1,2}(x)
                - \frac{3}{4} H_{2,0}(x) 
\nonumber\\
& &\hspace*{10mm}
- H_{2,1}(x) - H_{3}(x)\Bigr) p_{\rm{qg}}(x)
          \Biggr\}
\nonumber\\
&+\!\!&
         n_f C_A   \Biggl\{
            \frac{239}{9}
          + \frac{344}{27 x}
          + \frac{1072}{9} x
          - \frac{4493}{27} x^2
          - 4 \Bigl(1 - \frac{4}{3 x} - 20 x + \frac{67}{3} x^2\Bigr) H_{1,0}(x)
\nonumber\\
& &\hspace*{10mm}
          - 4 \Bigl(1 - \frac{4}{3 x} - 18 x + \frac{61}{3} x^2\Bigr) H_{1,1}(x)
          + 8 \Bigl(1 - \frac{2}{3 x} - 18 x + \frac{37}{2} x^2\Bigr) \z_2
\nonumber\\
& &\hspace*{10mm}
          + \frac{2}{3} \Bigl(31 - \frac{52}{3 x} + 227 x - \frac{785}{3} x^2\Bigr) H_{1}(x)
          - 24 \Bigl(1 + \frac{2}{9 x} - \frac{10}{9} x^2\Bigr) H_{-1,0}(x)
\nonumber\\
& &\hspace*{10mm}
          - 2 \Bigl(1 - 88 x + \frac{194}{3} x^2\Bigr) H_{0,0}(x)
          - 8 \Bigl(1 - 18 x + \frac{37}{2} x^2\Bigr) H_{2}(x)
          + 4 \Bigl(1 - 12 x + 6 x^2\Bigr) \z_3
\nonumber\\
& &\hspace*{10mm}
          + 4 (5 + 14 x) H_{0,0,0}(x)
          + 58 \Bigl(1 + \frac{292}{87} x - \frac{1045}{261} x^2\Bigr) H_{0}(x)
\nonumber\\
& &\hspace*{10mm}
          - 8 \Bigl(1 + 8 x - 2 x^2\Bigr) \Bigl(H_{0}(x) \z_2 - H_{3}(x)\Bigr)
          + 16 x (3 - x) \Bigl(H_{2,0}(x) + H_{2,1}(x)\Bigr)
\nonumber\\
& &\hspace*{10mm}
          + 16 x^2 \Bigl(H_{-2,0}(x) - \frac{1}{2} H_{-1}(x) \z_2 
                - H_{-1,-1,0}(x) 
                + \frac{1}{2} H_{-1,0,0}(x) - \frac{1}{2} H_{1}(x) \z_2 
\nonumber\\
& &\hspace*{10mm}
                + \frac{1}{2} H_{1,0,0}(x)\Bigr)
          - 4 \Bigl(H_{-1}(x) \z_2 - 2 H_{-1,-1,0}(x) - 2 H_{-1,0,0}(x) 
                - 2 H_{-1,2}(x)\Bigr) p_{\rm{qg}}( - x)
\nonumber\\
& &\hspace*{10mm}
          + 8 \Bigl(H_{1}(x) \z_2 - \frac{3}{2} H_{1,0,0}(x) - \frac{3}{2} H_{1,1,0}(x) 
                - \frac{1}{2} H_{1,1,1}(x) - \frac{1}{2} H_{1,2}(x)\Bigr) p_{\rm{qg}}(x)
          \Biggr\}
\, ,
\nonumber\\[2ex]
\lefteqn{
c^{(2),{\rm{ns}}}_{L,{\rm{q}}}(x)  \,=}\\
& & 
         n_f C_F   \Biggl\{
            \frac{8}{3}
          - \frac{100}{9} x
          - \frac{8}{3} x \Bigl(2 H_{0}(x) + H_{1}(x)\Bigr)
          \Biggr\}
\nonumber\\
&+\!\!&
         C_F \Bigl(C_F-\frac{C_A}{2}\Bigr)   \Biggl\{
            \frac{392}{15}
          + \frac{64}{5 x}
          - \frac{6916}{45} x
          + \frac{96}{5} x^2
          + 32 \Bigl(2 + \frac{2}{5 x^2} + x 
                - \frac{3}{5} x^3\Bigr) H_{-1,0}(x)
\nonumber\\
& &\hspace*{10mm}
          + \frac{32}{5} \Bigl(1 - \frac{2}{x} - \frac{47}{3} x + 3 x^2\Bigr) H_{0}(x)
          - \frac{184}{3} x H_{1}(x)
          + 32 x \Bigl(1 - \frac{3}{5} x^2\Bigr)  \Bigl(\z_2 - H_{0,0}(x) \Bigr)
\nonumber\\
& &\hspace*{10mm}
          + 32 x \Bigl(2 \z_3 
                 + \z_2 H_{1}(x) - \z_2 H_{-1}(x) 
          - 2 H_{-1,-1,0}(x) + H_{-1,0,0}(x) - H_{1,0,0}(x)  \Bigr)
          \Biggr\}
\nonumber\\
&+\!\!&
         C_F^2   \Biggl\{
          - \frac{52}{3}
          + \frac{710}{9} x
          - 8  \Bigl(1 - \frac{22}{3} x \Bigr) H_{0}(x)
          - 8  \Bigl(1 - \frac{25}{6} x \Bigr) H_{1}(x)
          - 16 x  \Bigl(\frac{3}{2} \z_2 
                - H_{0,0}(x)
\nonumber\\
& &\hspace*{10mm}
                - H_{1,0}(x) 
                - H_{1,1}(x) 
                - \frac{3}{2} H_{2}(x) \Bigr)
          \Biggr\}
\, ,
\nonumber\\[2ex]
\lefteqn{
c^{(2),{\rm{ps}}}_{L,{\rm{q}}}(x)  \,=}\\
& & 
         n_f C_F   \Biggl\{
            \frac{16}{3}
          - \frac{16}{9 x}
          - \frac{64}{3} x
          + \frac{160}{9} x^2
          + \frac{16}{3} \Bigl(3 - \frac{1}{x} - 2 x^2\Bigr) H_{1}(x)
          + 16  \Bigl(1 - x - 2 x^2 \Bigr) H_{0}(x)
\nonumber\\
& &\hspace*{10mm}
          - 16 x \Bigl(\z_2 - 2 H_{0,0}(x) - H_{2}(x) \Bigr)
          \Biggr\}
\, ,
\nonumber\\[2ex]
\lefteqn{
c^{(2)}_{L,{\rm{g}}}(x) \,=}\\
& &   
         n_f C_F   \Biggl\{
          - \frac{128}{15}
          + \frac{32}{15 x}
          - \frac{304}{5} x
          + \frac{336}{5} x^2
          - \frac{8}{15} \Bigl(13 + \frac{4}{x} + 78 x - 36 x^2\Bigr) H_{0}(x)
\nonumber\\
& &\hspace*{10mm}
          - 8 \Bigl(1 + 3 x - 4 x^2\Bigr) H_{1}(x)
          + \frac{32}{15} \Bigl(\frac{1}{x^2} - 5 x + 6 x^3\Bigr) H_{-1,0}(x)
          - \frac{64}{3} x \Bigl(1 + \frac{3}{5} x^2\Bigr) H_{0,0}(x)
\nonumber\\
& &\hspace*{10mm}
          + \frac{16}{3} x \Bigl(1 + \frac{12}{5} x^2\Bigr) \z_2
          - 16 x H_{2}(x)
          \Biggr\} 
\nonumber\\
&+\!\!&
         n_f C_A   \Biggl\{
            \frac{16}{3}
          - \frac{16}{9 x}
          + \frac{272}{3} x
          - \frac{848}{9} x^2
          + \frac{16}{3} \Bigl(3 - \frac{1}{x} + 27 x - 29 x^2\Bigr) H_{1}(x)
          - 32 x (2 - x) \z_2
\nonumber\\
& &\hspace*{10mm}
          + 16 \Bigl(1 + 8 x - 13 x^2\Bigr) H_{0}(x)
          + 32 x (1 - x) \Bigl(H_{1,0}(x) + H_{1,1}(x)\Bigr)
          + 32 x (3 - x) H_{2}(x)
\nonumber\\
& &\hspace*{10mm}
          + 32 x (1 + x) H_{-1,0}(x)
          + 96 x H_{0,0}(x)
          \Biggr\}
\nonumber        \, .
\end{eqnarray}

\end{document}